\documentclass[twocolumn]{aastex63}

\usepackage{IEEEtrantools} 
\usepackage{amsmath}

\newcolumntype{L}[1]{>{\raggedright\arraybackslash}p{#1}}
\newcolumntype{C}[1]{>{\centering\arraybackslash}p{#1}}
\newcolumntype{R}[1]{>{\raggedleft\arraybackslash}p{#1}}

\newcommand{\kepler}{\textit{Kepler}}

\newcommand{\gaia}{\textit{Gaia}}
\newcommand{\corot}{CoRoT}

\newcommand{\numax}{$\nu_{\rm max}$}
\newcommand{\dnu}{$\Delta \nu$}
\newcommand{\mstar}{$M_{\star}$}
\newcommand{\rstar}{$R_{\star}$}
\newcommand{\rhostar}{$\rho_{\star}$}
\newcommand{\lstar}{$L_{\star}$}
\newcommand{\teff}{$T_{\mathrm{eff}}$}
\newcommand{\logg}{$\log g$}

\newcommand{\vsini}{$v\sin i$}

\newcommand{\msol}{$M_{\odot}$}
\newcommand{\rsol}{$R_{\odot}$}

\newcommand{\lsol}{$L_{\odot}$}

\newcommand{\muhz}{$\mu$Hz}

\newcommand{\rom}[1]{{\scshape \romannumeral #1}}

\newcommand{\ambm}{$0.169 \pm 0.006$}
\newcommand{\ambr}{$0.19 \pm 0.01$}
\newcommand{\ambt}{$3054 \pm 44$}

\shorttitle{Asteroseismology of $\alpha$ Men A}
\shortauthors{Chontos et al.}

\begin{document}

\title{%
TESS Asteroseismology of $\alpha$ Mensae: Benchmark Ages for a G7 Dwarf and its M-dwarf Companion}

\author[0000-0003-1125-2564]{Ashley Chontos}
\altaffiliation{NSF Graduate Research Fellow}
\affiliation{Institute for Astronomy, University of Hawai'i, 2680 Woodlawn Drive, Honolulu, HI 96822, USA}

\author[0000-0001-8832-4488]{Daniel Huber}
\affiliation{Institute for Astronomy, University of Hawai'i, 2680 Woodlawn Drive, Honolulu, HI 96822, USA}

\author[0000-0002-2580-3614]{Travis A. Berger}
\affiliation{Institute for Astronomy, University of Hawai'i, 2680 Woodlawn Drive, Honolulu, HI 96822, USA}

\author[0000-0001-5222-4661]{Hans Kjeldsen}
\affiliation{Stellar Astrophysics Centre (SAC), Department of Physics and Astronomy, Aarhus University, Ny Munkegade 120, DK-8000 Aarhus C, Denmark}
\affiliation{Institute of Theoretical Physics and Astronomy, Vilnius University, Sauletekio av. 3, 10257 Vilnius, Lithuania}

\author[0000-0001-6359-2769]{Aldo M. Serenelli}
\affiliation{Institute of Space Sciences (ICE, CSIC) Campus UAB, Carrer de Can Magrans, s/n, E-08193, Barcelona, Spain}
\affiliation{Institut dEstudis Espacials de Catalunya (IEEC), C/Gran Capita, 2-4, E-08034, Barcelona, Spain}

\author[0000-0002-6137-903X]{Victor Silva Aguirre}
\affiliation{Stellar Astrophysics Centre (SAC), Department of Physics and Astronomy, Aarhus University, Ny Munkegade 120, DK-8000 Aarhus C, Denmark}


\author[0000-0002-4773-1017]{Warrick H. Ball}
\affiliation{School of Physics and Astronomy, University of Birmingham, Edgbaston, Birmingham, B15 2TT, UK}
\affiliation{Stellar Astrophysics Centre (SAC), Department of Physics and Astronomy, Aarhus University, Ny Munkegade 120, DK-8000 Aarhus C, Denmark}

\author[0000-0002-6163-3472]{Sarbani Basu}
\affiliation{Department of Astronomy, Yale University, PO Box 208101, New Haven, CT 06520-8101, USA}

\author[0000-0001-5222-4661]{Timothy R. Bedding}
\affiliation{Sydney Institute for Astronomy (SIfA), School of Physics, University of Sydney, NSW 2006, Australia}
\affiliation{Stellar Astrophysics Centre (SAC), Department of Physics and Astronomy, Aarhus University, Ny Munkegade 120, DK-8000 Aarhus C, Denmark}

\author[0000-0002-5714-8618]{William J. Chaplin}
\affiliation{School of Physics and Astronomy, University of Birmingham, Edgbaston, Birmingham, B15 2TT, UK}
\affiliation{Stellar Astrophysics Centre (SAC), Department of Physics and Astronomy, Aarhus University, Ny Munkegade 120, DK-8000 Aarhus C, Denmark}

\author[0000-0002-9879-3904]{Zachary R. Claytor}
\affiliation{Institute for Astronomy, University of Hawai`i, 2680 Woodlawn Drive, Honolulu, HI 96822, USA}

\author[0000-0001-8835-2075]{Enrico Corsaro}
\affiliation{INAF - Osservatorio Astrofisico di Catania, via S. Sofia 78, 95123, Catania, Italy}

\author[0000-0002-8854-3776]{Rafael A. Garcia}
\affiliation{IRFU, CEA, Universit\'e Paris-Saclay, F-91191 Gif-sur-Yvette, France}
\affiliation{AIM, CEA, CNRS, Universit\'e Paris-Saclay, Universit\'e Paris Diderot, Sorbonne Paris Cit\'e, F-91191 Gif-sur-Yvette, France}

\author[0000-0002-2532-2853]{Steve B. Howell}
\affiliation{NASA Ames Research Center, Moffett Field, CA, 94035, USA}

\author[0000-0002-8661-2571]{Mia S. Lundkvist}
\affiliation{Stellar Astrophysics Centre (SAC), Department of Physics and Astronomy, Aarhus University, Ny Munkegade 120, DK-8000 Aarhus C, Denmark}

\author[0000-0002-0129-0316]{Savita Mathur}
\affiliation{Intituto de Astrof\'isica de Canarias (IAC), E-38205 La Laguna, Tenerife, Spain}
\affiliation{Universidad de La Laguna (ULL), Departmento de Astrof\'isica, E-38206 La Laguna, Tenerife, Spain}

\author[0000-0003-4034-0416]{Travis S. Metcalfe}
\affiliation{Space Science Institute, 4765 Walnut St., Suite B, Boulder, CO 80301, USA}
\affiliation{White Dwarf Research Corporation, 3265 Foundry Pl., Unit 101, Boulder, CO 80301, USA}

\author[0000-0001-9169-2599]{Martin B. Nielsen}
\affiliation{School of Physics and Astronomy, University of Birmingham, Edgbaston, Birmingham, B15 2TT, UK}
\affiliation{Stellar Astrophysics Centre (SAC), Department of Physics and Astronomy, Aarhus University, Ny Munkegade 120, DK-8000 Aarhus C, Denmark}
\affiliation{Center for Space Science, NYUAD Institute, New York University Abu Dhabi, PO BOX 129188, Abu Dhabi, United Arab Emirates}

\author[0000-0001-7664-648X]{Jia Mian Joel Ong}
\affiliation{Department of Astronomy, Yale University, PO Box 208101, New Haven, CT 06520-8101, USA}

\author[0000-0002-9424-2339]{Zeynep \c{C}el\.ik Orhan}
\affiliation{Department of Astronomy and Space Sciences, Science Faculty, Ege University, 35100, Bornova, \.Izmir, Turkey}

\author[0000-0001-5759-7790]{S\.ibel \"Ortel}
\affiliation{Department of Astronomy and Space Sciences, Science Faculty, Ege University, 35100, Bornova, \.Izmir, Turkey}

\author[0000-0002-5082-6332]{Maissa Salama}
\affiliation{Institute for Astronomy, University of Hawai'i, 640 N Aohoku Pl \#209, Hilo, HI 96720, USA}

\author[0000-0002-3481-9052]{Keivan G. Stassun}
\affiliation{Vanderbilt University, Department of Physics \& Astronomy, 6301 Stevenson Center Lane, Nashville, TN 37235, USA}

\author[0000-0002-2522-8605]{R. H. D. Townsend}
\affiliation{Department of Astronomy, University of Wisconsin-Madison, 2535 Sterling Hall, 475 N. Charter Street, Madison, WI 53706, USA}
\affiliation{Kavli Institute for Theoretical Physics, University of California, Santa Barbara, CA 93106, USA}

\author[0000-0002-4284-8638]{Jennifer L. van Saders}
\affiliation{Institute for Astronomy, University of Hawai'i, 2680 Woodlawn Drive, Honolulu, HI 96822, USA}

\author{Mark Winther}
\affiliation{Stellar Astrophysics Centre (SAC), Department of Physics and Astronomy, Aarhus University, Ny Munkegade 120, DK-8000 Aarhus C, Denmark}

\author[0000-0002-7772-7641]{Mutlu Yildiz}
\affiliation{Department of Astronomy and Space Sciences, Science Faculty, Ege University, 35100, Bornova, \.Izmir, Turkey}


\author{R. Paul Butler}
\affiliation{Earth \& Planets Laboratory, Carnegie Institution for Science, 5241 Broad Branch Road NW, Washington, DC 20015, USA}

\author[0000-0002-7595-0970]{C. G. Tinney}
\affiliation{Exoplanetary Science at UNSW, School of Physics, UNSW Sydney, NSW 2052, Australia}

\author[0000-0001-9957-9304]{Robert A. Wittenmyer}
\affiliation{Centre for Astrophysics, University of Southern Queensland, USQ Toowoomba, QLD 4350, Australia}

\correspondingauthor{Ashley Chontos}
\email{achontos@hawaii.edu}


\begin{abstract}
\noindent
Asteroseismology of bright stars has become increasingly important as a method to determine fundamental properties (in particular ages) of stars. The \kepler\ Space Telescope initiated a revolution by detecting oscillations in more than 500 main-sequence and subgiant stars. However, most \kepler\ stars are faint, and therefore have limited constraints from independent methods such as long-baseline interferometry. Here, we present the discovery of solar-like oscillations in $\alpha$ Men A, a naked-eye (V$\,$=$\,$5.1) G7 dwarf in TESS's Southern Continuous Viewing Zone. Using a combination of astrometry, spectroscopy, and asteroseismology, we precisely characterize the solar analogue $\alpha$ Men A (\teff\ = 5569 $\pm$ 62 K, \rstar\ = 0.960 $\pm$ 0.016 \rsol, \mstar\ = 0.964 $\pm$ 0.045 \msol). To characterize the fully-convective M-dwarf companion, we derive empirical relations to estimate mass, radius and temperature given the absolute \gaia\ magnitude and metallicity, yielding \mstar\ = \ambm\ \msol, \rstar\ = \ambr\ \rsol\ and \teff\ = \ambt\ K. Our asteroseismic age of 6.2 $\pm$ 1.4 (stat) $\pm$ 0.6 (sys) Gyr for the primary places $\alpha$ Men B within a small population of M dwarfs with precisely measured ages. We combined multiple ground-based spectroscopy surveys to reveal an activity cycle of $P$ = 13.1 $\pm$ 1.1 years for $\alpha$ Men A, a period similar to that observed in the Sun. We used different gyrochronology models with the asteroseismic age to estimate a rotation period of $\sim$30 days for the primary. Alpha Men A is now the closest (d$\,$=$\,$10$\,$pc) solar analogue with a precise asteroseismic age from space-based photometry, making it a prime target for next-generation direct imaging missions searching for true Earth analogues.
\end{abstract}


\section{Introduction} \label{sec:intro}

Accurate ages are essential for stellar astrophysics but arguably the most difficult fundamental property to determine. Galactic archaeology uses stellar ages to reconstruct the formation history of the Milky Way Galaxy, while ages of exoplanet host stars are important to explain the diverse population of exoplanets observed today. Furthermore, ages will be important for next-generation space-based missions looking to image Earth-like planets orbiting Sun-like stars. For example, future imaging missions would greatly benefit from an age-based target selection when attempting to identify biosignatures in the context of exoplanet habitability \citep{bixel2020}.

There are many techniques to estimate stellar ages but no single method suitable for all spectral types \citep{soderblom2010}. The most widely used is isochrone fitting, which is most fruitful for stellar clusters, where the main-sequence turnoff provides an age for an ensemble of stars. Isochrones also typically produce reliable ages for massive stars ($\gtrsim$1.5$M_{\odot}$) or stars on the subgiant branch, for which stellar evolution is relatively quick. However, determining the ages of field stars is difficult, particularly for low-mass dwarfs that spend most of their lifetime on the main sequence. Consequently, many studies have focused on finding empirical relations between physically-motivated age indicators and other observables in lower main sequence stars. 

Early disk-integrated Ca \rom{2} H and K fluxes of the Sun revealed variations that correlated with the activity cycle,  leading to one of the first empirical age relations. Activity in the Sun is generated through the magnetic dynamo mechanism, whose efficiency depends on subsurface convection and differential rotation \citep{kraft1967}. Pioneering work by \citet{wilson1978} observed these two chromospheric emission lines for nearly 100 cool main-sequence stars and demonstrated that many stars have cyclic variations analogous to that found in the Sun. In addition, studies of open clusters revealed an inverse relationship between stellar age and activity \citep{wilson1963,wilson1966,skumanich1972,soderblom1991}. An empirical relation between chromospheric activity and age was established and would ultimately be the leading age indicator for later-type field stars for decades \citep{noyes1984a,baliunas1995,henry1996,wright2004}.

Another empirical relation uses stellar rotation periods to estimate ages based on the spindown of stars with time (gyrochronology). This mechanism is enabled by magnetic braking, where charged particles escape through magnetized winds, leading to mass and angular momentum loss \citep{skumanich1972}. Factoring in a mass (or color) dependence, \citet{barnes2007} derived an empirical rotation-age relation that successfully reproduced ages of young clusters to better than 20\%. Gyrochronology recently underwent a resurgence with \kepler\ \citep{borucki2010} through the measurement of rotation periods for more than 30,000 main-sequence stars \citep{nielsen2013,reinhold2013,mcquillan2014,santos2019}.

The success of empirical age relations makes it critical to verify them with independent calibrations. Recently, gyrochronology relations have failed to reproduce rotation rates for intermediate age clusters from K2, suggesting that the standard formalism needs to be adjusted \citep{curtis2019,douglas2019}. Moreover, \citet{vansaders2016} proposed a weakened braking law to explain the unexpected rapid rotation in older stars, indicating an additional source of uncertainty for rotation-based ages. This is further complicated for low-mass dwarfs that barely evolve over the nuclear time scale, and hence are also challenging to age through isochrones. Therefore, ages for lower main-sequence stars remain challenging and limited, which is largely due to the lack of calibrators in this regime.

\begin{figure*}
\centering
\includegraphics[width=\linewidth]{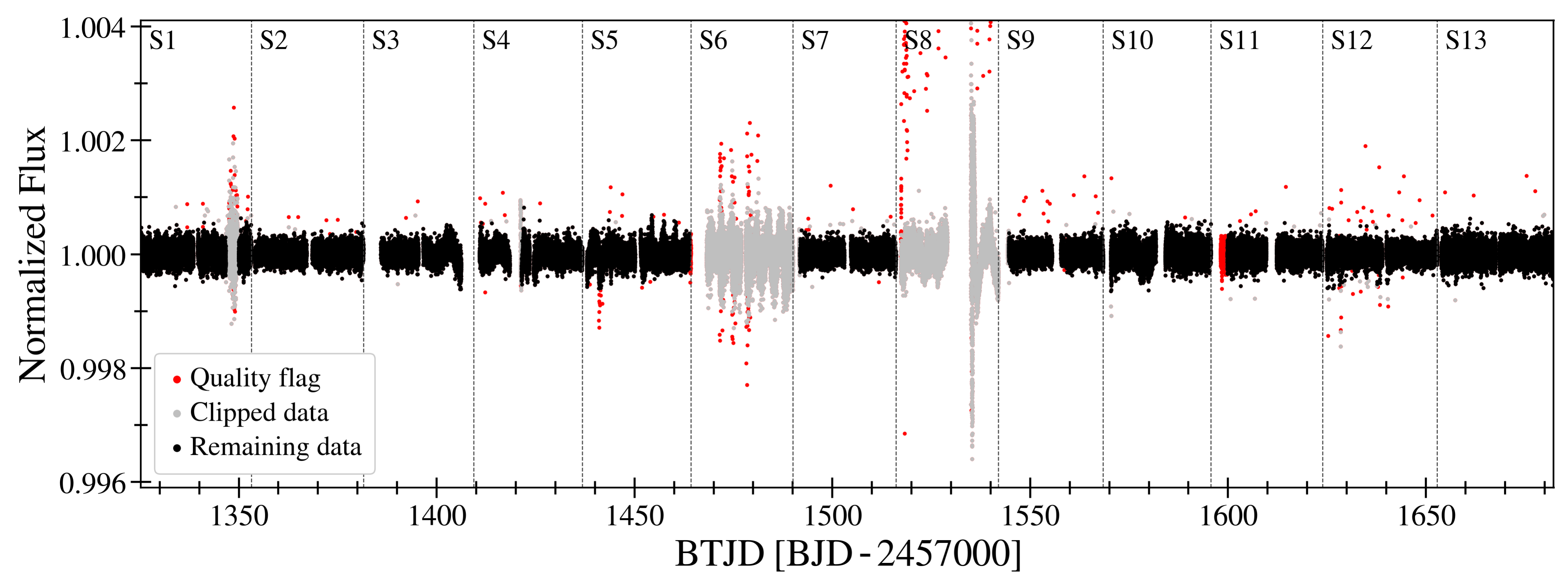}
\caption{Normalized TESS light curve of $\alpha$ Mensae A. Red points were removed based on quality flag information, grey points were clipped according to \cite{chontos2019}, and the remaining black points ($\sim$75\% of the original data) were used in the asteroseismic analysis. Dashed lines delineate the 13 sectors.}
\label{fig:LC}
\end{figure*}

A powerful method to determine accurate ages of field stars is asteroseismology, especially for solar-like oscillations driven by near-surface convection. \kepler\ revolutionized the field, but only detected oscillations in $\sim$500 main-sequence and subgiant stars, most of which are quite faint \citep{chaplin2014,garcia2019}. The Transiting Exoplanet Survey Satellite \citep[TESS;][]{ricker2015} is now targeting much brighter stars \citep[e.g.][]{chaplin2020} for which we also have long-term activity monitoring, enabling the opportunity to add benchmark calibrators for alternative age determination methods.

\begin{table}[]
\centering
\caption{Literature sources for spectroscopic \teff, \logg, and [Fe/H] values discussed in Section \ref{sec:spec}.}
\begin{tabular}{lccc}
\hline\hline
Source & \teff\ [K] & \logg\ [cgs] & [Fe/H] \\
\hline
\citet{santos2001}  & $5620$ & $4.56$ & $0.12$ \\
\citet{bensby2003} & $5550$ & $4.38$ &  $0.10$ \\
\citet{santos2004} & $5594$ & $4.41$ & $0.10$ \\
\citet{valenti2005} & $5587$ & $4.50$ & $0.09$ \\
\citet{bond2006} & $5557$ & $4.43$ & $0.06$ \\
\citet{ramirez2007} & $5536$ & $4.50$ & $0.12$ \\
\citet{bruntt2010} & $5570$ & $4.43$ & $0.15$ \\
\citet{casagrande2011} & $5605$ & & \\
\citet{dasilva2012} & $5630$ & $4.47$ & $0.11$ \\
\citet{maldonado2012} & $5649$ & $4.60$ & $0.12$ \\
\citet{ramirez2012} & $5569$ & $4.42$ & $0.11$ \\
\citet{bensby2014} & $5517$ & $4.48$ & $0.07$ \\
\citet{maldonaldo2015} & $5607$ & $4.51$ & \\
\citet{luck2018} & $5589$ & $4.44$ & $0.15$ \\
\hline
\end{tabular}
\label{tab:lit}
\end{table}

Here we present the discovery of solar-like oscillations in $\alpha$ Mensae using TESS, which is now the closest solar analog with an asteroseismic detection from space. Alpha Men A is a naked-eye G7 star (V = 5.1) in TESS's southern continuous viewing zone (SCVZ). It has an M-dwarf companion which we can now age-date using asteroseismology of the primary, making it an ideal target to age date two lower-main sequence stars and providing an invaluable nearby benchmark system.

\section{Observations} \label{sec:obs}

\subsection{TESS Photometry} \label{sec:tess}

Alpha Mensae falls in the TESS SCVZ and thus, was observed for the entire first year of the nominal mission. Alpha Men was observed in 2-minute cadence for all thirteen sectors, for a total baseline of 351 days. We used the light curve files produced by the TESS Science Processing Operations Center \citep[SPOC;][]{jenkins2016} that were made publicly available on the Mukulski Archive for Space Telescopes (MAST\footnote{\url{https://mast.stsci.edu/portal/Mashup/Clients/Mast/Portal.html}}).

We downloaded all SPOC 2-minute light curves and stitched individual sectors together using the SPOC-processed PDCSAP (Pre-Data Conditioning Standard Aperture Photometry) flux.  Upon initial inspection of the light curve (Figure \ref{fig:LC}), we noticed two sectors with increased scatter by a factor of at least two. We suspect that this is due to instrumental systematics and therefore removed these data before further analysis. To prepare the light curve for asteroseismic analysis, bad data points were removed as described in \citet{chontos2019}, including points with poor quality flags, $>$5$\sigma$ outliers, or sharp time domain artefacts, which ultimately accounted for $\sim$25\% of the data.

\subsection{High-Resolution Spectroscopy} \label{sec:spec}

Alpha Mensae is a well-studied star, with twenty-eight different sets of spectroscopic parameters available on Simbad\footnote{\url{http://simbad.u-strasbg.fr/simbad/}}. Retaining only results from  1980 onwards that used high-resolution instruments, a total of fourteen independent spectroscopic parameters remained and are listed in Table \ref{tab:lit}. We adopted the values from \cite{ramirez2012} and then added the standard deviation of literature values in quadrature with the reported formal uncertainties. The final set of atmospheric parameters for $\alpha$ Men A is \teff\ = 5569 $\pm$ 50 (stat) $\pm$ 36 (sys) K, \logg\ = 4.42 $\pm$ 0.03 (stat) $\pm$ 0.06 (sys) dex, and [Fe/H] = 0.11 $\pm$ 0.05 (stat) $\pm$ 0.03 (sys) dex (Table \ref{tab:primary}). We also checked the alpha abundances (in particular [Mg/Fe] and [Ca/Fe]) and found them to be consistent with solar values \citep{bensby2014}.

\begin{table}
\setlength{\tabcolsep}{0pt}
\begin{center}
\caption{Primary Stellar Parameters}\label{tab:primary}
\vspace{-0.3cm}
\begin{tabular}{lcc}
\hline\hline
\noalign{\smallskip}
\multicolumn{3}{c}{Other identifiers:} \\
\multicolumn{3}{c}{alf Men, HR 2261, HD 43834,} \\
\multicolumn{3}{c}{HIP 29271, \gaia\ DR2 5264749303461634816} \\
\multicolumn{3}{c}{\gaia\ eDR3 5264749303462961280, TIC 141810080} \\
\noalign{\smallskip}
\hline\hline
\noalign{\smallskip}
Parameter & Value & Source \\
\noalign{\smallskip}
\hline
\noalign{\smallskip}
Right ascension (RA), $\rm\alpha_{J2016}$	& $92.5624$ & 1, 2 \\
Declination (Dec), $\rm\delta_{J2016}$	& $-74.7540$  & 1, 2 \\
Parallax, $\pi$ (mas) & $97.9158 \pm 0.0573$ & 1, 2 \\
Distance, $d$ (pc) & $10.2129 \pm 0.0060$ & 1, 2 \\
Spectral type & G7V & 3 \\
\noalign{\smallskip}
\hline
\noalign{\smallskip}
\multicolumn{3}{c}{Photometry} \\
\noalign{\smallskip}
\hline
\noalign{\smallskip}
Tycho-2 B mag, $B_T$ & $5.968 \pm 0.014 $ & 4 \\
Tycho-2 V mag, $V_T$ & $5.151 \pm 0.009 $ & 4 \\
2MASS J mag, $J$ & $3.951 \pm 0.232$ & 5, 6 \\
2MASS H mag, $H$ & $3.508 \pm 0.228$ & 5, 6 \\
2MASS K$_{S}$ mag, $K_S$ & $3.446 \pm 0.200$ & 5, 6 \\
\gaia\ G mag, $G^{\dagger}$ & $4.8973 \pm 0.0025$ & 1, 2 \\
\gaia\ Bp mag, $Bp$ & $5.2783 \pm 0.0024$ & 1, 2 \\
\gaia\ Rp mag, $Rp$ & $4.3900 \pm 0.0023$ & 1, 2 \\
\noalign{\smallskip}
\hline
\noalign{\smallskip}
\multicolumn{3}{c}{Spectroscopy \& Gaia} \\
\noalign{\smallskip}
\hline
\noalign{\smallskip}
Effective temperature, \teff\ (K) & $5569 \pm 50(36)$ & 8 \\
Metallicity, [Fe/H] (dex) & $0.11 \pm 0.05(0.03) $ & 8 \\
Surface gravity, \logg\ (cgs) & $4.42 \pm 0.03(0.06)$ & 8 \\
Projected rotation speed, \vsini\ (km\,s$^{-1}$) & $0.6 \pm 0.6$ & 9 \\
Luminosity, \lstar\ (\lsol) &$0.81 \pm 0.02$ & \\
\noalign{\smallskip}
\hline
\noalign{\smallskip}
\multicolumn{3}{c}{Asteroseismology} \\
\noalign{\smallskip}
\hline
\noalign{\smallskip}
Stellar mass, \mstar\ (\msol)& $0.964 \pm 0.037 (0.026)$ & \\
Stellar radius, \rstar\ (\rsol)& $0.960 \pm 0.013 (0.009)$ & \\
Stellar density, \rhostar\ (gcc)& $1.531 \pm 0.018 (0.011)$ & \\
Surface gravity, \logg\ (cgs) & $4.459 \pm 0.006 (0.004)$ & \\
Age, $t$ (Gyr) & $6.2 \pm 1.4 (0.6)$ & \\
\noalign{\smallskip}
\hline
\noalign{\smallskip}
\end{tabular}
\textbf{References --} (1) \citet{gaia2016} (2) \citet{gaia2021} (3) \citet{gray2006} (4) \citet{hog2000} (5) \citet{cutri2003} (6) \citet{skrutskie2006} (7) \citet{stassun2019} (8) \citet{ramirez2012} (9) \citet{bruntt2010} \\
\end{center}
\vspace{-0.2cm}
{\sc \textbf{Note ---}} \\
$^{\dagger}$ Magnitude has been corrected for saturation according to \citet{evans2018}.
\end{table}

\subsection{Broadband Photometry \& \gaia\ Parallax} 
\label{sec:gaia}

Due to its brightness (V = 5.1) $\alpha$ Men A is saturated in many large photometric surveys. For optical magnitudes, we relied on $B_T$ and $V_T$ magnitudes from the Tycho-2 catalog \citep{hog2000}. \citet{cutri2012} reported reliable quality flags for $J,H,K_S$ photometry in the extended 2MASS catalog, although we note the higher uncertainties due to the choice of aperture needed for saturated stars. The European Space Agency's (ESA) \gaia\ Data Release 2 \citep[DR2\footnote{\url{https://www.cosmos.esa.int/web/gaia/home}};][]{gaia2018} estimated a mean \gaia\ magnitude $G$ = 4.850 for alpha Men. Brighter sources in DR2 photometry ($G$ $<$ 6) suffer from systematic errors due to saturation \citep{evans2018}. Using the empirical correction in \citet{evans2018}, we calculated a corrected \gaia\ magnitude, $G$ = 4.897. The new \gaia\ eDR3 catalog reported a value of $G$ = 4.900 for the primary and is therefore consistent with the corrected magnitude used in this analysis \citep{gaia2021,riello2021}.

\begin{figure*}
\centering
\includegraphics[width=\linewidth]{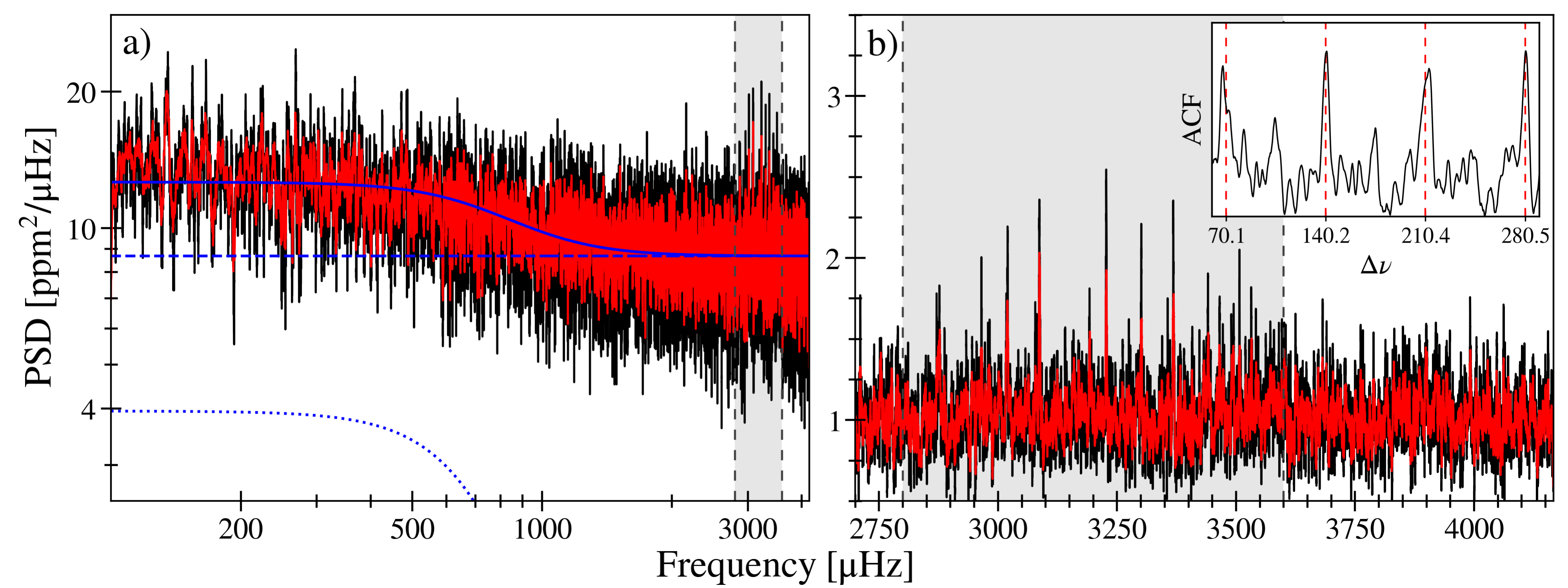}
\caption{\textit{Panel a):} Power spectrum of $\alpha$ Men A using box filters of 1.0 and 2.5 \muhz\ shown in black and red, respectively. The total background fit from \texttt{DIAMONDS} due to stellar contribution is shown by a solid blue line, which is the combination of a meso-granulation component (blue dotted line) and a white noise component (blue dashed line). \textit{Panel b):} The background-corrected power spectrum is centered on the power excess due to stellar oscillations, highlighted in the shaded region. The power in this region is used to calculate an autocorrelation (ACF), shown in the inset. Dashed lines in the inset represent expected peaks in the ACF due to the characteristic spacings of p-mode oscillations.}
\label{fig:PS}
\end{figure*}

Using the Tycho-2 $B_T$ and $V_T$ magnitudes, we derived two luminosities with \texttt{isoclassify}\footnote{\url{https://github.com/danxhuber/isoclassify}} \citep{huber2017}. The magnitude was combined with the Gaia parallax, bolometric corrections from MIST isochrones \citep{choi2016} and the composite reddening map \texttt{mwdust}\footnote{\url{https://github.com/jobovy/mwdust}} \citep{bovy2016}, yielding L$_{\star}$ = 0.800 $\pm$ 0.008 L$_{\odot}$ ($B_T$) and L$_{\star}$ = 0.812 $\pm$ 0.007 L$_{\odot}$ ($V_T$).

As an independent check on the derived luminosity, we analyzed the broadband spectral energy distribution (SED) together with the \gaia\ DR2 parallax following the procedures described by \citet{stassun2017,Stassun2018a}. We took the NUV flux from {\it GALEX}, the Johnson $U,B,V$ magnitudes from \citet{mermilliod2006}, the Str\"omgren $u,v,b,y$ magnitudes from \citet{paunzen2015}, the $B_T,V_T$ magnitudes from {\it Tycho-2}, the $J,H,K_S$ magnitudes from {\it 2MASS}, the W1--W4 magnitudes from {\it WISE}, and the $G,G_{BP},G_{RP}$ magnitudes from \gaia. Together, the available photometry spans a wavelength range 0.2--22~$\mu$m. 

We performed a fit using Kurucz stellar atmosphere models, adopting the effective temperature ($T_{\rm eff}$), surface gravity ($\log g$), and metallicity ([Fe/H]) from the spectroscopically determined values. The extinction ($A_V$) was set to zero due to the star being very nearby. The resulting fit has a reduced $\chi^2$ of 2.4. Integrating the (unreddened) model SED gives the bolometric flux at Earth of $F_{\rm bol} =  2.494 \pm 0.058 \times 10^{-7}$ erg~s$^{-1}$~cm$^{-2}$. Taking $F_{\rm bol}$ and $T_{\rm eff}$ together, with the \gaia\ parallax (adjusted by $+0.08$~mas to account for the systematic offset reported by \citealt{stassun2018b}), gives the stellar radius $R = 0.968 \pm 0.021$~R$_\odot$ and bolometric luminosity $L_{\rm bol} = 0.810 \pm 0.019$~L$_\odot$. We performed an additional fit excluding the \gaia\ magnitudes to test if the known systematics affected the derived properties but the results were unchanged. The derived values from the SED fit are in good agreement with those derived using \texttt{isoclassify}.

Similar to the method discussed in Section \ref{sec:spec} for the spectroscopic parameters, we performed a literature search for \gaia- and Hipparcos-derived luminosities to account for systematic differences. We used  bolometric luminosities from seven independent studies \citep{bruntt2010,casagrande2011,eiroa2013,heller2017,mcdonald2017,stevens2017,schofield2019} along with the three luminosities derived in our study to determine the scatter in the values. We adopted the \texttt{isoclassify} result using the Tycho-2 $V_T$ magnitude and added the standard deviation of the ten values ($\sigma_{\rm{L}_{\star}}$= 0.018 L$_{\odot}$) in quadrature with our derived uncertainty ($\sigma_{\rm{L}_{\star}}$= 0.007 L$_{\odot}$) yielding a bolometric luminosity L$_{\star}$ = 0.81 $\pm$ 0.02 L$_{\odot}$. The median value for the ten luminosities was slightly higher at L$_{\star}$ = 0.828, but is within 1$\sigma$ of our final reported value.

\section{Asteroseismology} \label{sec:astero}

\subsection{Background Fit} \label{sec:bg}

A high-pass filter was applied to the TESS 2-minute light curve (Figure \ref{fig:LC}) to remove long-period trends. The power spectrum was then calculated using a Lomb-Scargle periodogram \citep{lomb1976,scargle1982} through the publicly available \texttt{astropy} package. The power spectrum in Figure \ref{fig:PS}a shows a flat white noise component and a correlated red noise signal that rises at lower frequencies, indicative of stellar granulation. A roughly Gaussian power excess due to oscillations is clearly visible at $\sim$3200 \muhz\ (Figure \ref{fig:PS}b).

A common approach to model power spectra of solar-like oscillators typically has the form

\begin{equation} \label{equ:ps}
f(\nu) = W + R(\nu)[B(\nu)+G(\nu)],
\end{equation}
where $f(\nu)$ is the power density at frequency $\nu$ \citep{mathur2011,corsaro2018}. The frequency-independent term ($W$) is due to photon noise. The response function, $R(\nu)$, is an attenuation factor that affects the observed spectral amplitudes due to the sampling rate (or cadence) in a time series. The attenuation is greater for oscillations that occur near the Nyquist frequency, which for TESS 2-minute data is $\rm\nu_{Nyq}$ = 4166.67 $\rm\mu$Hz. The last two terms in Equation \ref{equ:ps} refer to contributions from the stellar granulation background $B(\nu)$ and the Gaussian envelope of oscillations $G(\nu)$. 

To determine the stellar background contribution, we used the publicly available \texttt{Background}\footnote{\url{https://github.com/EnricoCorsaro/Background}}, which is a software extension of \texttt{DIAMONDS}\footnote{\url{https://github.com/EnricoCorsaro/DIAMONDS}}. Initially created for more robust asteroseismic analyses, \texttt{DIAMONDS} is a nested sampling Monte Carlo (NSMC) algorithm for Bayesian parameter estimation and model comparison \citep{corsaro2014}. The background model built into this framework has the functional form:

\begin{equation} \label{equ:bgmodel}
B(\nu) = \zeta \sum_{i = 1}^{n}
\frac{\sigma_{i}^{2}/\nu_{i}^{}}{1 + (\nu/\nu_{i})^{4}},
\end{equation}
where $\zeta$ is a normalization factor ($2 \sqrt{2}/\pi$), $\sigma_{i}$ is the amplitude and $\nu_{i}$ is the characteristic frequency for $n$ Harvey-like components \citep{harvey1985}. Different stellar background contributions like granulation and meso-granulation have typical characteristic frequencies of $\sim$ \hspace{-0.63em} $\nu_{\mathrm{max}}$ and $\sim$ \hspace{-0.63em} $\nu_{\mathrm{max}}/3$ for solar-like oscillators \citep{corsaro2017}.

\begin{table}
\begin{center}
\caption{\texttt{DIAMONDS} background fit using $n=1$ Harvey-like component. Values are calculated by taking the median $\pm1\sigma$ (credible level of 68.3\%) from each parameter posterior distribution.} 
\label{tab:bg}
\begin{tabular}{l c}   
\hline\hline
Parameter & Value \\
\hline
White noise, $W$ & 8.67 $\pm$ 0.05 ppm$^2$ \muhz$^{-1}$ \\
Meso-granulation timescale, $\tau_{\rm{meso}}$ & 20.6 $\pm$ 0.8 minutes\\
Meso-granulation amplitude, $\sigma_{\rm{meso}}$ & 59.5 $\pm$ 1.1 ppm \\
Gaussian height, $H_{\rm{osc}}$ & 0.145 $\pm$ 0.129 ppm$^2$ \muhz$^{-1}$ \\
Gaussian center, \numax\ & 3134.28 $\pm$ 439.91 \muhz \\
Gaussian width, $\sigma_{\rm{osc}}$ & 403 $\pm$ 279 \muhz \\
\hline         
\end{tabular}
\end{center}
\end{table}

We used the following configuration for the NSMC analysis: shrinking rate, $\alpha$ = 0.02; enlargement fraction, $f_0$ = 1.43; number of live points, $N_{\rm{live}}$ = 500; number of clusters, 3 $\leq$ $N_{\rm{clust}}$ $\leq$ 6; max attempts when drawing a new sampling point, $M_{\rm{attempts}}$ = 5$\times10^4$; initial number of live points, $M_{\rm{init}}$ = $N_{\rm{live}}$; clustering only happens every N iterations or $M_{\rm{same}}$ = 50. Aside from minor changes to the shrinking rate ($\alpha$) and enlargement fraction ($f_0$), which control the sampling efficiency based on the number of free parameters in the model, the other parameters were the same as what was provided in the \texttt{DIAMONDS} documentation. We refer the reader to \citet{corsaro2014} for more details about the software.

\begin{figure}
\includegraphics[width=\linewidth]{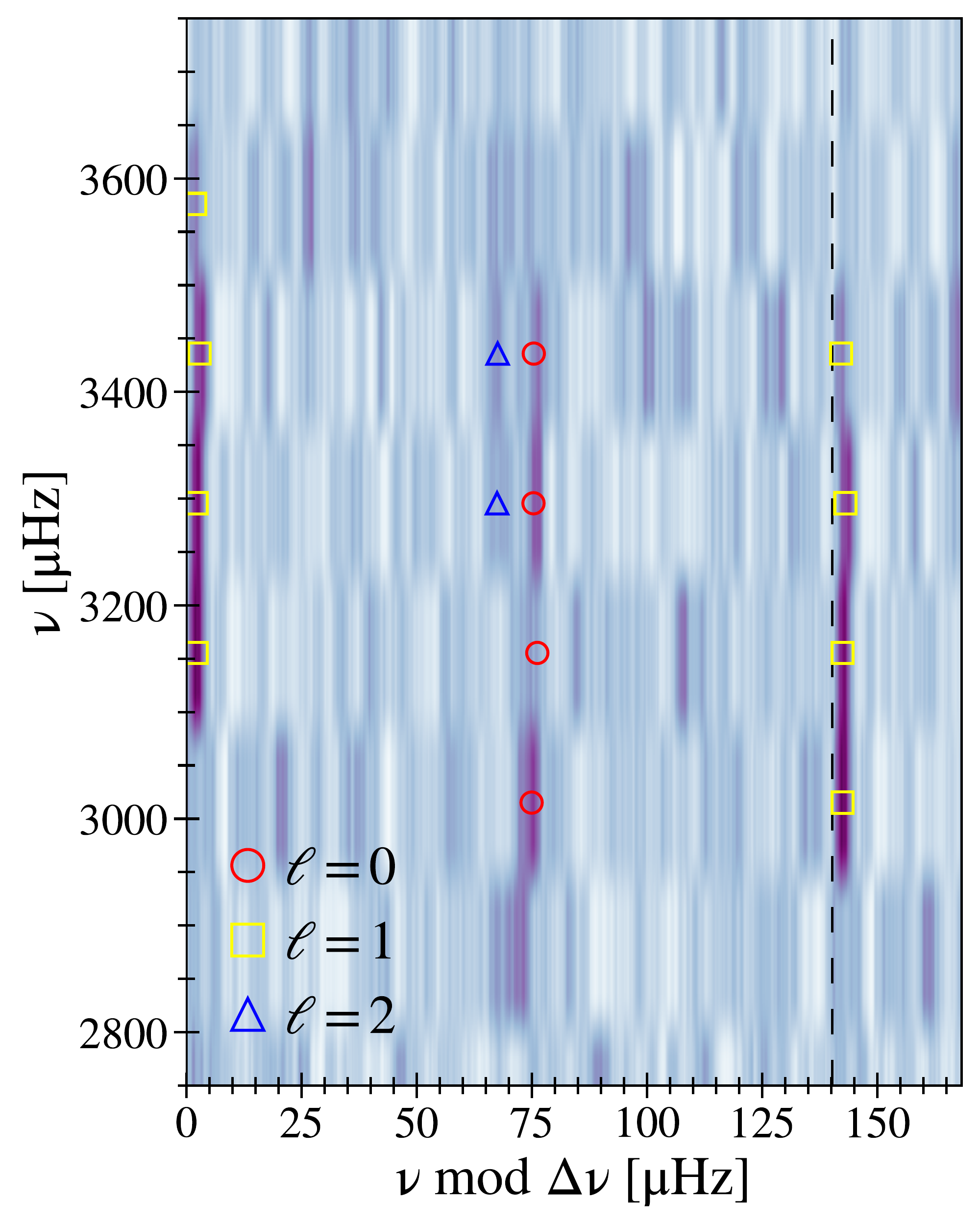}
\caption{\'Echelle diagram of $\alpha$ Men A from a smoothed power spectrum (with a boxcar filter width of 2.5 \muhz) using. \texttt{echelle} \citep{echelle}. Different oscillation modes are marked and colored by their spherical degree: radial modes ($\ell=0$) by red circles, dipole modes ($\ell=1$) by yellow squares, and quadrupole modes ($\ell=2$) by blue triangles. The large frequency separation derived from radial modes is \dnu\ = $140.24 \pm 1.98$ \muhz\ and is delineated by the black dashed line.}
\label{fig:echelle}
\end{figure}

Ultimately, the data did not provide enough evidence for \texttt{DIAMONDS} to converge on reliable results for more complex models (i.e. multiple Harvey-like terms), and therefore no model comparison was needed. We attempted to model the granulation component but it was mostly unconstrained or resulted in very small amplitudes. This is likely because the amplitude of the granulation signal is comparable to or less than the white noise level in the power spectrum. The final background fit is shown in Figure \ref{fig:PS}a as a solid blue line, which is the summed contributions from a white noise component (blue dashed line) and a meso-granulation term (blue dotted line).

\begin{figure*}
\centering
\includegraphics[width=\linewidth]{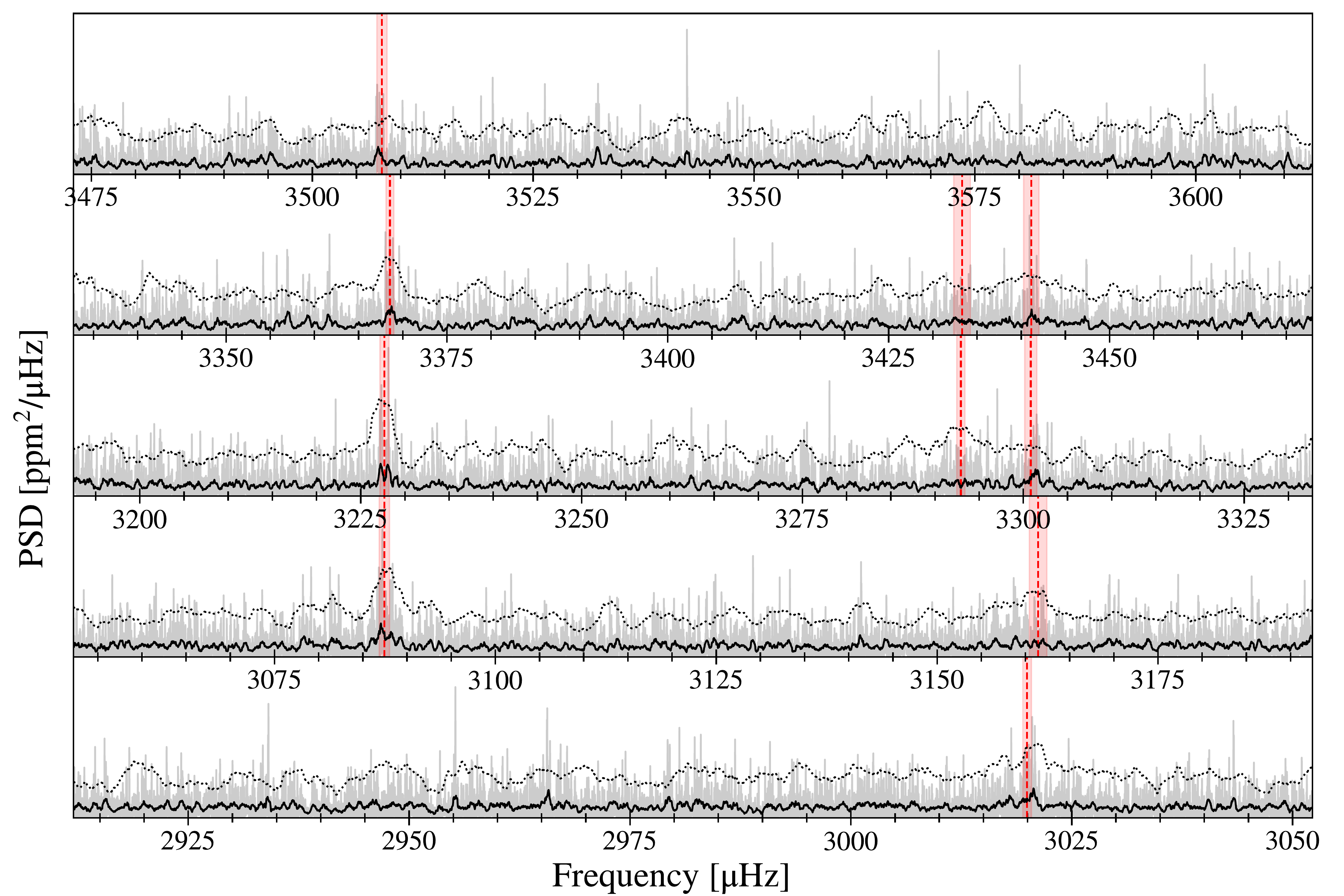}
\caption{Power spectrum of $\alpha$ Men A arranged in \'echelle format. The original power spectrum calculated using TESS data (Section \ref{sec:tess}) is shown in grey, as well as a smoothed version of the same power spectrum (using a boxcar width = 1.5 \muhz) in black. The weighted power spectrum discussed in Section \ref{sec:model} is overplotted by a dotted black line with a vertical offset for a direct comparison. The final frequencies from Table \ref{tab:freqs} are added in red, where shaded regions are equal to $\pm 1 \sigma$ for each mode.}
\label{fig:stackedPS}
\end{figure*}

\subsection{Global Asteroseismic Parameters} \label{sec:global}

The shaded region in Figure \ref{fig:PS}b shows the power excess due to oscillations. Within the \texttt{DIAMONDS} framework, this power excess is modeled by a Gaussian

\begin{equation} \label{equ:gaussian}
G(\nu) = H_{\rm{osc}}\, \rm{exp}\bigg[- \frac{(\nu - \nu_{\rm{max}})^2}{2\sigma_{\rm{osc}}^{2}}\bigg],
\end{equation}
centered at \numax with height $H_{\rm{osc}}$ and width $\sigma_{\rm{osc}}$ \citep{corsaro2014}. The resulting parameters of the global \texttt{DIAMONDS} analysis for $\alpha$ Men A are listed in Table \ref{tab:bg}.

We also derived an independent value for the frequency corresponding to maximum power using the SYD pipeline \citep{huber2009}, yielding \numax\ $\sim3267$ \muhz, consistent with the results from \texttt{DIAMONDS}. Two independent analyses additionally confirmed power excess in the same region, with \numax\ = 3230 \muhz\ \citep[A2Z;][]{mathur2010} and \numax\ = 3216 \muhz\ \citep{lundkvist2015}, both consistent to $\lesssim\,$1$\sigma$ from our derived values. Our derived \numax\ is larger than that in the Sun ($\nu_{\rm{max},\odot}$ = 3090 \muhz). Therefore, $\alpha$ Men A joins only a handful of other stars such as $\tau$ Cet \citep{teixeira2009}, $\alpha$ Cen B \citep{carrier2003,kjeldsen2005} and Kepler-444 \citep{campante2015} that have a higher \numax\ than the Sun.

To estimate a preliminary value for the large frequency separation, we calculated an \'echelle diagram. In the case of solar-like oscillators, modes of different radial order ($n$) with the same spherical degree ($\ell$) should form vertical ridges if the correct spacing is used. We calculated the best-fitting value by taking small steps in frequency space until the ridges lined up vertically, yielding \dnu\ $\sim$140 \muhz. Figure 4 shows the resulting \'echelle diagram created using  \texttt{echelle}\footnote{\url{https://github.com/danhey/echelle}} \citep{echelle}, which clearly confirms the detection of solar-like oscillations.

\subsection{Individual Frequencies} \label{sec:pb}

We extracted frequencies from the background-corrected power spectrum using three independent methods, which are based on fitting Lorentzian profiles to individual modes \citep{garcia2001,garcia2009,handberg2011,nielsen2015,nielsen2017}. A second approach used an alternative power spectrum calculated using weights to account for different noise levels across time series \citep{arentoft2008}. 

To compare the two approaches, Figure \ref{fig:stackedPS} shows both the unweighted (solid grey/original, black lines/smoothed) and weighted (dotted black line) power spectrum stacked by radial order $n$ about \numax. The figure clearly exhibits the consistency between the two independently-calculated spectra, especially for the higher signal-to-noise (SNR) modes. 

\begin{figure}
\includegraphics[width=\linewidth]{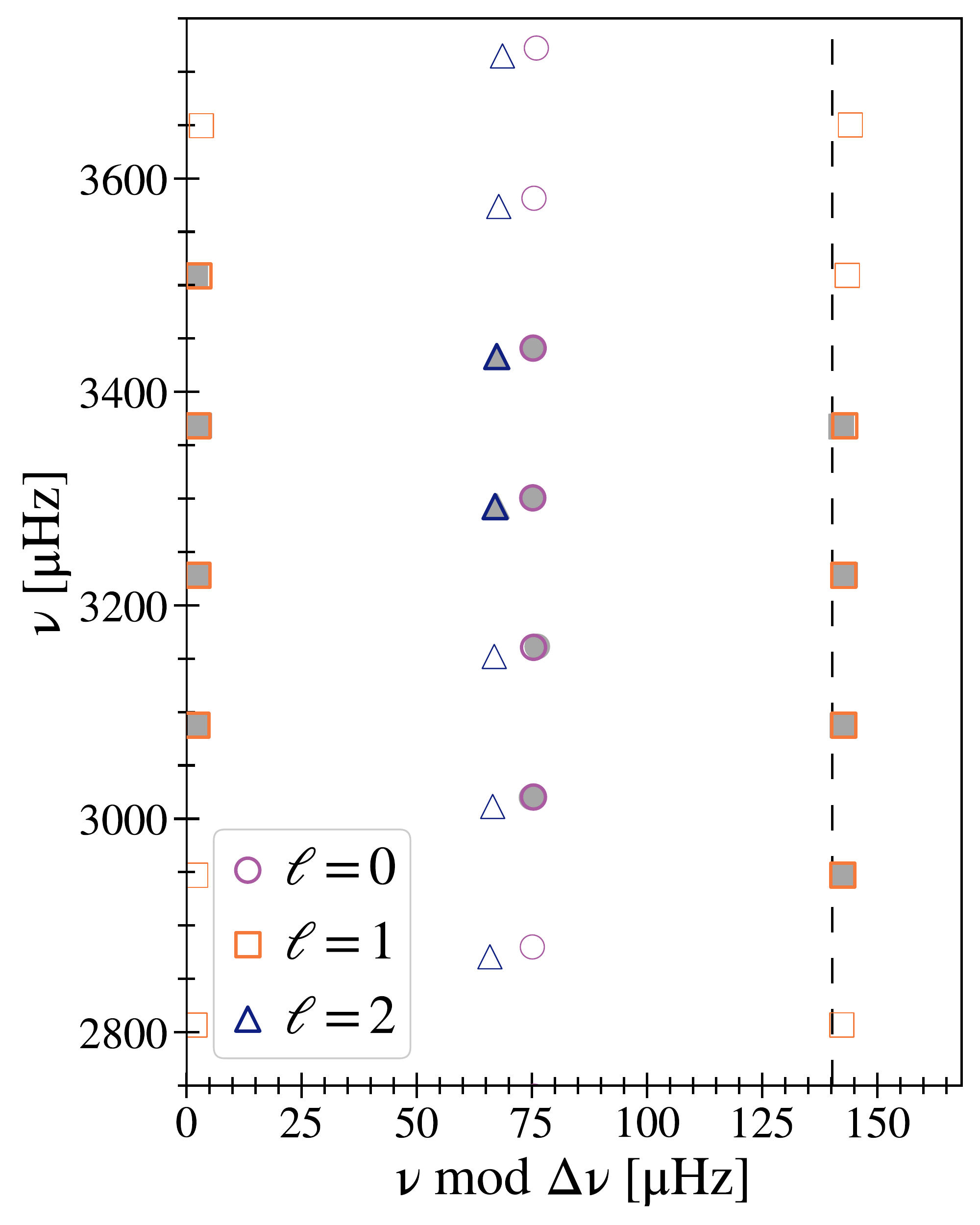}
\caption{\'Echelle diagram of observed frequencies (filled gray symbols) and best-fit model frequencies (open colored symbols) using \texttt{BeSPP} \citep{serenelli2013,serenelli2017}. Modeled frequencies that correspond to an observed frequency are highlighted with a thicker marker outline.}
\label{fig:dupechelle}
\end{figure}

Our final frequency list was constructed by taking modes for which both approaches reported a detection. We report 4 radial ($\ell=0$) modes, 4 dipole ($\ell=1$) modes, and 2 quadrupole ($\ell=2$) modes in Table \ref{tab:freqs}. Formal uncertainties were adopted using the frequencies calculated from the weighted power spectrum and adding in quadrature the scatter in frequencies derived from independent methods to account for systemic uncertainties. The final set of frequencies are plotted on the \'echelle diagram (Figure \ref{fig:echelle}, marked by their spherical degree $\ell$) and on the stacked power spectrum (Figure \ref{fig:stackedPS}).

\subsection{Frequency Modeling} \label{sec:model}

To properly account for systematic uncertainties, we derived fundamental stellar properties of $\alpha$ Men A using \textbf{nine} independent modeling pipelines, including \texttt{BASTA} \citep{silvaaguirre2015}, \texttt{YB} \citep{basu2010,gai2011,basu2012}, \texttt{AMP}\footnote{\url{https://github.com/travismetcalfe/amp2}} \citep{metcalfe2003,metcalfe2009a,metcalfe2012b}, \texttt{BeSPP} \citep{serenelli2013,serenelli2017}, \texttt{Izmir} \citep{yildiz2019}, \texttt{GOE} \citep{silvaaguirre2017} and \texttt{YALE-M} \citep{tasoulis2004,mier2017,ball2020}. Model grids were calculated from various stellar evolution codes (\texttt{YY}, \citealt{demarque2004}; \texttt{MESA} r10398, r12115, \textbf{r15140}, \citealt{paxton2011,paxton2013,paxton2015,paxton2018,paxton2019}; \texttt{GARSTEC}, \citealt{weiss2008}; \texttt{YREC}, \citealt{demarque2008}; \texttt{BaSTI}, \citealt{pietrinferni2004}; \texttt{DSEP}, \citealt{dotter2007,dotter2008}; \texttt{CESAM2k}, \citealt{morel2008}; \texttt{YREC2}, \citealt{basu2012}; \texttt{ASTEC}, \citealt{cd2008}; \texttt{CESTAM}, \citealt{marques2013}; and Padova, \citealt{marigo2008,girardi2000}) using different assumptions about \textit{input physics}. Oscillation frequencies were generated from oscillation codes (\texttt{ADIPLS}, \citealt{cd2008}; \texttt{GYRE}, \citealt{townsend2013}), where most of the methods listed here also applied corrections for near-surface effects \citep{kjeldsen2008,ball2014}.

\begin{table}
\begin{center}
\caption{Extracted mode identifications and oscillation frequencies for $\alpha$ Men A sorted by spherical degree $\ell$.}
\label{tab:freqs}
\begin{tabular}{c c c c}
\hline\hline
$\nu$ (\muhz)  & $\sigma_{\nu}$ (\muhz) & $n$ & $\ell$ \\
\hline
3019.95 & 0.50 & 23 & 0 \\
3161.43 & 0.99 & 24 & 0 \\
3300.84 & 0.69 & 25 & 0 \\
3441.14 & 0.87 & 26 & 0 \\
\hline
3087.44 & 0.59 & 23 & 1 \\
3227.69 & 0.52 & 24 & 1 \\
3368.55 & 0.44 & 25 & 1 \\
3507.88 & 0.58 & 26 & 1 \\
\hline
3292.93 & 0.47 & 24 & 2 \\
3433.30 & 0.95 & 25 & 2 \\
\hline         
\end{tabular}
\end{center}
\end{table}

Each method derived four sets of stellar parameters  based on the following sets of constraints:

\small
\begin{enumerate}
\item \{\teff, [Fe/H], \lstar, \numax, \dnu, $\nu(n,0)$, $\nu(n,1)$, $\nu(n,2)$\}
\item \{\teff, [Fe/H], \numax, \dnu, $\nu(n,0)$, $\nu(n,1)$, $\nu(n,2)$\}
\item \{\teff, [Fe/H], \lstar, \numax, \dnu, $\nu(n,0)$, $\nu(n,1)$\}
\item \{\teff, [Fe/H], \numax, \dnu, $\nu(n,0)$, $\nu(n,1)$\}
\end{enumerate}
\normalsize

\noindent The main purpose for all four runs was to test for inconsistencies between the luminosity derived from asteroseismology and the independent $Gaia$-derived luminosity, as well as to check if the weaker quadrupole ($\ell=2$) modes had any affect on the final age estimates. Results from each pipeline were self-consistent in that the runs that excluded the quadrupole modes generally preferred younger ages but ultimately the differences were not significant ($\leq$10\%) and $\lesssim$1$\sigma$. Moreover, across the numerous methods and model inputs mentioned, the derived stellar parameters between all pipelines agreed to within 1$\sigma$.

For final stellar parameters, we adopted the results from \texttt{BeSPP} \citep[Bellaterra Stellar Properties Pipeline,][]{serenelli2013,serenelli2017}, which was closest to the median values for fundamental stellar parameters (mass and age) in case 1. \texttt{BeSPP} constructed a grid of stellar models with \texttt{GARSTEC} \citep{weiss2008} using a gray model atmosphere based \citet{vernazza1981}, the solar mixture model from \citet{grevesse1993}, and the diffusion of elements according to \citet{thoul1994}. We refer the reader to \citet{weiss2008} for more details on the input physics of stellar models computed with \texttt{GARSTEC}. \texttt{BeSPP} yielded a bimodal solution as a result of a bimodal surface correction, which at a fixed mass, was older ($\sim$2 Gyr) and more metal rich ($\sim$0.1 dex). The surface correction for the younger solution was unexpectedly large ($\sim$50\muhz) for a solar analogue and hence provided strong support in favor of the older model. 

To account for systematic differences between various methods, uncertainties were calculated by adding the standard deviation for each parameter \{$M$, $R$, $\rho$, $\tau$, $\log g$\} from all pipelines in quadrature with the formal uncertainty from \texttt{BeSPP}. Corrected model frequencies are plotted with the observed frequencies in an \'echelle diagram in Figure \ref{fig:dupechelle}. Stellar parameters are listed in Table \ref{tab:primary} (\textbf{i.e. see} the asteroseismology section), with fractional uncertainties of 1.4\% (1.2\% stat $\pm$ 0.7\% sys) in density, 1.7\% (1.4\% stat $\pm$ 0.9\% sys) in radius, 4.7\% (3.9\% stat $\pm$ 2.7\% sys) in mass, and 24.2\% (21.8\% stat $\pm$ 10.4\% sys) in age.

\begin{figure}
\centering
\includegraphics[width=\linewidth]{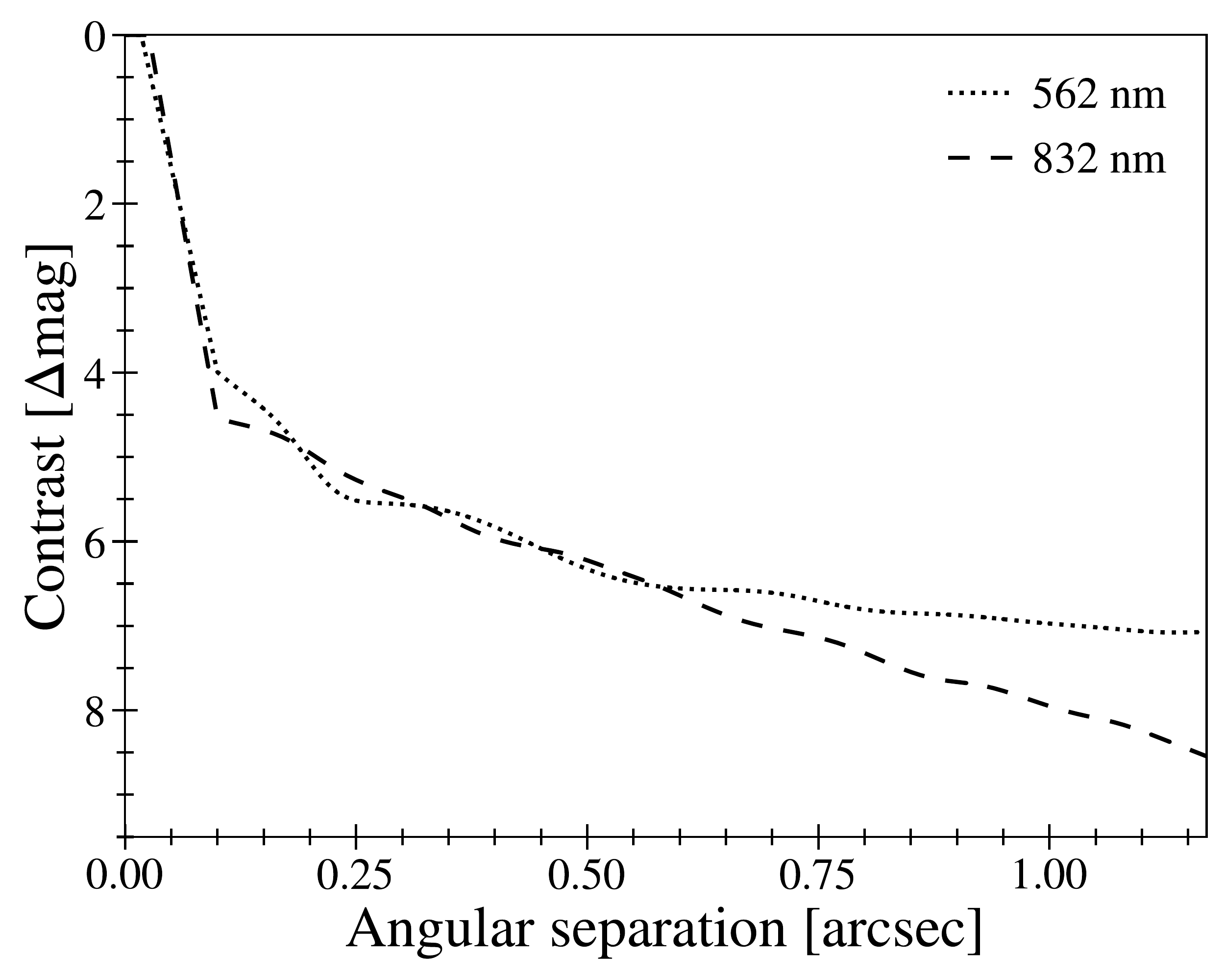}
\caption{Detection limits from speckle imager Zorro on the Gemini South telescope in r- and z-band (562 nm and 832 nm, respectively), ruling out any additional close-in companions to the contrast levels obtained.}
\label{fig:speckle}
\end{figure}

\begin{table}
\setlength{\tabcolsep}{3pt}
{\small %
\begin{center}
\caption{Secondary Stellar Parameters}\label{tab:secondary}
\vspace{-0.2cm}
\begin{tabular}{lcc}
\hline\hline
\noalign{\smallskip}
\multicolumn{3}{c}{\textbf{Other identifiers:}} \\
\multicolumn{3}{c}{$\alpha$ Men B, HD 43834 B} \\
\multicolumn{3}{c}{Gaia DR2/eDR3 5264749303457104384} \\
\noalign{\smallskip}
\hline\hline
\noalign{\smallskip}
Parameter & Value & Method \\
\noalign{\smallskip}
\hline
\noalign{\smallskip}
\multicolumn{3}{c}{Gaia$^{1,2}$} \\
\noalign{\smallskip}
\hline
\noalign{\smallskip}
Right ascension (RA), $\rm\alpha_{J2016}$	& $92.5590$ & A \\
Declination (Dec), $\rm\delta_{J2016}$	& $-74.7542$ & A \\
Parallax, $\pi$ (mas) & $97.8666 \pm 0.0898$ & A \\
Distance, $d$ (pc) & $10.2180 \pm 0.0094$ & A \\
Gaia G mag, $G$ & $12.3653 \pm 0.0044$ & P \\
Gaia G contrast, $\Delta G$ & $7.4652 \pm 0.0053$ & P \\
\noalign{\smallskip}
\hline
\noalign{\smallskip}
\multicolumn{3}{c}{Other Work$^{3,4,5}$} \\
\noalign{\smallskip}
\hline
\noalign{\smallskip}
Projected separation, $\rm \rho_{K}$ ($''$) & 3.02 $\pm$ 0.01 & A \\
Position angle, $\rm PA_{K}$ ($\rm ^{o}$) & 250.87 $\pm$ 0.11 & A \\
NACO K contrast, $\rm \Delta m_{K}$ & 4.97 $\pm$ 0.05 & P \\
2MASS K$_{S}$ mag, $\rm K_S^{*}$ & $8.476 \pm 0.200$ & P \\
Spectral type & M3.5--M6.5 & R1 \\
Stellar mass, \mstar\ (\msol) & 0.14 $\pm$ 0.01 & R2 \\
Orbital period, $P$ (years) & 162.04 & - \\
\noalign{\smallskip}
\hline
\noalign{\smallskip}
\multicolumn{3}{c}{This Work} \\
\noalign{\smallskip}
\hline
\noalign{\smallskip}
Effective temperature, \teff\ (K) & \ambt\ & E \\
Stellar radius, \rstar\ (\rsol) & \ambr\ & E \\
Stellar mass, \mstar\ (\msol)& \ambm\ & E \\
Age, $t$ (Gyr) & $6.2 \pm 1.4$ & F \\
Orbital period, $P$ (years) & 157.44 & - \\
\noalign{\smallskip}
\hline
\noalign{\smallskip}
\end{tabular}
\textbf{References --} (1) \citet{gaia2016} (2) \citet{gaia2021} (3) \citet{eggenberger2007} (4) \citet{cutri2012} (5) \citet{tokovinin2014} \\ \textbf{Methods --} (A) Astrometry, (E) empirical relations derived from \cite{mann2015} and \cite{mann2019}, (R1) relationship between absolute magnitude and spectral type from \citet{eggenberger2007} (using data from \citet{delfosse2000}, \citet{leggett2001}, \citet{dahn2002}, and \citet{vrba2004}) (R2) infrared mass-luminosity relation for low-mass stars \citep{delfosse2000}, (P) photometry, or (F) frequency modeling via asteroseismology. \\
\end{center}
\vspace{-0.2cm}
{\sc \textbf{Notes --}} \\
$^{*}$ The provided Ks magnitude uncertainty for the companion was less than that reported by \citet{cutri2012} for the primary. Therefore, we inflated the uncertainty in the Ks magnitude for HD 43834 B to reflect that.}\\
\smallskip
\end{table}

\begin{figure*}
\includegraphics[width=\linewidth]{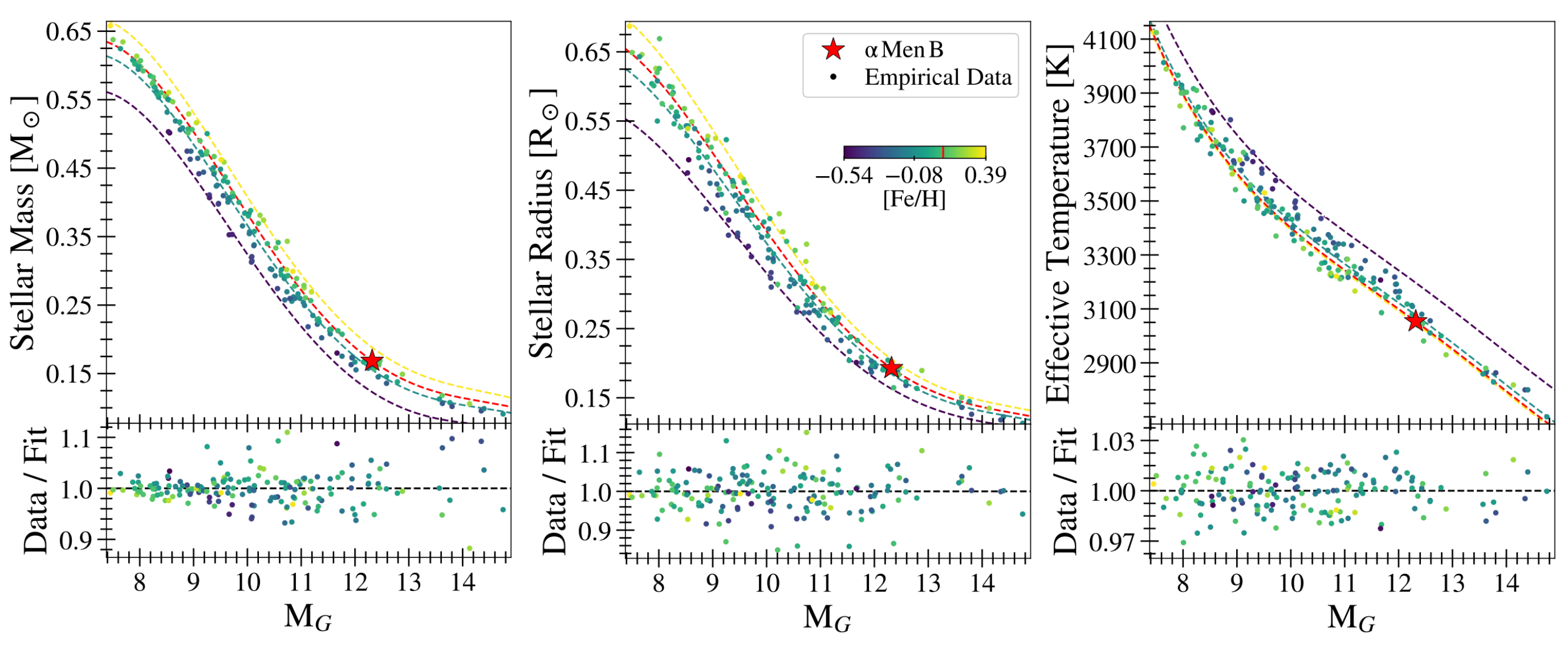}
\caption{Empirical best-fit relations for mass (left), radius (middle) and temperature (right) as a function of absolute \gaia\ magnitude for the sample in \citet{mann2015} color-coded by metallicity. The bottom plot shows the residuals of the data subtracted by the best fit polynomial (black dashed line). The red star represents the location of $\alpha$ Men B.}
\label{fig:companion}
\end{figure*}

\section{M-dwarf Companion}

\subsection{Discovery \& Initial Characterization}

A bound M-dwarf companion to $\alpha$ Men A was first identified by \cite{eggenberger2007} in a study investigating the impact of stellar duplicity on planet occurrence rates using adaptive optics imaging with NACO/VLT. \citet{eggenberger2007} ruled out the possibility of HD 43834 B being a background star, stating that the astrometry was compatible with orbital motion. In addition, they added that the physical association was further supported by a linear drift present in CORALIE data. 

\citet{eggenberger2007} reported a magnitude difference of $\Delta$m = 4.97 $\pm$ 0.05 in the narrowband K filter ($\lambda_{c} = 2.166 \, \mu$m). After a correction to account for the differences in relative photometric systems, they reported an absolute 2MASS Ks magnitude, M$_{K_{\rm s}}$ = 8.43 $\pm$ 0.05 for the companion. They concluded that HD 43834 B is consistent with an M3.5-M6.5 dwarf companion with a mass of M$\rm _{\star, B}$ = 0.14 $\pm$ 0.01 \msol\ at a projected separation of 3" from the primary, which corresponds to a physical separation of $\sim$30 AU. \citet{tokovinin2014} characterized nearby multiple star systems and, using the literature mass of M$\rm _{\star, A}$ = 1.01 \msol for the primary, estimated an orbital period of $\sim$162 years for the wide companion.

\subsection{A Search for Additional Companions} \label{sec:image}

To search for additional close companions, we observed
$\alpha$ Mensae with Zorro\footnote{\url{https://www.gemini.edu/instrumentation/current-instruments/alopeke-zorro}}, a dual-channel imager on the 8.1-m telescope at the Gemini South Observatory (Cerro Pachon, Chile). Zorro provides simultaneous diffraction-limited optical imaging (FWHM 0.02" at 650nm) in 2 channels. We observed $\alpha$ Men A in speckle mode to search for close-in companions between UT December 22 2019 and December 23 2019. The images were subjected to the standard Fourier analysis as described in \citet{howell2011} and were used to produce reconstructed images in each color providing high-resolution angular results. In addition to detecting the M dwarf companion at 3" distance, no other companions to Mensae were found. Figure \ref{fig:speckle} shows the contrast curves from the reduced speckle data in both bands, indicating that there are no additional close companions ($<$1.2") from the diffraction limits down to contrasts of $\Delta\mathrm{m}\sim7$ in r-band (562 nm) and $\Delta\mathrm{m}\sim8$ in z-band (832 nm). At the distance of $\alpha$ Mensae, these angular limits correspond to spatial limits of 0.2 to 1.2 AU.

\subsection{Revised Properties of $\alpha$ Men B}
\label{sec:secondary}

The wide companion was resolved in \gaia\ DR2, which reported a magnitude of $G = 11.8057$ and a contrast of $\Delta G = 6.955$. However, the early \gaia\ Data Release 3 \citep[eDR3;][]{gaia2021,riello2021} reported a significantly fainter companion with $G$ = 12.365, corresponding to a contrast of $\Delta G$ = 7.465. Additionally, eDR3 provided a parallax for the companion, which was unavailable in DR2.

In order to estimate the companion properties, we derived empirical relations for masses, effective temperatures, and radii of M dwarfs given the absolute \emph{Gaia} magnitude ($\mathrm{M}_G$). We adopted the \teff\ and radius values from \citet{mann2015} and used the $\mathrm{M}_K$-mass relation from \citet{mann2019} to estimate the M-dwarf masses. We computed the absolute \gaia\ magnitudes using eDR3 \citep{gaia2021} photometry \citep{riello2021} and parallaxes \citep{lindegren2021}, including the relevant corrections when applicable. We removed two stars from the \cite{mann2015} sample for our relations:  Gl 896 B, due to its largely discrepant measurements in mass, \teff, and radius, and FBS L 10-72 due to its discrepant [Fe/H] given its other measured values. For each of the three parameters (\teff, \rstar, \mstar), we optimized polynomial coefficients using a least-squares minimization method available with \texttt{scipy}. Finally, to select the optimal order, we chose the curve that minimized the Bayesian Information Criterion \citep[BIC;][]{schwarz1978}.

Figure \ref{fig:companion} shows the resulting relations, which are:

\begin{equation}\label{equ:mass}
\begin{split}
\mathrm{M}_\star = -7.0653 + 3.0293 x  - 0.4218 x^2 \\
+ 0.0244 x^3 - 0.0005 x^4 + 0.1771 y - 0.0087 x y,
\end{split}
\end{equation}

\begin{equation}\label{equ:teff}
\begin{split}
T_{\mathrm{eff}} = 19938.4 - 4999.7 x + 582.0 x^2 \\
- 30.8 x^3 + 0.6 x^4 - 131.0 y + 208.0 y^2,
\end{split}
\end{equation}

\noindent \textbf{and}

\begin{equation}\label{equ:radius}
\begin{split}
\mathrm{R}_\star = (-4.1305 + 1.9735 x - 0.2832 x^2 \\
+ 0.0166 x^3 - 0.0003 x^4) \times (1 + 0.2450 y),
\end{split}
\end{equation}

\noindent where $x$ is the absolute Gaia magnitude $\rm M_{\mathit{G}}$ from eDR3 and $y$ is the metallicity from \cite{mann2015}.

Using the \gaia\ absolute magnitude for $\alpha$ Men B and metallicity for $\alpha$ Men A, the relations yield a mass of \mstar\ = \ambm\ \msol, radius of \rstar\ = \ambr\ \rsol\ and \teff\ = \ambt\ K for $\alpha$ Men B. Uncertainties were calculated from the residual scatter between the models and data, which are 3.7\% in mass, 4.4\% in radius and 44 K in \teff. The mass uncertainty was calculated using the scatter of 2.2\% in our derived relation (Equation \ref{equ:mass}) added in quadrature with the conservative estimate of 3\% from the \citet{mann2019} $\mathrm{M}_{Ks}-\mathrm{M}_\star$ relation. Notably, our derived mass of \ambm\ \msol\ for the fully-convective M dwarf is slightly higher than the value of 0.14 $\pm$ 0.01 \msol\ from \citet{eggenberger2007}, which was based on an infrared mass-luminosity relation for low-mass stars. All observed and derived properties of $\alpha$ Men B are summarized in Table \ref{tab:secondary}.

\begin{figure*}
\includegraphics[width=\linewidth]{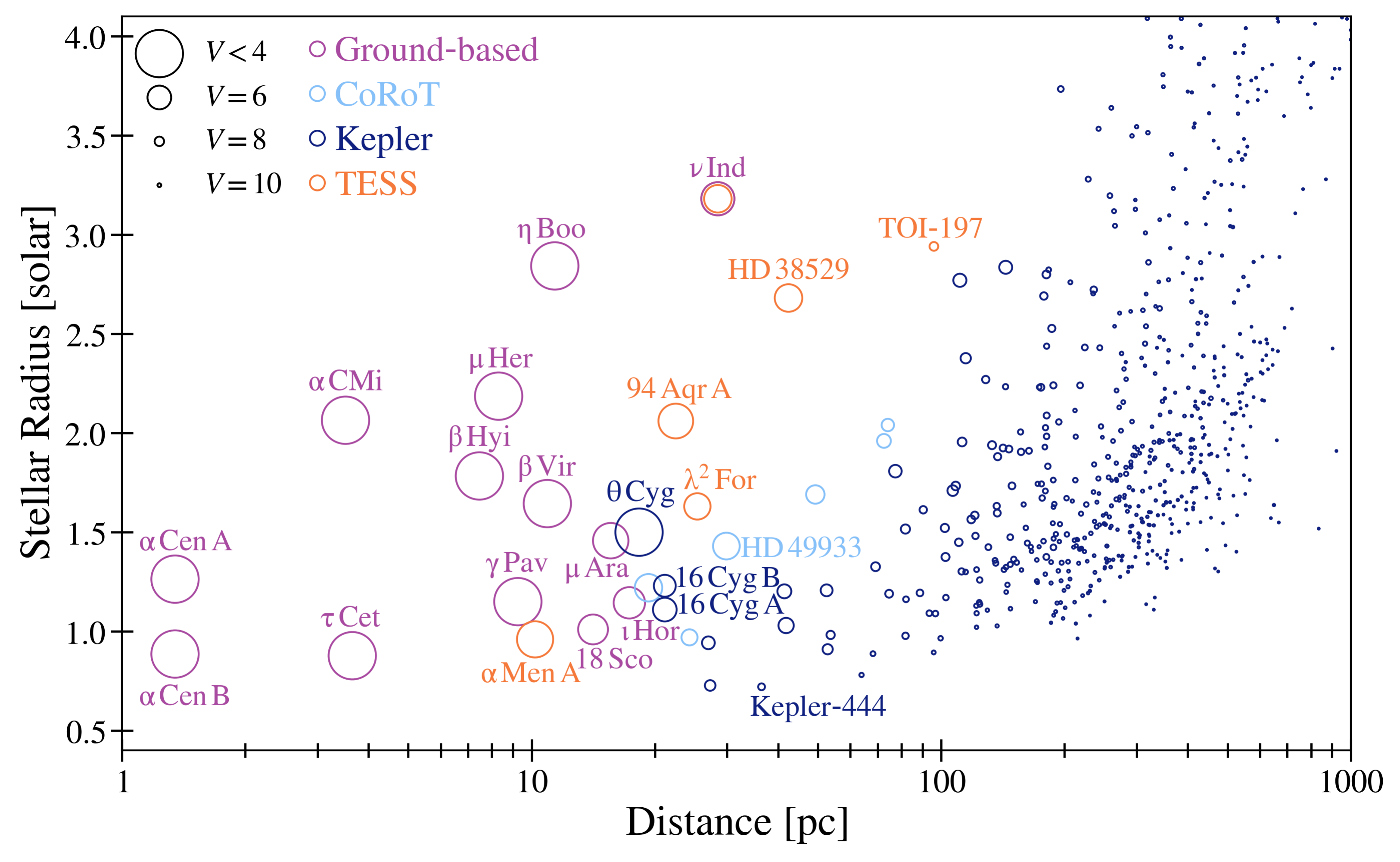}
\caption{Current population of asteroseismic detections with confirmed \dnu\ measurements plotted as the stellar size (in solar radii) as a function of distance, where markers are sized by their visual magnitude. Ground-based seismic detections (purple) are limited to close, bright stars while \kepler\ (navy blue) found hundreds of oscillations in faint, distant stars. TESS (orange) is beginning to close the gap between these two regimes and hence, providing bright nearby targets to follow up with ground-based instruments. In fact, $\nu$ Ind is the first example where an early ground-based asteroseismic detection \citep{bedding2006} was redetected later with TESS \citep{chaplin2020}.  \textbf{Label references (in order of proximity) --} $\alpha$ Cen A \citep{brown1991,bouchy2001,bouchy2002,butler2004,bedding2004}, $\alpha$ Cen B \citep{carrier2003,kjeldsen2005}, $\alpha$ CMi \citep{mosser1998,martic1999,arentoft2008,bedding2010}, $\tau$ Cet \citep{teixeira2009}, $\beta$ Hyi \citep{bedding2001,carrier2001,bedding2006}, $\mu$ Her \citep{bonanno2008,grundahl2017}, $\gamma$ Pav \citep{mosser2008}, $\alpha$ Men A (this work), $\beta$ Vir \citep{carrier2005a}, $\eta$ Boo \citep{kjeldsen1995,kjeldsen2003,carrier2005b}, 18 Sco \citep{bazot2011}, $\mu$ Ara \citep{bouchy2005}, $\iota$ Hor \citep{vauclair2008}, $\theta$ Cyg \citep{guzik2011}, 16 Cyg A \& B \citep{metcalfe2012a}, 94 Aqr A \citep{metcalfe2020}, $\lambda^2$ For \citep{nielsen2020}, $\nu$ Ind \citep{bedding2006,carrier2007,chaplin2020}, $\rm HD\,49933$ \citep{appourchaux2008,benomar2009}, Kepler-444 \citep{campante2015}, $\rm HD\,38529$ \citep{ball2020}, and TOI-197 \citep{huber2019}.}
\label{fig:ensemble}
\end{figure*}

\section{Discussion}

\subsection{Testing Stellar Physics with Asteroseismology}

Asteroseismology of nearby bright stars allows us to test stellar models by combining with other high-resolution stellar classification techniques like spectroscopy, interferometry, and astrometry \citep{bruntt2010,silvaaguirre2012,hawkins2016}. An example is \citet{bruntt2010}, who combined interferometry, asteroseismology, and spectroscopy to derive the most accurate and precise fundamental properties for 23 bright stars. 

Figure \ref{fig:ensemble} shows radii and distances of stars in which solar-like oscillations have been studied. The  brightest detections were discovered prior to the launches of \kepler\ and \corot\ \citep[Convection Rotation and planetary Transits;][]{baglin2006}. This means that most bright stars only have asteroseismology from ground-based radial velocity measurements, which suffer from  aliasing problems due to gaps in data. Examples of well-known ground-based asteroseismic detections include $\beta$ Hyi \citep{bedding2001,carrier2001,bedding2006}, $\alpha$ Cen A \citep{bouchy2002,bedding2004}, and $\alpha$ Cen B \citep{carrier2003,kjeldsen2005}.

The flood of continuous high-precision high-cadence photometry from \corot\ and \kepler\ marked the start of the so-called ``space-based photometry revolution". \kepler\ revolutionized the field of asteroseismology by detecting oscillations in $\sim$500 main-sequence and subgiant stars. However, most \kepler\ targets are faint and distant, and thus do not have information from complementary techniques such as interferometry. This limitation was only partially solved by novel techniques called ``halo" and ``smear" photometry, which allowed the production of high-precision light curves for even heavily saturated \kepler\ stars \citep{pope2016,white2017,pope2019a,pope2019b}. 

The TESS mission provides an ideal solution to this problem. Several early asteroseismic detections by TESS have been made for bright nearby stars such as $\nu$ Ind \citep{chaplin2020}, HD 38529 \citep{ball2020}, $\lambda^2$ For \citep{nielsen2020}, 94 Aqr A \citep{metcalfe2020}, and the first new TESS asteroseismic host, TOI-197 \citep{huber2019}. Alpha Men A is now the closest solar analogue with an asteroseismic detection from space, making it a prime example of a bright benchmark system from the nominal TESS mission. In fact, $\alpha$ Men A was included in \citet{bruntt2010}, but was the only star in the sample without an asteroseismic detection.

\begin{figure*}
\centering
\includegraphics[width=\linewidth]{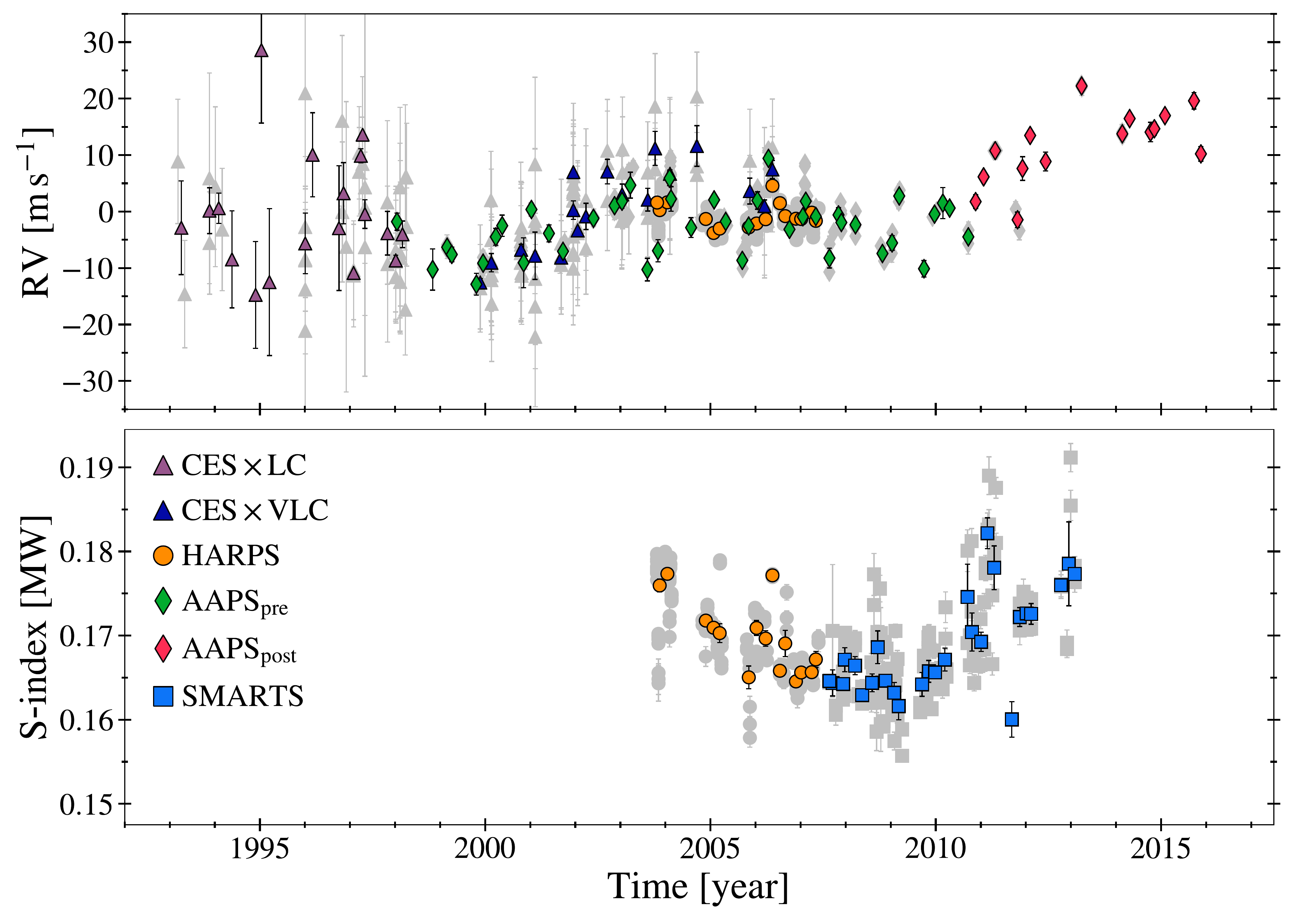}
\caption{Long-term ground-based RV (top) and Mount Wilson calibrated S-index (bottom) time series for $\alpha$ Men A using combined data from 5 different surveys. Each instrument is shown by a different marker, where offsets and/or upgrades of a similar instrument are shown by a different color. Grey points are original data from each survey and colored points are binned over 60 days. We find a long-term cyclic trend of $\sim$13 years in both RVs and chromospheric emission, indicative of a stellar activity cycle.}
\label{fig:activity}
\end{figure*}

\subsection{Stellar Activity}

The connection between oscillations and activity cycles are important to understand the long-term magnetic evolution of stars. For example, observations in the Sun have shown a strong correlation between oscillation frequencies and amplitudes with the solar activity cycle \citep{broomhall2011}. Currently there are only a handful of examples that exist for stars other than the Sun \citep[e.g.][]{garcia2010}, and therefore expanding this sample to more stars would be very valuable. Fortunately TESS is already well-positioned for this, as demonstrated by \citet{metcalfe2020} who combined TESS asteroseismology with 35 years of activity measurements to study the evolution of rotation and magnetic activity in 94 Aqr Aa.

Stellar activity is traditionally observed indirectly through long-term monitoring of chromospheric emission in the Ca \rom{2} H and K lines \citep{noyes1984a,baliunas1995,henry1996}. More recently, \citet{santos2010} and \citet{lovis2011} used RVs derived from the cross-correlation function (CCF) method to show that stellar activity also correlates with parameters from the CCF like the FWHM and the bisector inverse slope (BIS). Indeed a study by \citet{zechmeister2013} compared archival HARPS RVs with three different indicators for $\sim$30 well-studied stars and found a positive correlation with all three ($\log R'_{\rm{HK}}$, BIS, FWHM) for alpha Men, indicative of a magnetic cycle as a cause of the RV variations. However, the main goal of the study was to observationally confirm the existence of correlations in RV indicators and they did not report any activity cycle periods.

To investigate the stellar activity cycle for $\alpha$ Men A, we collected publicly available RV data from two instruments on the ESO 3.6m telescope: the Coud\'e Echelle Spectrograph (CES, pre- and post-upgrade) and HARPS, which was already corrected for systemic instrumental offsets in \citet{zechmeister2013}. We also collected data from the Anglo-Australian Planet Search (AAPS), which perfectly overlapped with the CES and HARPS data. Figure \ref{fig:activity} shows the complete RV time series, which covers 22 years. AAPS data after 2011 is plotted as a separate instrument because of an unexplained RV offset. We used a conservative bin size of 60 days to average over the scatter due to stellar rotation, revealing a period which is similar to that observed in the Sun. Using the publicly available \texttt{GLS} code \citep{zechmeister2009}, a generalized Lomb-Scargle periodogram analysis that is better suited for unevenly (and sparsely) sampled data, we detect a period $P=$ 13.1 $\pm$ 1.1 years in the RV time series. There is evidence for a long-term linear trend which is likely from the companion, depending on the inclination of the system. Note that we did not correct for any additional systematic offsets when we combined the time series from multiple instruments.

To confirm that the RV variations are due to an activity cycle, we analyzed the Mount Wilson calibrated S-index time series available from the HARPS DRS pipeline. Chromospheric emission in alpha Men was also observed as part of the SMARTS southern HK project from 2007-2013 \citep{metcalfe2009b}. The bottom of Figure \ref{fig:activity} shows the combined S-index time series, which span roughly one activity cycle and follow a similar trend to that seen in the RVs. This suggests that the observed RV variations are intrinsic to the star and not from a long-period planetary mass companion, providing additional evidence in support of an activity cycle detection.

\subsection{Gyrochronology}

Recent observations of stellar rotation periods have challenged commonly adopted age-rotation relationships in two distinct parameter spaces. Specifically, the observation of slow rotation periods in middle-aged solar-type stars has been proposed to be related to weakened magnetic braking due to stellar winds \citep{vansaders2016}, while the stalled spin-down observed in lower mass cluster members \citep{curtis2019} has been hypothesized to be related to reduced angular momentum transport caused by a decoupling of the convective core and the radiative envelope \citep{spada2020}. At an age of $\sim$6 Gyr, $\alpha$ Men A is in the latter half of its main sequence life and therefore provides a valuable test for the weakened braking hypothesis. Alpha Men B, on the other hand, is a fully convective M dwarf and thus provides an excellent test of whether core-envelope decoupling is indeed responsible for the stalled spin-down in M dwarfs with radiative envelopes. Consequently, rotation periods for either star in the $\alpha$ Mensae system would be extremely valuable to calibrate gyrochronology, which is currently the most promising method for ages for field dwarfs.

\begin{figure}
\centering
\includegraphics[width=\linewidth]{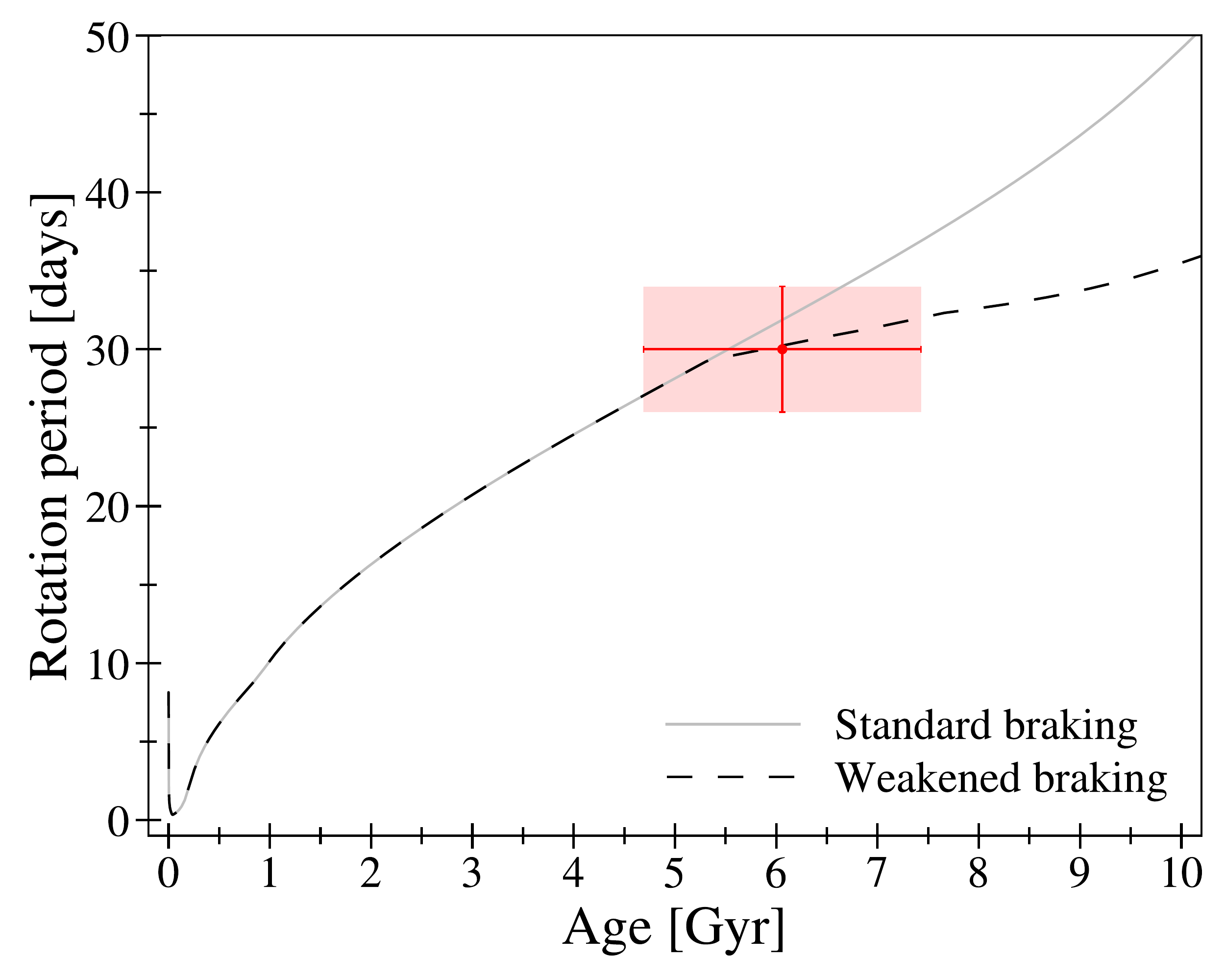}
\caption{Predicted rotation periods of $\alpha$ Men A from gyrochronology models using a standard braking law \citep[solid grey line,][]{vansaders2013} and a weakened braking law \citep[dashed black line,][]{vansaders2016}. The rotation period uncertainties were calculated by taking the average of the lower and upper quantiles from the corner plots derived by \texttt{kiauhoku} \citep{claytor2020}, where the more conservative uncertainty of 4.5 days is plotted in this figure. The shaded red area represents the 1$\sigma$ uncertainty region for the location of $\alpha$ Men A with respect to theoretical gyrochronological models, which predicts that the solar analogue has recently undergone or is currently in transition in its rotational behavior.}
\label{fig:rotation}
\end{figure}

\citet{saar1997} reported a rotation period ($P_{\rm{rot}}$) of 32 days for alpha Men based on Ca \rom{2} flux measurements. Futher investigation showed that the rotation period was not directly observed but empirically derived. A relationship between chromospheric activity and the Rossby number (Ro) of a star, a parametrization of the rotation period and convective turnover time ($\tau_{\rm{conv}}$), was established by \citet{noyes1984a} and updated by \citet{mamajek2008} using a larger sample of stars. For $\alpha$ Men A, a mean activity level log$R'_{\rm{HK}}$ = -4.94 from \citet{henry1996} with Eq.5 \citep{mamajek2008} yields Ro = 2.05. Using Eq.4 from \citet{noyes1984a} and the Johnson color index $(B-V)_{J}$ = 0.72 for the primary, estimate a turnover time of log$\,\tau_c$ = 1.192. We arrive at an approximate value of $P_{\rm{rot}}$ $\sim$32 days, in agreement with the value reported by \citet{saar1997}. 

We also calculated the color index from the \textit{Tycho-2} catalog \citep{hog2000} using the transformation from  B$\rm{_T}$ and V$\rm{_T}$ to Johnson indices (see Section 1.3 Appendex 4 from Hipparcos catalog, ESA 1997). We used this color index $(B-V)_{J}$ = 0.69 to obtain log$\,\tau_c$ = 1.16 and $P_{\rm{rot}}$ $\sim$30 days. It is also worth pointing out that the updated empirical age-activity-rotation relation from \citet{mamajek2008} estimated an age of 5.5 Gyr for $\alpha$ Men, which is consistent with our asteroseismic age. 

To search for stellar rotational modulation we analyzed the TESS SAP light curve, which is more conservative in preserving long-term variability than the PDCSAP light curve shown in Figure \ref{fig:LC}. We identified a period around 36 days, consistent with the estimates from activity indicators. We caution however that this period is highly uncertain due to intra-sector TESS systematics, which are non-negligible and therefore make it difficult to detect reliable rotation periods $\gtrsim$13 days.

Using the asteroseismic age, we calculated rotation periods of $\alpha$ Men A using different spindown models. Figure \ref{fig:rotation} shows the rotation period for the primary as a function of age using YREC \citep[Yale Rotating stellar Evolution Code;][]{demarque2008} models with a standard braking law \citep{vansaders2013} and a stalled braking law \citep{vansaders2016}, as implemented in \texttt{kiauhoku} \citep{claytor2020}. The models predict rotation periods of 30.4 $\pm$ 4.5 days and 29.6 $\pm$ 3.0 days respectively, indicating that alpha Men A may be close to the critical Rossby number, which is suggested to mark a transition in its rotational behavior \citep{metcalfe2016}, leading to a weakened spindown \citep{vansaders2016}.

Gyrochronology of M dwarfs remains challenging, mostly because the required constraints (e.g. $P_{\mathrm{rot}}$, ages) are not readily available. In particular, $\alpha$ Men B is below the convective boundary, where braking laws are uncertain. While the rotation period for the M dwarf is currently unknown, measuring a period in combination with the asteroseismic age would be valuable to place better constraints on gyrochronology models in low-mass stars.

\subsection{Exoplanet Synergies}

The future of exoplanet characterization will be heavily focused on direct imaging, which provides direct information about the planet composition and atmosphere. The next generation of space-based imaging missions (e.g., LUVOIR \citealt{luvoir2019}; HabEx, \citealt{gaudi2020}) will be equipped with instruments capable of imaging Earth-like planets around nearby stars.

A critical challenge for future direct imaging missions will be target selection. Historically, lower luminosity M-dwarfs have been popular targets when searching for rocky potentially habitable planets, since the habitable zones (HZs) are close to the host star. While this is ideal for methods such as transits and RVs that yield larger signals for planets that are closer in, the smaller separation is challenging for direct imaging. Consequently, the prime targets for missions like LUVOIR and HabEx will be nearby, well-characterized Sun-like stars, whose HZ is further from the host star. \citet{bixel2020} discussed the importance of age-based target selection specifically in the context of understanding planet habitability, noting that the presence of oxygen in the Earth's atmosphere has had a rich dynamic history. An Earth analogue around $\alpha$ Men A has a predicted separation of $\sim$100 mas and contrast $\sim5\times10^{-9}$, in reach for next-generation space-based imaging missions.

Additionally, new spectrographs have recently achieved the sub-meter-per-second precision that is needed to detect an Earth-like planet around a Sun-like star through precise radial velocities (PRVs). A major limitation for these efforts has been the background from stellar signals that have comparable periods and amplitudes to low-mass planetary companions, which can lead to spurious detections. Several newer techniques have been developed to help mitigate the effects induced on PRVs as a result of stellar activity \citep[e.g. ][]{damasso2017,feng2017,dumusque2018,zhao2020}. Our newly reported activity cycle for $\alpha$ Men A (P = 13.1 +/- 1.1 years with an amplitude of $\sim$5.5 $\rm m\,s^{-1}$) is consistent with the long-term RV scatter found in \citet{wittenmyer2016}) and could potentially help disentangle smaller planet-like signals with ground-based PRV surveys. 

\section{Conclusions}

We have used asteroseismology to precisely characterize the solar-analogue $\alpha$ Men A and its M-dwarf companion. Our main conclusions can be summarized as follows:

\begin{itemize}
\item Alpha Men A is a naked-eye G7 dwarf in TESS's Southern Continuous Viewing Zone. Combined astrometric, spectroscopic and asteroseismic modeling confirmed the  solar-analogue nature, with \rstar\ = 0.960 $\pm$ 0.016 \rsol, \mstar\ = 0.961 $\pm$ 0.045 \msol, and an age of 6.1 $\pm$ 1.4 Gyr. Alpha Men A is the closest star cooler than the Sun with an asteroseismic detection from space-based photometry and demonstrates the power of TESS for cool dwarf asteroseismology.
\item Alpha Men A has a bound companion, which was previously characterized as a mid-to-late M dwarf using 2MASS photometry. Using Gaia eDR3 photometry, we derived empirical M-dwarf relations for mass, effective temperature and radius as a function of metallicity. Using the relations, we provide revised properties of the fully-convective late M dwarf (\rstar\ = \ambr\ \rsol, \mstar\ = \ambm\ \msol, \teff\ = \ambt\ K). Through the asteroseismic characterization of the primary,  Alpha Men B joins a very small population of M dwarfs with a precisely measured age.
\item We used a combination of multiple radial velocity surveys to measure an activity cycle of P = 13.1 $\pm$ 1.1 years in $\alpha$ Men A, making it a prime target to  investigate the interplay of long-term magnetic evolution and stellar oscillations in a solar-type star.
\item Using the asteroseismic age, we used gyrochronology models to estimate rotation periods of 30.4 $\pm$ 4.5 days and 29.6 $\pm$ 3.0 days using a standard braking law and weakened braking law, respectively. Asteroseismic ages in two low-mass main-sequence stars makes the $\alpha$ Mensae system a benchmark calibrator for gyrochronology relations, which is currently the most promising age-dating method for late type stars.

\end{itemize}

With a precisely measured age and activity cycle, $\alpha$ Men A is now one of the best characterized nearby solar analogues, a useful calibrator for stellar astrophysics, and a prime target for next generation direct imaging missions to search for Earth-like planets. Continued all-sky TESS observations, in particular using 20-second cadence observations started in the extended mission, will enable asteroseismic detections in other solar analogues and continue the powerful synergies between stellar astrophysics and exoplanet science enabled by space-based photometry.

\section*{Acknowledgements}

We acknowledge the traditional owners of the land on which the Anglo-Australian Telescope stands, the Gamilaraay people, and pay our respects to elders past, present, and emerging. The authors would like to thank the staff at the Gemini South Observatory for follow-up observations.

A.C. acknowledges support from the National Science Foundation under the Graduate Research Fellowship Program (DGE 1842402). D.H. acknowledges support from the Alfred P. Sloan Foundation, the National Aeronautics and Space Administration (80NSSC18K1585, 80NSSC19K0379), and the National Science Foundation (AST-1717000).  T.A.B. acknowledges support by a NASA FINESST award (80NSSC19K1424). A.S. is partially supported MICINN project PRPPID2019-108709GB-I00. V.S.A. acknowledges support from the Independent Research Fund Denmark (Research grant 7027-00096B) and the Carlsberg foundation (grant agreement CF19-0649). T.R.B. acknowledges support from the Australian Research Council (DP210103119). W.H.B., W.J.C. and M.B.N. thank the UK Science and Technology Facilities Council (STFC) for support under grant ST/R0023297/1. R.A.G. acknowledge the support of the PLATO and GOLF CNES grants. M.S.L. is supported by the Carlsberg Foundation (Grant agreement no.: CF17-0760). Funding for the Stellar Astrophysics Centre is provided by The Danish National Research Foundation (Grant DNRF106). S.M. acknowledges support from the Spanish Ministry of Science and Innovation with the Ramon y Cajal fellowship number RYC-2015-17697 and from the grant number PID2019-107187GB-I00. T.S.M. acknowledges support from NASA grant 80NSSC20K0458. Computational time at the Texas Advanced Computing Center was provided through XSEDE allocation TG-AST090107. R.H.D.T. acknowledges support from NSF grants ACI-1663696, AST-1716436 and PHY-1748958, and NASA grant 80NSSC20K0515.

Some of the observations in the paper made use of the High-Resolution Imaging Instrument Zorro obtained under Gemini LLP Proposal Number: GN/S-2021A-LP-105. Zorro was funded by the NASA Exoplanet Exploration Program and built at the NASA AMES Research Center by Steve B. Howell, Nic Scott, Elliott P. Horch, and Emmett Quigley. This work has made use of data from the European Space Agency (ESA) mission
{\it Gaia} (\url{https://www.cosmos.esa.int/gaia}), processed by the {\it Gaia}
Data Processing and Analysis Consortium (DPAC,
\url{https://www.cosmos.esa.int/web/gaia/dpac/consortium}). Funding for the DPAC
has been provided by national institutions, in particular the institutions
participating in the {\it Gaia} Multilateral Agreement.

\facilities{\href{https://archive.stsci.edu/index.html}{MAST}, TESS, Gemini:South (Zorro), \href{https://www.cosmos.esa.int/gaia}{Gaia}}

\software{\texttt{ADIPLS} \citep{cd2008}, \texttt{AMP} \citep{metcalfe2003,metcalfe2009a,metcalfe2012b}, \texttt{ASTEC} \citep{cd2008}, \texttt{astropy} \citep{astropy}, \texttt{BASTA} \citep{silvaaguirre2015}, \texttt{BaSTI} \citep{pietrinferni2004,silvaaguirre2013}, \texttt{BeSPP} \citep{serenelli2013,serenelli2017}, \texttt{CESAM2k} \citep{morel2008}, \texttt{CESTAM} \citep{marques2013}, \texttt{DIAMONDS} \citep{corsaro2014}, \texttt{DSEP} \citep{dotter2007,dotter2008} \texttt{echelle} \citep{echelle}, \texttt{emcee} \citep{emcee}, \texttt{GARSTEC} \citep{weiss2008,silvaaguirre2012}, \texttt{GLS} \citep{gls}, \texttt{GYRE} \citep{townsend2013}, \texttt{isoclassify} \citep{huber2017,berger2020}, \texttt{Izmir} \citep{yildiz2019}, \texttt{kiauhoku} \citep{claytor2020}, \texttt{MESA} \citep[r10398, r12115,r15140;][]{paxton2011,paxton2013,paxton2015,paxton2018,paxton2019} \texttt{YB} \citep{basu2010,gai2011}, \texttt{YREC} \citep{demarque2008}, \texttt{YREC2} \citep{basu2012}, \texttt{YY} \citep{demarque2004}}

\bibliography{main.bib}

\begin{thebibliography}{}
\expandafter\ifx\csname natexlab\endcsname\relax\def\natexlab#1{#1}\fi
\providecommand{\url}[1]{\href{#1}{#1}}

\bibitem[{{Appourchaux} {et~al.}(2008){Appourchaux}, {Michel}, {Auvergne},
  {Baglin}, {Toutain}, {Baudin}, {Benomar}, {Chaplin}, {Deheuvels}, {Samadi},
  {Verner}, {Boumier}, {Garc{\'\i}a}, {Mosser}, {Hulot}, {Ballot}, {Barban},
  {Elsworth}, {Jim{\'e}nez-Reyes}, {Kjeldsen}, {R{\'e}gulo}, \&
  {Roxburgh}}]{appourchaux2008}
{Appourchaux}, T., {Michel}, E., {Auvergne}, M., {et~al.} 2008, \aap, 488, 705

\bibitem[{{Arentoft} {et~al.}(2008){Arentoft}, {Kjeldsen}, {Bedding}, {Bazot},
  {Christensen-Dalsgaard}, {Dall}, {Karoff}, {Carrier}, {Eggenberger},
  {Sosnowska}, {Wittenmyer}, {Endl}, {Metcalfe}, {Hekker}, {Reffert}, {Butler},
  {Bruntt}, {Kiss}, {O'Toole}, {Kambe}, {Ando}, {Izumiura}, {Sato}, {Hartmann},
  {Hatzes}, {Bouchy}, {Mosser}, {Appourchaux}, {Barban}, {Berthomieu},
  {Garcia}, {Michel}, {Provost}, {Turck-Chi{\`e}ze}, {Marti{\'c}}, {Lebrun},
  {Schmitt}, {Bertaux}, {Bonanno}, {Benatti}, {Claudi}, {Cosentino}, {Leccia},
  {Frandsen}, {Brogaard}, {Glowienka}, {Grundahl}, \&
  {Stempels}}]{arentoft2008}
{Arentoft}, T., {Kjeldsen}, H., {Bedding}, T.~R., {et~al.} 2008, \apj, 687,
  1180

\bibitem[{{Astropy Collaboration} {et~al.}(2013){Astropy Collaboration},
  {Robitaille}, {Tollerud}, {Greenfield}, {Droettboom}, {Bray}, {Aldcroft},
  {Davis}, {Ginsburg}, {Price-Whelan}, {Kerzendorf}, {Conley}, {Crighton},
  {Barbary}, {Muna}, {Ferguson}, {Grollier}, {Parikh}, {Nair}, {Unther},
  {Deil}, {Woillez}, {Conseil}, {Kramer}, {Turner}, {Singer}, {Fox}, {Weaver},
  {Zabalza}, {Edwards}, {Azalee Bostroem}, {Burke}, {Casey}, {Crawford},
  {Dencheva}, {Ely}, {Jenness}, {Labrie}, {Lim}, {Pierfederici}, {Pontzen},
  {Ptak}, {Refsdal}, {Servillat}, \& {Streicher}}]{astropy}
{Astropy Collaboration}, {Robitaille}, T.~P., {Tollerud}, E.~J., {et~al.} 2013,
  \aap, 558, A33

\bibitem[{{Baglin} {et~al.}(2006){Baglin}, {Auvergne}, {Boisnard}, {Lam-Trong},
  {Barge}, {Catala}, {Deleuil}, {Michel}, \& {Weiss}}]{baglin2006}
{Baglin}, A., {Auvergne}, M., {Boisnard}, L., {et~al.} 2006, in 36th COSPAR
  Scientific Assembly, Vol.~36, 3749

\bibitem[{{Baliunas} {et~al.}(1995){Baliunas}, {Donahue}, {Soon}, {Horne},
  {Frazer}, {Woodard-Eklund}, {Bradford}, {Rao}, {Wilson}, {Zhang}, {Bennett},
  {Briggs}, {Carroll}, {Duncan}, {Figueroa}, {Lanning}, {Misch}, {Mueller},
  {Noyes}, {Poppe}, {Porter}, {Robinson}, {Russell}, {Shelton}, {Soyumer},
  {Vaughan}, \& {Whitney}}]{baliunas1995}
{Baliunas}, S.~L., {Donahue}, R.~A., {Soon}, W.~H., {et~al.} 1995, \apj, 438,
  269

\bibitem[{{Ball} \& {Gizon}(2014)}]{ball2014}
{Ball}, W.~H., \& {Gizon}, L. 2014, \aap, 568, A123

\bibitem[{{Ball} {et~al.}(2020){Ball}, {Chaplin}, {Nielsen},
  {Gonz{\'a}lez-Cuesta}, {Mathur}, {Santos}, {Garc{\'\i}a}, {Buzasi}, {Mosser},
  {Deal}, {Stokholm}, {Mosumgaard}, {Silva Aguirre}, {Nsamba}, {Campante},
  {Cunha}, {Ong}, {Basu}, {{\"O}rtel}, {{\c{C}}elik Orhan}, {Y{\i}ld{\i}z},
  {Stassun}, {Kane}, \& {Huber}}]{ball2020}
{Ball}, W.~H., {Chaplin}, W.~J., {Nielsen}, M.~B., {et~al.} 2020, \mnras, 499,
  6084

\bibitem[{{Barnes}(2007)}]{barnes2007}
{Barnes}, S.~A. 2007, \apj, 669, 1167

\bibitem[{{Basu} {et~al.}(2010){Basu}, {Chaplin}, \& {Elsworth}}]{basu2010}
{Basu}, S., {Chaplin}, W.~J., \& {Elsworth}, Y. 2010, \apj, 710, 1596

\bibitem[{{Basu} {et~al.}(2012){Basu}, {Verner}, {Chaplin}, \&
  {Elsworth}}]{basu2012}
{Basu}, S., {Verner}, G.~A., {Chaplin}, W.~J., \& {Elsworth}, Y. 2012, \apj,
  746, 76

\bibitem[{{Bazot} {et~al.}(2011){Bazot}, {Ireland}, {Huber}, {Bedding},
  {Broomhall}, {Campante}, {Carfantan}, {Chaplin}, {Elsworth}, {Mel{\'e}ndez},
  {Petit}, {Th{\'e}ado}, {Van Grootel}, {Arentoft}, {Asplund}, {Castro},
  {Christensen-Dalsgaard}, {Do Nascimento}, {Dintrans}, {Dumusque}, {Kjeldsen},
  {McAlister}, {Metcalfe}, {Monteiro}, {Santos}, {Sousa}, {Sturmann},
  {Sturmann}, {ten Brummelaar}, {Turner}, \& {Vauclair}}]{bazot2011}
{Bazot}, M., {Ireland}, M.~J., {Huber}, D., {et~al.} 2011, \aap, 526, L4

\bibitem[{{Bedding} {et~al.}(2004){Bedding}, {Kjeldsen}, {Butler}, {McCarthy},
  {Marcy}, {O'Toole}, {Tinney}, \& {Wright}}]{bedding2004}
{Bedding}, T.~R., {Kjeldsen}, H., {Butler}, R.~P., {et~al.} 2004, \apj, 614,
  380

\bibitem[{{Bedding} {et~al.}(2001){Bedding}, {Butler}, {Kjeldsen}, {Baldry},
  {O'Toole}, {Tinney}, {Marcy}, {Kienzle}, \& {Carrier}}]{bedding2001}
{Bedding}, T.~R., {Butler}, R.~P., {Kjeldsen}, H., {et~al.} 2001, \apjl, 549,
  L105

\bibitem[{{Bedding} {et~al.}(2006){Bedding}, {Butler}, {Carrier}, {Bouchy},
  {Brewer}, {Eggenberger}, {Grundahl}, {Kjeldsen}, {McCarthy}, {Nielsen},
  {Retter}, \& {Tinney}}]{bedding2006}
{Bedding}, T.~R., {Butler}, R.~P., {Carrier}, F., {et~al.} 2006, \apj, 647, 558

\bibitem[{{Bedding} {et~al.}(2010){Bedding}, {Kjeldsen}, {Campante},
  {Appourchaux}, {Bonanno}, {Chaplin}, {Garcia}, {Marti{\'c}}, {Mosser},
  {Butler}, {Bruntt}, {Kiss}, {O'Toole}, {Kambe}, {Ando}, {Izumiura}, {Sato},
  {Hartmann}, {Hatzes}, {Barban}, {Berthomieu}, {Michel}, {Provost},
  {Turck-Chi{\`e}ze}, {Lebrun}, {Schmitt}, {Bertaux}, {Benatti}, {Claudi},
  {Cosentino}, {Leccia}, {Frandsen}, {Brogaard}, {Glowienka}, {Grundahl},
  {Stempels}, {Arentoft}, {Bazot}, {Christensen-Dalsgaard}, {Dall}, {Karoff},
  {Lundgreen-Nielsen}, {Carrier}, {Eggenberger}, {Sosnowska}, {Wittenmyer},
  {Endl}, {Metcalfe}, {Hekker}, \& {Reffert}}]{bedding2010}
{Bedding}, T.~R., {Kjeldsen}, H., {Campante}, T.~L., {et~al.} 2010, \apj, 713,
  935

\bibitem[{{Benomar} {et~al.}(2009){Benomar}, {Baudin}, {Campante}, {Chaplin},
  {Garc{\'\i}a}, {Gaulme}, {Toutain}, {Verner}, {Appourchaux}, {Ballot},
  {Barban}, {Elsworth}, {Mathur}, {Mosser}, {R{\'e}gulo}, {Roxburgh},
  {Auvergne}, {Baglin}, {Catala}, {Michel}, \& {Samadi}}]{benomar2009}
{Benomar}, O., {Baudin}, F., {Campante}, T.~L., {et~al.} 2009, \aap, 507, L13

\bibitem[{{Bensby} {et~al.}(2003){Bensby}, {Feltzing}, \&
  {Lundstr{\"o}m}}]{bensby2003}
{Bensby}, T., {Feltzing}, S., \& {Lundstr{\"o}m}, I. 2003, \aap, 410, 527

\bibitem[{{Bensby} {et~al.}(2014){Bensby}, {Feltzing}, \& {Oey}}]{bensby2014}
{Bensby}, T., {Feltzing}, S., \& {Oey}, M.~S. 2014, \aap, 562, A71

\bibitem[{{Berger} {et~al.}(2020){Berger}, {Huber}, {van Saders}, {Gaidos},
  {Tayar}, \& {Kraus}}]{berger2020}
{Berger}, T.~A., {Huber}, D., {van Saders}, J.~L., {et~al.} 2020, \aj, 159, 280

\bibitem[{{Bixel} \& {Apai}(2020)}]{bixel2020}
{Bixel}, A., \& {Apai}, D. 2020, \apj, 896, 131

\bibitem[{{Bonanno} {et~al.}(2008){Bonanno}, {Benatti}, {Claudi}, {Desidera},
  {Gratton}, {Leccia}, \& {Patern{\`o}}}]{bonanno2008}
{Bonanno}, A., {Benatti}, S., {Claudi}, R., {et~al.} 2008, \apj, 676, 1248

\bibitem[{{Bond} {et~al.}(2006){Bond}, {Tinney}, {Butler}, {Jones}, {Marcy},
  {Penny}, \& {Carter}}]{bond2006}
{Bond}, J.~C., {Tinney}, C.~G., {Butler}, R.~P., {et~al.} 2006, \mnras, 370,
  163

\bibitem[{{Borucki} {et~al.}(2010){Borucki}, {Koch}, {Basri}, {Batalha},
  {Brown}, {Caldwell}, {Caldwell}, {Christensen-Dalsgaard}, {Cochran},
  {DeVore}, {Dunham}, {Dupree}, {Gautier}, {Geary}, {Gilliland}, {Gould},
  {Howell}, {Jenkins}, {Kondo}, {Latham}, {Marcy}, {Meibom}, {Kjeldsen},
  {Lissauer}, {Monet}, {Morrison}, {Sasselov}, {Tarter}, {Boss}, {Brownlee},
  {Owen}, {Buzasi}, {Charbonneau}, {Doyle}, {Fortney}, {Ford}, {Holman},
  {Seager}, {Steffen}, {Welsh}, {Rowe}, {Anderson}, {Buchhave}, {Ciardi},
  {Walkowicz}, {Sherry}, {Horch}, {Isaacson}, {Everett}, {Fischer}, {Torres},
  {Johnson}, {Endl}, {MacQueen}, {Bryson}, {Dotson}, {Haas}, {Kolodziejczak},
  {Van Cleve}, {Chandrasekaran}, {Twicken}, {Quintana}, {Clarke}, {Allen},
  {Li}, {Wu}, {Tenenbaum}, {Verner}, {Bruhweiler}, {Barnes}, \&
  {Prsa}}]{borucki2010}
{Borucki}, W.~J., {Koch}, D., {Basri}, G., {et~al.} 2010, Science, 327, 977

\bibitem[{{Bouchy} {et~al.}(2005){Bouchy}, {Bazot}, {Santos}, {Vauclair}, \&
  {Sosnowska}}]{bouchy2005}
{Bouchy}, F., {Bazot}, M., {Santos}, N.~C., {Vauclair}, S., \& {Sosnowska}, D.
  2005, \aap, 440, 609

\bibitem[{{Bouchy} \& {Carrier}(2001)}]{bouchy2001}
{Bouchy}, F., \& {Carrier}, F. 2001, \aap, 374, L5

\bibitem[{{Bouchy} \& {Carrier}(2002)}]{bouchy2002}
---. 2002, \aap, 390, 205

\bibitem[{{Bovy} {et~al.}(2016){Bovy}, {Rix}, {Green}, {Schlafly}, \&
  {Finkbeiner}}]{bovy2016}
{Bovy}, J., {Rix}, H.-W., {Green}, G.~M., {Schlafly}, E.~F., \& {Finkbeiner},
  D.~P. 2016, \apj, 818, 130

\bibitem[{{Broomhall} {et~al.}(2011){Broomhall}, {Chaplin}, {Elsworth}, \&
  {New}}]{broomhall2011}
{Broomhall}, A.~M., {Chaplin}, W.~J., {Elsworth}, Y., \& {New}, R. 2011,
  \mnras, 413, 2978

\bibitem[{{Brown} {et~al.}(1991){Brown}, {Gilliland}, {Noyes}, \&
  {Ramsey}}]{brown1991}
{Brown}, T.~M., {Gilliland}, R.~L., {Noyes}, R.~W., \& {Ramsey}, L.~W. 1991,
  \apj, 368, 599

\bibitem[{{Bruntt} {et~al.}(2010){Bruntt}, {Bedding}, {Quirion}, {Lo Curto},
  {Carrier}, {Smalley}, {Dall}, {Arentoft}, {Bazot}, \& {Butler}}]{bruntt2010}
{Bruntt}, H., {Bedding}, T.~R., {Quirion}, P.~O., {et~al.} 2010, \mnras, 405,
  1907

\bibitem[{{Butler} {et~al.}(2004){Butler}, {Bedding}, {Kjeldsen}, {McCarthy},
  {O'Toole}, {Tinney}, {Marcy}, \& {Wright}}]{butler2004}
{Butler}, R.~P., {Bedding}, T.~R., {Kjeldsen}, H., {et~al.} 2004, \apjl, 600,
  L75

\bibitem[{{Campante} {et~al.}(2015){Campante}, {Barclay}, {Swift}, {Huber},
  {Adibekyan}, {Cochran}, {Burke}, {Isaacson}, {Quintana}, {Davies}, {Silva
  Aguirre}, {Ragozzine}, {Riddle}, {Baranec}, {Basu}, {Chaplin},
  {Christensen-Dalsgaard}, {Metcalfe}, {Bedding}, {Handberg}, {Stello},
  {Brewer}, {Hekker}, {Karoff}, {Kolbl}, {Law}, {Lundkvist}, {Miglio}, {Rowe},
  {Santos}, {Van Laerhoven}, {Arentoft}, {Elsworth}, {Fischer}, {Kawaler},
  {Kjeldsen}, {Lund}, {Marcy}, {Sousa}, {Sozzetti}, \& {White}}]{campante2015}
{Campante}, T.~L., {Barclay}, T., {Swift}, J.~J., {et~al.} 2015, \apj, 799, 170

\bibitem[{{Carrier} \& {Bourban}(2003)}]{carrier2003}
{Carrier}, F., \& {Bourban}, G. 2003, \aap, 406, L23

\bibitem[{{Carrier} {et~al.}(2005{\natexlab{a}}){Carrier}, {Eggenberger}, \&
  {Bouchy}}]{carrier2005b}
{Carrier}, F., {Eggenberger}, P., \& {Bouchy}, F. 2005{\natexlab{a}}, \aap,
  434, 1085

\bibitem[{{Carrier} {et~al.}(2005{\natexlab{b}}){Carrier}, {Eggenberger},
  {D'Alessandro}, \& {Weber}}]{carrier2005a}
{Carrier}, F., {Eggenberger}, P., {D'Alessandro}, A., \& {Weber}, L.
  2005{\natexlab{b}}, \na, 10, 315

\bibitem[{{Carrier} {et~al.}(2001){Carrier}, {Bouchy}, {Kienzle}, {Bedding},
  {Kjeldsen}, {Butler}, {Baldry}, {O'Toole}, {Tinney}, \&
  {Marcy}}]{carrier2001}
{Carrier}, F., {Bouchy}, F., {Kienzle}, F., {et~al.} 2001, \aap, 378, 142

\bibitem[{{Carrier} {et~al.}(2007){Carrier}, {Kjeldsen}, {Bedding}, {Brewer},
  {Butler}, {Eggenberger}, {Grundahl}, {McCarthy}, {Retter}, \&
  {Tinney}}]{carrier2007}
{Carrier}, F., {Kjeldsen}, H., {Bedding}, T.~R., {et~al.} 2007, \aap, 470, 1059

\bibitem[{{Casagrande} {et~al.}(2011){Casagrande}, {Sch{\"o}nrich}, {Asplund},
  {Cassisi}, {Ram{\'\i}rez}, {Mel{\'e}ndez}, {Bensby}, \&
  {Feltzing}}]{casagrande2011}
{Casagrande}, L., {Sch{\"o}nrich}, R., {Asplund}, M., {et~al.} 2011, \aap, 530,
  A138

\bibitem[{{Chaplin} {et~al.}(2014){Chaplin}, {Basu}, {Huber}, {Serenelli},
  {Casagrande}, {Silva Aguirre}, {Ball}, {Creevey}, {Gizon}, {Handberg},
  {Karoff}, {Lutz}, {Marques}, {Miglio}, {Stello}, {Suran}, {Pricopi},
  {Metcalfe}, {Monteiro}, {Molenda-{\.Z}akowicz}, {Appourchaux},
  {Christensen-Dalsgaard}, {Elsworth}, {Garc{\'{\i}}a}, {Houdek}, {Kjeldsen},
  {Bonanno}, {Campante}, {Corsaro}, {Gaulme}, {Hekker}, {Mathur}, {Mosser},
  {R{\'e}gulo}, \& {Salabert}}]{chaplin2014}
{Chaplin}, W.~J., {Basu}, S., {Huber}, D., {et~al.} 2014, \apjs, 210, 1

\bibitem[{{Chaplin} {et~al.}(2020){Chaplin}, {Serenelli}, {Miglio}, {Morel},
  {Mackereth}, {Vincenzo}, {Kjeldsen}, {Basu}, {Ball}, {Stokholm}, {Verma},
  {Mosumgaard}, {Silva Aguirre}, {Mazumdar}, {Ranadive}, {Antia}, {Lebreton},
  {Ong}, {Appourchaux}, {Bedding}, {Christensen-Dalsgaard}, {Creevey},
  {Garc{\'\i}a}, {Handberg}, {Huber}, {Kawaler}, {Lund}, {Metcalfe}, {Stassun},
  {Bazot}, {Beck}, {Bell}, {Bergemann}, {Buzasi}, {Benomar}, {Bossini},
  {Bugnet}, {Campante}, {Orhan}, {Corsaro}, {Gonz{\'a}lez-Cuesta}, {Davies},
  {Di Mauro}, {Egeland}, {Elsworth}, {Gaulme}, {Ghasemi}, {Guo}, {Hall},
  {Hasanzadeh}, {Hekker}, {Howe}, {Jenkins}, {Jim{\'e}nez}, {Kiefer},
  {Kuszlewicz}, {Kallinger}, {Latham}, {Lundkvist}, {Mathur}, {Montalb{\'a}n},
  {Mosser}, {Bed{\'o}n}, {Nielsen}, {{\"O}rtel}, {Rendle}, {Ricker},
  {Rodrigues}, {Roxburgh}, {Safari}, {Schofield}, {Seager}, {Smalley},
  {Stello}, {Szab{\'o}}, {Tayar}, {Theme{\ss}l}, {Thomas}, {Vanderspek}, {van
  Rossem}, {Vrard}, {Weiss}, {White}, {Winn}, \& {Y{\i}ld{\i}z}}]{chaplin2020}
{Chaplin}, W.~J., {Serenelli}, A.~M., {Miglio}, A., {et~al.} 2020, Nature
  Astronomy, 4, 382

\bibitem[{{Choi} {et~al.}(2016){Choi}, {Dotter}, {Conroy}, {Cantiello},
  {Paxton}, \& {Johnson}}]{choi2016}
{Choi}, J., {Dotter}, A., {Conroy}, C., {et~al.} 2016, \apj, 823, 102

\bibitem[{{Chontos} {et~al.}(2019){Chontos}, {Huber}, {Latham}, {Bieryla}, {Van
  Eylen}, {Bedding}, {Berger}, {Buchhave}, {Campante}, {Chaplin}, {Colman},
  {Coughlin}, {Davies}, {Hirano}, {Howard}, \& {Isaacson}}]{chontos2019}
{Chontos}, A., {Huber}, D., {Latham}, D.~W., {et~al.} 2019, \aj, 157, 192

\bibitem[{{Christensen-Dalsgaard}(2008)}]{cd2008}
{Christensen-Dalsgaard}, J. 2008, \apss, 316, 13

\bibitem[{{Claytor} {et~al.}(2020){Claytor}, {van Saders}, {Santos},
  {Garc{\'\i}a}, {Mathur}, {Tayar}, {Pinsonneault}, \&
  {Shetrone}}]{claytor2020}
{Claytor}, Z.~R., {van Saders}, J.~L., {Santos}, {\^A}. R.~G., {et~al.} 2020,
  \apj, 888, 43

\bibitem[{{Corsaro} \& {De Ridder}(2014)}]{corsaro2014}
{Corsaro}, E., \& {De Ridder}, J. 2014, \aap, 571, A71

\bibitem[{{Corsaro} {et~al.}(2018){Corsaro}, {De Ridder}, \&
  {Garc{\'\i}a}}]{corsaro2018}
{Corsaro}, E., {De Ridder}, J., \& {Garc{\'\i}a}, R.~A. 2018, \aap, 612, C2

\bibitem[{{Corsaro} {et~al.}(2017){Corsaro}, {Mathur}, {Garc{\'\i}a}, {Gaulme},
  {Pinsonneault}, {Stassun}, {Stello}, {Tayar}, {Trampedach}, {Jiang},
  {Nitschelm}, \& {Salabert}}]{corsaro2017}
{Corsaro}, E., {Mathur}, S., {Garc{\'\i}a}, R.~A., {et~al.} 2017, \aap, 605, A3

\bibitem[{{Curtis} {et~al.}(2019){Curtis}, {Ag{\"u}eros}, {Douglas}, \&
  {Meibom}}]{curtis2019}
{Curtis}, J.~L., {Ag{\"u}eros}, M.~A., {Douglas}, S.~T., \& {Meibom}, S. 2019,
  \apj, 879, 49

\bibitem[{{Cutri} {et~al.}(2003){Cutri}, {Skrutskie}, {van Dyk}, {Beichman},
  {Carpenter}, {Chester}, {Cambresy}, {Evans}, {Fowler}, {Gizis}, {Howard},
  {Huchra}, {Jarrett}, {Kopan}, {Kirkpatrick}, {Light}, {Marsh}, {McCallon},
  {Schneider}, {Stiening}, {Sykes}, {Weinberg}, {Wheaton}, {Wheelock}, \&
  {Zacarias}}]{cutri2003}
{Cutri}, R.~M., {Skrutskie}, M.~F., {van Dyk}, S., {et~al.} 2003, {2MASS All
  Sky Catalog of point sources.}

\bibitem[{{Cutri} {et~al.}(2012){Cutri}, {Skrutskie}, {van Dyk}, {Beichman},
  {Carpenter}, {Chester}, {Cambresy}, {Evans}, {Fowler}, {Gizis}, {Howard},
  {Huchra}, {Jarrett}, {Kopan}, {Kirkpatrick}, {Light}, {Marsh}, {McCallon},
  {Schneider}, {Stiening}, {Sykes}, {Weinberg}, {Wheaton}, {Wheelock}, \&
  {Zacharias}}]{cutri2012}
---. 2012, VizieR Online Data Catalog, II/281

\bibitem[{{da Silva} {et~al.}(2012){da Silva}, {Porto de Mello}, {Milone}, {da
  Silva}, {Ribeiro}, \& {Rocha-Pinto}}]{dasilva2012}
{da Silva}, R., {Porto de Mello}, G.~F., {Milone}, A.~C., {et~al.} 2012, \aap,
  542, A84

\bibitem[{{Dahn} {et~al.}(2002){Dahn}, {Harris}, {Vrba}, {Guetter}, {Canzian},
  {Henden}, {Levine}, {Luginbuhl}, {Monet}, {Monet}, {Pier}, {Stone}, {Walker},
  {Burgasser}, {Gizis}, {Kirkpatrick}, {Liebert}, \& {Reid}}]{dahn2002}
{Dahn}, C.~C., {Harris}, H.~C., {Vrba}, F.~J., {et~al.} 2002, \aj, 124, 1170

\bibitem[{{Damasso} \& {Del Sordo}(2017)}]{damasso2017}
{Damasso}, M., \& {Del Sordo}, F. 2017, \aap, 599, A126

\bibitem[{{Delfosse} {et~al.}(2000){Delfosse}, {Forveille}, {S{\'e}gransan},
  {Beuzit}, {Udry}, {Perrier}, \& {Mayor}}]{delfosse2000}
{Delfosse}, X., {Forveille}, T., {S{\'e}gransan}, D., {et~al.} 2000, \aap, 364,
  217

\bibitem[{{Demarque} {et~al.}(2008){Demarque}, {Guenther}, {Li}, {Mazumdar}, \&
  {Straka}}]{demarque2008}
{Demarque}, P., {Guenther}, D.~B., {Li}, L.~H., {Mazumdar}, A., \& {Straka},
  C.~W. 2008, \apss, 316, 31

\bibitem[{{Demarque} {et~al.}(2004){Demarque}, {Woo}, {Kim}, \&
  {Yi}}]{demarque2004}
{Demarque}, P., {Woo}, J.-H., {Kim}, Y.-C., \& {Yi}, S.~K. 2004, \apjs, 155,
  667

\bibitem[{{Dotter} {et~al.}(2007){Dotter}, {Chaboyer}, {Jevremovi{\'c}},
  {Baron}, {Ferguson}, {Sarajedini}, \& {Anderson}}]{dotter2007}
{Dotter}, A., {Chaboyer}, B., {Jevremovi{\'c}}, D., {et~al.} 2007, \aj, 134,
  376

\bibitem[{{Dotter} {et~al.}(2008){Dotter}, {Chaboyer}, {Jevremovi{\'c}},
  {Kostov}, {Baron}, \& {Ferguson}}]{dotter2008}
---. 2008, \apjs, 178, 89

\bibitem[{{Douglas} {et~al.}(2019){Douglas}, {Curtis}, {Ag{\"u}eros},
  {Cargile}, {Brewer}, {Meibom}, \& {Jansen}}]{douglas2019}
{Douglas}, S.~T., {Curtis}, J.~L., {Ag{\"u}eros}, M.~A., {et~al.} 2019, \apj,
  879, 100

\bibitem[{{Dumusque}(2018)}]{dumusque2018}
{Dumusque}, X. 2018, \aap, 620, A47

\bibitem[{{Eggenberger} {et~al.}(2007){Eggenberger}, {Udry}, {Chauvin},
  {Beuzit}, {Lagrange}, {S{\'e}gransan}, \& {Mayor}}]{eggenberger2007}
{Eggenberger}, A., {Udry}, S., {Chauvin}, G., {et~al.} 2007, \aap, 474, 273

\bibitem[{{Eiroa} {et~al.}(2013){Eiroa}, {Marshall}, {Mora}, {Montesinos},
  {Absil}, {Augereau}, {Bayo}, {Bryden}, {Danchi}, {del Burgo}, {Ertel},
  {Fridlund}, {Heras}, {Krivov}, {Launhardt}, {Liseau}, {L{\"o}hne},
  {Maldonado}, {Pilbratt}, {Roberge}, {Rodmann}, {Sanz-Forcada}, {Solano},
  {Stapelfeldt}, {Th{\'e}bault}, {Wolf}, {Ardila}, {Ar{\'e}valo}, {Beichmann},
  {Faramaz}, {Gonz{\'a}lez-Garc{\'\i}a}, {Guti{\'e}rrez}, {Lebreton},
  {Mart{\'\i}nez-Arn{\'a}iz}, {Meeus}, {Montes}, {Olofsson}, {Su}, {White},
  {Barrado}, {Fukagawa}, {Gr{\"u}n}, {Kamp}, {Lorente}, {Morbidelli},
  {M{\"u}ller}, {Mutschke}, {Nakagawa}, {Ribas}, \& {Walker}}]{eiroa2013}
{Eiroa}, C., {Marshall}, J.~P., {Mora}, A., {et~al.} 2013, \aap, 555, A11

\bibitem[{{Evans} {et~al.}(2018){Evans}, {Riello}, {De Angeli}, {Carrasco},
  {Montegriffo}, {Fabricius}, {Jordi}, {Palaversa}, {Diener}, {Busso},
  {Cacciari}, {van Leeuwen}, {Burgess}, {Davidson}, {Harrison}, {Hodgkin},
  {Pancino}, {Richards}, {Altavilla}, {Balaguer-N{\'u}{\~n}ez}, {Barstow},
  {Bellazzini}, {Brown}, {Castellani}, {Cocozza}, {De Luise}, {Delgado},
  {Ducourant}, {Galleti}, {Gilmore}, {Giuffrida}, {Holl}, {Kewley}, {Koposov},
  {Marinoni}, {Marrese}, {Osborne}, {Piersimoni}, {Portell}, {Pulone},
  {Ragaini}, {Sanna}, {Terrett}, {Walton}, {Wevers}, \&
  {Wyrzykowski}}]{evans2018}
{Evans}, D.~W., {Riello}, M., {De Angeli}, F., {et~al.} 2018, \aap, 616, A4

\bibitem[{{Feng} {et~al.}(2017){Feng}, {Tuomi}, \& {Jones}}]{feng2017}
{Feng}, F., {Tuomi}, M., \& {Jones}, H.~R.~A. 2017, \mnras, 470, 4794

\bibitem[{{Foreman-Mackey} {et~al.}(2013){Foreman-Mackey}, {Hogg}, {Lang}, \&
  {Goodman}}]{emcee}
{Foreman-Mackey}, D., {Hogg}, D.~W., {Lang}, D., \& {Goodman}, J. 2013, PASP,
  125, 306

\bibitem[{{Gai} {et~al.}(2011){Gai}, {Basu}, {Chaplin}, \&
  {Elsworth}}]{gai2011}
{Gai}, N., {Basu}, S., {Chaplin}, W.~J., \& {Elsworth}, Y. 2011, \apj, 730, 63

\bibitem[{{Gaia Collaboration} {et~al.}(2016){Gaia Collaboration}, {Prusti},
  {de Bruijne}, {Brown}, {Vallenari}, {Babusiaux}, {Bailer-Jones}, {Bastian},
  {Biermann}, {Evans}, {Eyer}, {Jansen}, {Jordi}, {Klioner}, {Lammers},
  {Lindegren}, {Luri}, {Mignard}, {Milligan}, {Panem}, {Poinsignon},
  {Pourbaix}, {Randich}, {Sarri}, {Sartoretti}, {Siddiqui}, {Soubiran},
  {Valette}, {van Leeuwen}, {Walton}, {Aerts}, {Arenou}, {Cropper}, {Drimmel},
  {H{\o}g}, {Katz}, {Lattanzi}, {O'Mullane}, {Grebel}, {Holland}, {Huc},
  {Passot}, {Bramante}, {Cacciari}, {Casta{\~n}eda}, {Chaoul}, {Cheek}, {De
  Angeli}, {Fabricius}, {Guerra}, {Hern{\'a}ndez}, {Jean-Antoine-Piccolo},
  {Masana}, {Messineo}, {Mowlavi}, {Nienartowicz}, {Ord{\'o}{\~n}ez-Blanco},
  {Panuzzo}, {Portell}, {Richards}, {Riello}, {Seabroke}, {Tanga},
  {Th{\'e}venin}, {Torra}, {Els}, {Gracia-Abril}, {Comoretto},
  {Garcia-Reinaldos}, {Lock}, {Mercier}, {Altmann}, {Andrae}, {Astraatmadja},
  {Bellas-Velidis}, {Benson}, {Berthier}, {Blomme}, {Busso}, {Carry},
  {Cellino}, {Clementini}, {Cowell}, {Creevey}, {Cuypers}, {Davidson}, {De
  Ridder}, {de Torres}, {Delchambre}, {Dell'Oro}, {Ducourant}, {Fr{\'e}mat},
  {Garc{\'\i}a-Torres}, {Gosset}, {Halbwachs}, {Hambly}, {Harrison}, {Hauser},
  {Hestroffer}, {Hodgkin}, {Huckle}, {Hutton}, {Jasniewicz}, {Jordan},
  {Kontizas}, {Korn}, {Lanzafame}, {Manteiga}, {Moitinho}, {Muinonen},
  {Osinde}, {Pancino}, {Pauwels}, {Petit}, {Recio-Blanco}, {Robin}, {Sarro},
  {Siopis}, {Smith}, {Smith}, {Sozzetti}, {Thuillot}, {van Reeven}, {Viala},
  {Abbas}, {Abreu Aramburu}, {Accart}, {Aguado}, {Allan}, {Allasia},
  {Altavilla}, {{\'A}lvarez}, {Alves}, {Anderson}, {Andrei}, {Anglada Varela},
  {Antiche}, {Antoja}, {Ant{\'o}n}, {Arcay}, {Atzei}, {Ayache}, {Bach},
  {Baker}, {Balaguer-N{\'u}{\~n}ez}, {Barache}, {Barata}, {Barbier}, {Barblan},
  {Baroni}, {Barrado y Navascu{\'e}s}, {Barros}, {Barstow}, {Becciani},
  {Bellazzini}, {Bellei}, {Bello Garc{\'\i}a}, {Belokurov}, {Bendjoya},
  {Berihuete}, {Bianchi}, {Bienaym{\'e}}, {Billebaud}, {Blagorodnova},
  {Blanco-Cuaresma}, {Boch}, {Bombrun}, {Borrachero}, {Bouquillon}, {Bourda},
  {Bouy}, {Bragaglia}, {Breddels}, {Brouillet}, {Br{\"u}semeister},
  {Bucciarelli}, {Budnik}, {Burgess}, {Burgon}, {Burlacu}, {Busonero}, {Buzzi},
  {Caffau}, {Cambras}, {Campbell}, {Cancelliere}, {Cantat-Gaudin}, {Carlucci},
  {Carrasco}, {Castellani}, {Charlot}, {Charnas}, {Charvet}, {Chassat},
  {Chiavassa}, {Clotet}, {Cocozza}, {Collins}, {Collins}, {Costigan}, {Crifo},
  {Cross}, {Crosta}, {Crowley}, {Dafonte}, {Damerdji}, {Dapergolas}, {David},
  {David}, {De Cat}, {de Felice}, {de Laverny}, {De Luise}, {De March}, {de
  Martino}, {de Souza}, {Debosscher}, {del Pozo}, {Delbo}, {Delgado},
  {Delgado}, {di Marco}, {Di Matteo}, {Diakite}, {Distefano}, {Dolding}, {Dos
  Anjos}, {Drazinos}, {Dur{\'a}n}, {Dzigan}, {Ecale}, {Edvardsson}, {Enke},
  {Erdmann}, {Escolar}, {Espina}, {Evans}, {Eynard Bontemps}, {Fabre},
  {Fabrizio}, {Faigler}, {Falc{\~a}o}, {Farr{\`a}s Casas}, {Faye}, {Federici},
  {Fedorets}, {Fern{\'a}ndez-Hern{\'a}ndez}, {Fernique}, {Fienga}, {Figueras},
  {Filippi}, {Findeisen}, {Fonti}, {Fouesneau}, {Fraile}, {Fraser}, {Fuchs},
  {Furnell}, {Gai}, {Galleti}, {Galluccio}, {Garabato}, {Garc{\'\i}a-Sedano},
  {Gar{\'e}}, {Garofalo}, {Garralda}, {Gavras}, {Gerssen}, {Geyer}, {Gilmore},
  {Girona}, {Giuffrida}, {Gomes}, {Gonz{\'a}lez-Marcos},
  {Gonz{\'a}lez-N{\'u}{\~n}ez}, {Gonz{\'a}lez-Vidal}, {Granvik}, {Guerrier},
  {Guillout}, {Guiraud}, {G{\'u}rpide}, {Guti{\'e}rrez-S{\'a}nchez}, {Guy},
  {Haigron}, {Hatzidimitriou}, {Haywood}, {Heiter}, {Helmi}, {Hobbs},
  {Hofmann}, {Holl}, {Holland }, {Hunt}, {Hypki}, {Icardi}, {Irwin}, {Jevardat
  de Fombelle}, {Jofr{\'e}}, {Jonker}, {Jorissen}, {Julbe}, {Karampelas},
  {Kochoska}, {Kohley}, {Kolenberg}, {Kontizas}, {Koposov}, {Kordopatis},
  {Koubsky}, {Kowalczyk}, {Krone-Martins}, {Kudryashova}, {Kull}, {Bachchan},
  {Lacoste-Seris}, {Lanza}, {Lavigne}, {Le Poncin-Lafitte}, {Lebreton},
  {Lebzelter}, {Leccia}, {Leclerc}, {Lecoeur-Taibi}, {Lemaitre}, {Lenhardt},
  {Leroux}, {Liao}, {Licata}, {Lindstr{\o}m}, {Lister}, {Livanou}, {Lobel},
  {L{\"o}ffler}, {L{\'o}pez}, {Lopez-Lozano}, {Lorenz}, {Loureiro},
  {MacDonald}, {Magalh{\~a}es Fernandes}, {Managau}, {Mann}, {Mantelet},
  {Marchal}, {Marchant}, {Marconi}, {Marie}, {Marinoni}, {Marrese},
  {Marschalk{\'o}}, {Marshall}, {Mart{\'\i}n-Fleitas}, {Martino}, {Mary},
  {Matijevi{\v{c}}}, {Mazeh}, {McMillan}, {Messina}, {Mestre}, {Michalik},
  {Millar}, {Miranda}, {Molina}, {Molinaro}, {Molinaro}, {Moln{\'a}r},
  {Moniez}, {Montegriffo}, {Monteiro}, {Mor}, {Mora}, {Morbidelli}, {Morel},
  {Morgenthaler}, {Morley}, {Morris}, {Mulone}, {Muraveva}, {Musella},
  {Narbonne}, {Nelemans}, {Nicastro}, {Noval}, {Ord{\'e}novic},
  {Ordieres-Mer{\'e}}, {Osborne}, {Pagani}, {Pagano}, {Pailler}, {Palacin},
  {Palaversa}, {Parsons}, {Paulsen}, {Pecoraro}, {Pedrosa}, {Pentik{\"a}inen},
  {Pereira}, {Pichon}, {Piersimoni}, {Pineau}, {Plachy}, {Plum}, {Poujoulet},
  {Pr{\v{s}}a}, {Pulone}, {Ragaini}, {Rago}, {Rambaux}, {Ramos-Lerate},
  {Ranalli}, {Rauw}, {Read}, {Regibo}, {Renk}, {Reyl{\'e}}, {Ribeiro},
  {Rimoldini}, {Ripepi}, {Riva}, {Rixon}, {Roelens}, {Romero-G{\'o}mez},
  {Rowell}, {Royer}, {Rudolph}, {Ruiz-Dern}, {Sadowski}, {Sagrist{\`a}
  Sell{\'e}s}, {Sahlmann}, {Salgado}, {Salguero}, {Sarasso}, {Savietto},
  {Schnorhk}, {Schultheis}, {Sciacca}, {Segol}, {Segovia}, {Segransan},
  {Serpell}, {Shih}, {Smareglia}, {Smart}, {Smith}, {Solano}, {Solitro},
  {Sordo}, {Soria Nieto}, {Souchay}, {Spagna}, {Spoto}, {Stampa}, {Steele},
  {Steidelm{\"u}ller}, {Stephenson}, {Stoev}, {Suess}, {S{\"u}veges}, {Surdej},
  {Szabados}, {Szegedi-Elek}, {Tapiador}, {Taris}, {Tauran}, {Taylor},
  {Teixeira}, {Terrett}, {Tingley}, {Trager}, {Turon}, {Ulla}, {Utrilla},
  {Valentini}, {van Elteren}, {Van Hemelryck}, {van Leeuwen}, {Varadi},
  {Vecchiato}, {Veljanoski}, {Via}, {Vicente}, {Vogt}, {Voss}, {Votruba},
  {Voutsinas}, {Walmsley}, {Weiler}, {Weingrill}, {Werner}, {Wevers},
  {Whitehead}, {Wyrzykowski}, {Yoldas}, {{\v{Z}}erjal}, {Zucker}, {Zurbach},
  {Zwitter}, {Alecu}, {Allen}, {Allende Prieto}, {Amorim},
  {Anglada-Escud{\'e}}, {Arsenijevic}, {Azaz}, {Balm}, {Beck}, {Bernstein},
  {Bigot}, {Bijaoui}, {Blasco}, {Bonfigli}, {Bono}, {Boudreault}, {Bressan},
  {Brown}, {Brunet}, {Bunclark}, {Buonanno}, {Butkevich}, {Carret}, {Carrion},
  {Chemin}, {Ch{\'e}reau}, {Corcione}, {Darmigny}, {de Boer}, {de Teodoro}, {de
  Zeeuw}, {Delle Luche}, {Domingues}, {Dubath}, {Fodor}, {Fr{\'e}zouls},
  {Fries}, {Fustes}, {Fyfe}, {Gallardo}, {Gallegos}, {Gardiol}, {Gebran},
  {Gomboc}, {G{\'o}mez}, {Grux}, {Gueguen}, {Heyrovsky}, {Hoar}, {Iannicola},
  {Isasi Parache}, {Janotto}, {Joliet}, {Jonckheere}, {Keil}, {Kim},
  {Klagyivik}, {Klar}, {Knude}, {Kochukhov}, {Kolka}, {Kos}, {Kutka}, {Lainey},
  {LeBouquin}, {Liu}, {Loreggia}, {Makarov}, {Marseille}, {Martayan},
  {Martinez-Rubi}, {Massart}, {Meynadier}, {Mignot}, {Munari}, {Nguyen},
  {Nordlander}, {Ocvirk}, {O'Flaherty}, {Olias Sanz}, {Ortiz}, {Osorio},
  {Oszkiewicz}, {Ouzounis}, {Palmer}, {Park}, {Pasquato}, {Peltzer}, {Peralta},
  {P{\'e}turaud}, {Pieniluoma}, {Pigozzi}, {Poels}, {Prat}, {Prod'homme},
  {Raison}, {Rebordao}, {Risquez}, {Rocca-Volmerange}, {Rosen}, {Ruiz-Fuertes},
  {Russo}, {Sembay}, {Serraller Vizcaino}, {Short}, {Siebert}, {Silva},
  {Sinachopoulos}, {Slezak}, {Soffel}, {Sosnowska}, {Strai{\v{z}}ys}, {ter
  Linden}, {Terrell}, {Theil}, {Tiede}, {Troisi}, {Tsalmantza}, {Tur},
  {Vaccari}, {Vachier}, {Valles}, {Van Hamme}, {Veltz}, {Virtanen}, {Wallut},
  {Wichmann}, {Wilkinson}, {Ziaeepour}, \& {Zschocke}}]{gaia2016}
{Gaia Collaboration}, {Prusti}, T., {de Bruijne}, J.~H.~J., {et~al.} 2016,
  \aap, 595, A1

\bibitem[{{Gaia Collaboration} {et~al.}(2018){Gaia Collaboration}, {Brown},
  {Vallenari}, {Prusti}, {de Bruijne}, {Babusiaux}, {Bailer-Jones}, {Biermann},
  {Evans}, {Eyer}, {Jansen}, {Jordi}, {Klioner}, {Lammers}, {Lindegren},
  {Luri}, {Mignard}, {Panem}, {Pourbaix}, {Randich}, {Sartoretti}, {Siddiqui},
  {Soubiran}, {van Leeuwen}, {Walton}, {Arenou}, {Bastian}, {Cropper},
  {Drimmel}, {Katz}, {Lattanzi}, {Bakker}, {Cacciari}, {Casta{\~n}eda},
  {Chaoul}, {Cheek}, {De Angeli}, {Fabricius}, {Guerra}, {Holl}, {Masana},
  {Messineo}, {Mowlavi}, {Nienartowicz}, {Panuzzo}, {Portell}, {Riello},
  {Seabroke}, {Tanga}, {Th{\'e}venin}, {Gracia-Abril}, {Comoretto},
  {Garcia-Reinaldos}, {Teyssier}, {Altmann}, {Andrae}, {Audard},
  {Bellas-Velidis}, {Benson}, {Berthier}, {Blomme}, {Burgess}, {Busso},
  {Carry}, {Cellino}, {Clementini}, {Clotet}, {Creevey}, {Davidson}, {De
  Ridder}, {Delchambre}, {Dell'Oro}, {Ducourant},
  {Fern{\'a}ndez-Hern{\'a}ndez}, {Fouesneau}, {Fr{\'e}mat}, {Galluccio},
  {Garc{\'\i}a-Torres}, {Gonz{\'a}lez-N{\'u}{\~n}ez}, {Gonz{\'a}lez-Vidal},
  {Gosset}, {Guy}, {Halbwachs}, {Hambly}, {Harrison}, {Hern{\'a}ndez},
  {Hestroffer}, {Hodgkin}, {Hutton}, {Jasniewicz}, {Jean-Antoine-Piccolo},
  {Jordan}, {Korn}, {Krone-Martins}, {Lanzafame}, {Lebzelter}, {L{\"o}ffler},
  {Manteiga}, {Marrese}, {Mart{\'\i}n-Fleitas}, {Moitinho}, {Mora}, {Muinonen},
  {Osinde}, {Pancino}, {Pauwels}, {Petit}, {Recio-Blanco}, {Richards},
  {Rimoldini}, {Robin}, {Sarro}, {Siopis}, {Smith}, {Sozzetti}, {S{\"u}veges},
  {Torra}, {van Reeven}, {Abbas}, {Abreu Aramburu}, {Accart}, {Aerts},
  {Altavilla}, {{\'A}lvarez}, {Alvarez}, {Alves}, {Anderson}, {Andrei},
  {Anglada Varela}, {Antiche}, {Antoja}, {Arcay}, {Astraatmadja}, {Bach},
  {Baker}, {Balaguer-N{\'u}{\~n}ez}, {Balm}, {Barache}, {Barata}, {Barbato},
  {Barblan}, {Barklem}, {Barrado}, {Barros}, {Barstow}, {Bartholom{\'e}
  Mu{\~n}oz}, {Bassilana}, {Becciani}, {Bellazzini}, {Berihuete}, {Bertone},
  {Bianchi}, {Bienaym{\'e}}, {Blanco-Cuaresma}, {Boch}, {Boeche}, {Bombrun},
  {Borrachero}, {Bossini}, {Bouquillon}, {Bourda}, {Bragaglia}, {Bramante},
  {Breddels}, {Bressan}, {Brouillet}, {Br{\"u}semeister}, {Brugaletta},
  {Bucciarelli}, {Burlacu}, {Busonero}, {Butkevich}, {Buzzi}, {Caffau},
  {Cancelliere}, {Cannizzaro}, {Cantat-Gaudin}, {Carballo}, {Carlucci},
  {Carrasco}, {Casamiquela}, {Castellani}, {Castro-Ginard}, {Charlot},
  {Chemin}, {Chiavassa}, {Cocozza}, {Costigan}, {Cowell}, {Crifo}, {Crosta},
  {Crowley}, {Cuypers}, {Dafonte}, {Damerdji}, {Dapergolas}, {David}, {David},
  {de Laverny}, {De Luise}, {De March}, {de Martino}, {de Souza}, {de Torres},
  {Debosscher}, {del Pozo}, {Delbo}, {Delgado}, {Delgado}, {Di Matteo},
  {Diakite}, {Diener}, {Distefano}, {Dolding}, {Drazinos}, {Dur{\'a}n},
  {Edvardsson}, {Enke}, {Eriksson}, {Esquej}, {Eynard Bontemps}, {Fabre},
  {Fabrizio}, {Faigler}, {Falc{\~a}o}, {Farr{\`a}s Casas}, {Federici},
  {Fedorets}, {Fernique}, {Figueras}, {Filippi}, {Findeisen}, {Fonti},
  {Fraile}, {Fraser}, {Fr{\'e}zouls}, {Gai}, {Galleti}, {Garabato},
  {Garc{\'\i}a-Sedano}, {Garofalo}, {Garralda}, {Gavel}, {Gavras}, {Gerssen},
  {Geyer}, {Giacobbe}, {Gilmore}, {Girona}, {Giuffrida}, {Glass}, {Gomes},
  {Granvik}, {Gueguen}, {Guerrier}, {Guiraud}, {Guti{\'e}rrez-S{\'a}nchez},
  {Haigron}, {Hatzidimitriou}, {Hauser}, {Haywood}, {Heiter}, {Helmi}, {Heu},
  {Hilger}, {Hobbs}, {Hofmann}, {Holland}, {Huckle}, {Hypki}, {Icardi},
  {Jan{\ss}en}, {Jevardat de Fombelle}, {Jonker}, {Juh{\'a}sz}, {Julbe},
  {Karampelas}, {Kewley}, {Klar}, {Kochoska}, {Kohley}, {Kolenberg},
  {Kontizas}, {Kontizas}, {Koposov}, {Kordopatis}, {Kostrzewa-Rutkowska},
  {Koubsky}, {Lambert}, {Lanza}, {Lasne}, {Lavigne}, {Le Fustec}, {Le
  Poncin-Lafitte}, {Lebreton}, {Leccia}, {Leclerc}, {Lecoeur-Taibi},
  {Lenhardt}, {Leroux}, {Liao}, {Licata}, {Lindstr{\o}m}, {Lister}, {Livanou},
  {Lobel}, {L{\'o}pez}, {Managau}, {Mann}, {Mantelet}, {Marchal}, {Marchant},
  {Marconi}, {Marinoni}, {Marschalk{\'o}}, {Marshall}, {Martino}, {Marton},
  {Mary}, {Massari}, {Matijevi{\v{c}}}, {Mazeh}, {McMillan}, {Messina},
  {Michalik}, {Millar}, {Molina}, {Molinaro}, {Moln{\'a}r}, {Montegriffo},
  {Mor}, {Morbidelli}, {Morel}, {Morris}, {Mulone}, {Muraveva}, {Musella},
  {Nelemans}, {Nicastro}, {Noval}, {O'Mullane}, {Ord{\'e}novic},
  {Ord{\'o}{\~n}ez-Blanco}, {Osborne}, {Pagani}, {Pagano}, {Pailler},
  {Palacin}, {Palaversa}, {Panahi}, {Pawlak}, {Piersimoni}, {Pineau}, {Plachy},
  {Plum}, {Poggio}, {Poujoulet}, {Pr{\v{s}}a}, {Pulone}, {Racero}, {Ragaini},
  {Rambaux}, {Ramos-Lerate}, {Regibo}, {Reyl{\'e}}, {Riclet}, {Ripepi}, {Riva},
  {Rivard}, {Rixon}, {Roegiers}, {Roelens}, {Romero-G{\'o}mez}, {Rowell},
  {Royer}, {Ruiz-Dern}, {Sadowski}, {Sagrist{\`a} Sell{\'e}s}, {Sahlmann},
  {Salgado}, {Salguero}, {Sanna}, {Santana-Ros}, {Sarasso}, {Savietto},
  {Schultheis}, {Sciacca}, {Segol}, {Segovia}, {S{\'e}gransan}, {Shih},
  {Siltala}, {Silva}, {Smart}, {Smith}, {Solano}, {Solitro}, {Sordo}, {Soria
  Nieto}, {Souchay}, {Spagna}, {Spoto}, {Stampa}, {Steele},
  {Steidelm{\"u}ller}, {Stephenson}, {Stoev}, {Suess}, {Surdej}, {Szabados},
  {Szegedi-Elek}, {Tapiador}, {Taris}, {Tauran}, {Taylor}, {Teixeira},
  {Terrett}, {Teyssand ier}, {Thuillot}, {Titarenko}, {Torra Clotet}, {Turon},
  {Ulla}, {Utrilla}, {Uzzi}, {Vaillant}, {Valentini}, {Valette}, {van Elteren},
  {Van Hemelryck}, {van Leeuwen}, {Vaschetto}, {Vecchiato}, {Veljanoski},
  {Viala}, {Vicente}, {Vogt}, {von Essen}, {Voss}, {Votruba}, {Voutsinas},
  {Walmsley}, {Weiler}, {Wertz}, {Wevers}, {Wyrzykowski}, {Yoldas},
  {{\v{Z}}erjal}, {Ziaeepour}, {Zorec}, {Zschocke}, {Zucker}, {Zurbach}, \&
  {Zwitter}}]{gaia2018}
{Gaia Collaboration}, {Brown}, A.~G.~A., {Vallenari}, A., {et~al.} 2018, \aap,
  616, A1

\bibitem[{{Gaia Collaboration} {et~al.}(2021){Gaia Collaboration}, {Brown},
  {Vallenari}, {Prusti}, {de Bruijne}, {Babusiaux}, {Biermann}, {Creevey},
  {Evans}, {Eyer}, {Hutton}, {Jansen}, {Jordi}, {Klioner}, {Lammers},
  {Lindegren}, {Luri}, {Mignard}, {Panem}, {Pourbaix}, {Randich}, {Sartoretti},
  {Soubiran}, {Walton}, {Arenou}, {Bailer-Jones}, {Bastian}, {Cropper},
  {Drimmel}, {Katz}, {Lattanzi}, {van Leeuwen}, {Bakker}, {Cacciari},
  {Casta{\~n}eda}, {De Angeli}, {Ducourant}, {Fabricius}, {Fouesneau},
  {Fr{\'e}mat}, {Guerra}, {Guerrier}, {Guiraud}, {Jean-Antoine Piccolo},
  {Masana}, {Messineo}, {Mowlavi}, {Nicolas}, {Nienartowicz}, {Pailler},
  {Panuzzo}, {Riclet}, {Roux}, {Seabroke}, {Sordo}, {Tanga}, {Th{\'e}venin},
  {Gracia-Abril}, {Portell}, {Teyssier}, {Altmann}, {Andrae}, {Bellas-Velidis},
  {Benson}, {Berthier}, {Blomme}, {Brugaletta}, {Burgess}, {Busso}, {Carry},
  {Cellino}, {Cheek}, {Clementini}, {Damerdji}, {Davidson}, {Delchambre},
  {Dell'Oro}, {Fern{\'a}ndez-Hern{\'a}ndez}, {Galluccio}, {Garc{\'\i}a-Lario},
  {Garcia-Reinaldos}, {Gonz{\'a}lez-N{\'u}{\~n}ez}, {Gosset}, {Haigron},
  {Halbwachs}, {Hambly}, {Harrison}, {Hatzidimitriou}, {Heiter},
  {Hern{\'a}ndez}, {Hestroffer}, {Hodgkin}, {Holl}, {Jan{\ss}en}, {Jevardat de
  Fombelle}, {Jordan}, {Krone-Martins}, {Lanzafame}, {L{\"o}ffler}, {Lorca},
  {Manteiga}, {Marchal}, {Marrese}, {Moitinho}, {Mora}, {Muinonen}, {Osborne},
  {Pancino}, {Pauwels}, {Petit}, {Recio-Blanco}, {Richards}, {Riello},
  {Rimoldini}, {Robin}, {Roegiers}, {Rybizki}, {Sarro}, {Siopis}, {Smith},
  {Sozzetti}, {Ulla}, {Utrilla}, {van Leeuwen}, {van Reeven}, {Abbas}, {Abreu
  Aramburu}, {Accart}, {Aerts}, {Aguado}, {Ajaj}, {Altavilla}, {{\'A}lvarez},
  {{\'A}lvarez Cid-Fuentes}, {Alves}, {Anderson}, {Anglada Varela}, {Antoja},
  {Audard}, {Baines}, {Baker}, {Balaguer-N{\'u}{\~n}ez}, {Balbinot}, {Balog},
  {Barache}, {Barbato}, {Barros}, {Barstow}, {Bartolom{\'e}}, {Bassilana},
  {Bauchet}, {Baudesson-Stella}, {Becciani}, {Bellazzini}, {Bernet}, {Bertone},
  {Bianchi}, {Blanco-Cuaresma}, {Boch}, {Bombrun}, {Bossini}, {Bouquillon},
  {Bragaglia}, {Bramante}, {Breedt}, {Bressan}, {Brouillet}, {Bucciarelli},
  {Burlacu}, {Busonero}, {Butkevich}, {Buzzi}, {Caffau}, {Cancelliere},
  {C{\'a}novas}, {Cantat-Gaudin}, {Carballo}, {Carlucci}, {Carnerero},
  {Carrasco}, {Casamiquela}, {Castellani}, {Castro-Ginard}, {Castro Sampol},
  {Chaoul}, {Charlot}, {Chemin}, {Chiavassa}, {Cioni}, {Comoretto}, {Cooper},
  {Cornez}, {Cowell}, {Crifo}, {Crosta}, {Crowley}, {Dafonte}, {Dapergolas},
  {David}, {David}, {de Laverny}, {De Luise}, {De March}, {De Ridder}, {de
  Souza}, {de Teodoro}, {de Torres}, {del Peloso}, {del Pozo}, {Delbo},
  {Delgado}, {Delgado}, {Delisle}, {Di Matteo}, {Diakite}, {Diener},
  {Distefano}, {Dolding}, {Eappachen}, {Edvardsson}, {Enke}, {Esquej}, {Fabre},
  {Fabrizio}, {Faigler}, {Fedorets}, {Fernique}, {Fienga}, {Figueras},
  {Fouron}, {Fragkoudi}, {Fraile}, {Franke}, {Gai}, {Garabato},
  {Garcia-Gutierrez}, {Garc{\'\i}a-Torres}, {Garofalo}, {Gavras}, {Gerlach},
  {Geyer}, {Giacobbe}, {Gilmore}, {Girona}, {Giuffrida}, {Gomel}, {Gomez},
  {Gonzalez-Santamaria}, {Gonz{\'a}lez-Vidal}, {Granvik},
  {Guti{\'e}rrez-S{\'a}nchez}, {Guy}, {Hauser}, {Haywood}, {Helmi}, {Hidalgo},
  {Hilger}, {H{\l}adczuk}, {Hobbs}, {Holland}, {Huckle}, {Jasniewicz},
  {Jonker}, {Juaristi Campillo}, {Julbe}, {Karbevska}, {Kervella}, {Khanna},
  {Kochoska}, {Kontizas}, {Kordopatis}, {Korn}, {Kostrzewa-Rutkowska},
  {Kruszy{\'n}ska}, {Lambert}, {Lanza}, {Lasne}, {Le Campion}, {Le Fustec},
  {Lebreton}, {Lebzelter}, {Leccia}, {Leclerc}, {Lecoeur-Taibi}, {Liao},
  {Licata}, {Lindstr{\o}m}, {Lister}, {Livanou}, {Lobel}, {Madrero Pardo},
  {Managau}, {Mann}, {Marchant}, {Marconi}, {Marcos Santos}, {Marinoni},
  {Marocco}, {Marshall}, {Martin Polo}, {Mart{\'\i}n-Fleitas}, {Masip},
  {Massari}, {Mastrobuono-Battisti}, {Mazeh}, {McMillan}, {Messina},
  {Michalik}, {Millar}, {Mints}, {Molina}, {Molinaro}, {Moln{\'a}r},
  {Montegriffo}, {Mor}, {Morbidelli}, {Morel}, {Morris}, {Mulone}, {Munoz},
  {Muraveva}, {Murphy}, {Musella}, {Noval}, {Ord{\'e}novic}, {Orr{\`u}},
  {Osinde}, {Pagani}, {Pagano}, {Palaversa}, {Palicio}, {Panahi}, {Pawlak},
  {Pe{\~n}alosa Esteller}, {Penttil{\"a}}, {Piersimoni}, {Pineau}, {Plachy},
  {Plum}, {Poggio}, {Poretti}, {Poujoulet}, {Pr{\v{s}}a}, {Pulone}, {Racero},
  {Ragaini}, {Rainer}, {Raiteri}, {Rambaux}, {Ramos}, {Ramos-Lerate}, {Re
  Fiorentin}, {Regibo}, {Reyl{\'e}}, {Ripepi}, {Riva}, {Rixon}, {Robichon},
  {Robin}, {Roelens}, {Rohrbasser}, {Romero-G{\'o}mez}, {Rowell}, {Royer},
  {Rybicki}, {Sadowski}, {Sagrist{\`a} Sell{\'e}s}, {Sahlmann}, {Salgado},
  {Salguero}, {Samaras}, {Sanchez Gimenez}, {Sanna}, {Santove{\~n}a},
  {Sarasso}, {Schultheis}, {Sciacca}, {Segol}, {Segovia}, {S{\'e}gransan},
  {Semeux}, {Shahaf}, {Siddiqui}, {Siebert}, {Siltala}, {Slezak}, {Smart},
  {Solano}, {Solitro}, {Souami}, {Souchay}, {Spagna}, {Spoto}, {Steele},
  {Steidelm{\"u}ller}, {Stephenson}, {S{\"u}veges}, {Szabados}, {Szegedi-Elek},
  {Taris}, {Tauran}, {Taylor}, {Teixeira}, {Thuillot}, {Tonello}, {Torra},
  {Torra}, {Turon}, {Unger}, {Vaillant}, {van Dillen}, {Vanel}, {Vecchiato},
  {Viala}, {Vicente}, {Voutsinas}, {Weiler}, {Wevers}, {Wyrzykowski}, {Yoldas},
  {Yvard}, {Zhao}, {Zorec}, {Zucker}, {Zurbach}, \& {Zwitter}}]{gaia2021}
---. 2021, \aap, 649, A1

\bibitem[{{Garc{\'\i}a} \& {Ballot}(2019)}]{garcia2019}
{Garc{\'\i}a}, R.~A., \& {Ballot}, J. 2019, Living Reviews in Solar Physics,
  16, 4

\bibitem[{{Garc{\'\i}a} {et~al.}(2010){Garc{\'\i}a}, {Mathur}, {Salabert},
  {Ballot}, {R{\'e}gulo}, {Metcalfe}, \& {Baglin}}]{garcia2010}
{Garc{\'\i}a}, R.~A., {Mathur}, S., {Salabert}, D., {et~al.} 2010, Science,
  329, 1032

\bibitem[{{Garc{\'I}a} {et~al.}(2001){Garc{\'I}a}, {R{\'e}gulo},
  {Turck-Chi{\`e}ze}, {Bertello}, {Kosovichev}, {Brun}, {Couvidat}, {Henney},
  {Lazrek}, {Ulrich}, \& {Varadi}}]{garcia2001}
{Garc{\'I}a}, R.~A., {R{\'e}gulo}, C., {Turck-Chi{\`e}ze}, S., {et~al.} 2001,
  \solphys, 200, 361

\bibitem[{{Garc{\'\i}a} {et~al.}(2009){Garc{\'\i}a}, {R{\'e}gulo}, {Samadi},
  {Ballot}, {Barban}, {Benomar}, {Chaplin}, {Gaulme}, {Appourchaux}, {Mathur},
  {Mosser}, {Toutain}, {Verner}, {Auvergne}, {Baglin}, {Baudin}, {Boumier},
  {Bruntt}, {Catala}, {Deheuvels}, {Elsworth}, {Jim{\'e}nez-Reyes}, {Michel},
  {P{\'e}rez Hern{\'a}ndez}, {Roxburgh}, \& {Salabert}}]{garcia2009}
{Garc{\'\i}a}, R.~A., {R{\'e}gulo}, C., {Samadi}, R., {et~al.} 2009, \aap, 506,
  41

\bibitem[{{Gaudi} {et~al.}(2020){Gaudi}, {Seager}, {Mennesson}, {Kiessling},
  {Warfield}, {Cahoy}, {Clarke}, {Domagal-Goldman}, {Feinberg}, {Guyon},
  {Kasdin}, {Mawet}, {Plavchan}, {Robinson}, {Rogers}, {Scowen}, {Somerville},
  {Stapelfeldt}, {Stark}, {Stern}, {Turnbull}, {Amini}, {Kuan}, {Martin},
  {Morgan}, {Redding}, {Stahl}, {Webb}, {Alvarez-Salazar}, {Arnold}, {Arya},
  {Balasubramanian}, {Baysinger}, {Bell}, {Below}, {Benson}, {Blais}, {Booth},
  {Bourgeois}, {Bradford}, {Brewer}, {Brooks}, {Cady}, {Caldwell}, {Calvet},
  {Carr}, {Chan}, {Cormarkovic}, {Coste}, {Cox}, {Danner}, {Davis}, {Dewell},
  {Dorsett}, {Dunn}, {East}, {Effinger}, {Eng}, {Freebury}, {Garcia}, {Gaskin},
  {Greene}, {Hennessy}, {Hilgemann}, {Hood}, {Holota}, {Howe}, {Huang}, {Hull},
  {Hunt}, {Hurd}, {Johnson}, {Kissil}, {Knight}, {Kolenz}, {Kraus}, {Krist},
  {Li}, {Lisman}, {Mandic}, {Mann}, {Marchen}, {Marrese-Reading}, {McCready},
  {McGown}, {Missun}, {Miyaguchi}, {Moore}, {Nemati}, {Nikzad}, {Nissen},
  {Novicki}, {Perrine}, {Pineda}, {Polanco}, {Putnam}, {Qureshi}, {Richards},
  {Eldorado Riggs}, {Rodgers}, {Rud}, {Saini}, {Scalisi}, {Scharf}, {Schulz},
  {Serabyn}, {Sigrist}, {Sikkia}, {Singleton}, {Shaklan}, {Smith}, {Southerd},
  {Stahl}, {Steeves}, {Sturges}, {Sullivan}, {Tang}, {Taras}, {Tesch},
  {Therrell}, {Tseng}, {Valente}, {Van Buren}, {Villalvazo}, {Warwick}, {Webb},
  {Westerhoff}, {Wofford}, {Wu}, {Woo}, {Wood}, {Ziemer}, {Arney}, {Anderson},
  {Ma{\'\i}z-Apell{\'a}niz}, {Bartlett}, {Belikov}, {Bendek}, {Cenko},
  {Douglas}, {Dulz}, {Evans}, {Faramaz}, {Feng}, {Ferguson}, {Follette},
  {Ford}, {Garc{\'\i}a}, {Geha}, {Gelino}, {G{\"o}tberg}, {Hildebrand t}, {Hu},
  {Jahnke}, {Kennedy}, {Kreidberg}, {Isella}, {Lopez}, {Marchis}, {Macri},
  {Marley}, {Matzko}, {Mazoyer}, {McCandliss}, {Meshkat}, {Mordasini},
  {Morris}, {Nielsen}, {Newman}, {Petigura}, {Postman}, {Reines}, {Roberge},
  {Roederer}, {Ruane}, {Schwieterman}, {Sirbu}, {Spalding}, {Teplitz},
  {Tumlinson}, {Turner}, {Werk}, {Wofford}, {Wyatt}, {Young}, \&
  {Zellem}}]{gaudi2020}
{Gaudi}, B.~S., {Seager}, S., {Mennesson}, B., {et~al.} 2020, arXiv e-prints,
  arXiv:2001.06683

\bibitem[{{Girardi} {et~al.}(2000){Girardi}, {Bressan}, {Bertelli}, \&
  {Chiosi}}]{girardi2000}
{Girardi}, L., {Bressan}, A., {Bertelli}, G., \& {Chiosi}, C. 2000, \aaps, 141,
  371

\bibitem[{Gray {et~al.}(2006)Gray, Corbally, Garrison, McFadden, Bubar,
  McGahee, O\'Donoghue, \& Knox}]{gray2006}
Gray, R.~O., Corbally, C.~J., Garrison, R.~F., {et~al.} 2006, The Astronomical
  Journal, 132, 161.
\newblock \url{https://doi.org/10.1086%2F504637}

\bibitem[{{Grevesse} \& {Noels}(1993)}]{grevesse1993}
{Grevesse}, N., \& {Noels}, A. 1993, Physica Scripta Volume T, 47, 133

\bibitem[{{Grundahl} {et~al.}(2017){Grundahl}, {Fredslund Andersen},
  {Christensen-Dalsgaard}, {Antoci}, {Kjeldsen}, {Handberg}, {Houdek},
  {Bedding}, {Pall{\'e}}, {Jessen-Hansen}, {Silva Aguirre}, {White},
  {Frandsen}, {Albrecht}, {Andersen}, {Arentoft}, {Brogaard}, {Chaplin},
  {Harps{\o}e}, {J{\o}rgensen}, {Karovicova}, {Karoff}, {Kj{\ae}rgaard
  Rasmussen}, {Lund}, {Sloth Lundkvist}, {Skottfelt}, {Norup S{\o}rensen},
  {Tronsgaard}, \& {Weiss}}]{grundahl2017}
{Grundahl}, F., {Fredslund Andersen}, M., {Christensen-Dalsgaard}, J., {et~al.}
  2017, \apj, 836, 142

\bibitem[{{Guzik} {et~al.}(2011){Guzik}, {Houdek}, {Chaplin}, {Kurtz},
  {Gilliland}, {Mullally}, {Rowe}, {Haas}, {Bryson}, {Still}, \&
  {Boyajian}}]{guzik2011}
{Guzik}, J.~A., {Houdek}, G., {Chaplin}, W.~J., {et~al.} 2011, arXiv e-prints,
  arXiv:1110.2120

\bibitem[{{Handberg} \& {Campante}(2011)}]{handberg2011}
{Handberg}, R., \& {Campante}, T.~L. 2011, \aap, 527, A56

\bibitem[{{Harvey}(1985)}]{harvey1985}
{Harvey}, J. 1985, in ESA Special Publication, Vol. 235, Future Missions in
  Solar, Heliospheric \& Space Plasma Physics, ed. E.~{Rolfe} \& B.~{Battrick},
  199

\bibitem[{{Hawkins} {et~al.}(2016){Hawkins}, {Jofr{\'e}}, {Heiter}, {Soubiran},
  {Blanco-Cuaresma}, {Casagrande}, {Gilmore}, {Lind}, {Magrini}, {Masseron},
  {Pancino}, {Randich}, \& {Worley}}]{hawkins2016}
{Hawkins}, K., {Jofr{\'e}}, P., {Heiter}, U., {et~al.} 2016, \aap, 592, A70

\bibitem[{{Heller} {et~al.}(2017){Heller}, {Hippke}, \&
  {Kervella}}]{heller2017}
{Heller}, R., {Hippke}, M., \& {Kervella}, P. 2017, \aj, 154, 115

\bibitem[{{Henry} {et~al.}(1996){Henry}, {Soderblom}, {Donahue}, \&
  {Baliunas}}]{henry1996}
{Henry}, T.~J., {Soderblom}, D.~R., {Donahue}, R.~A., \& {Baliunas}, S.~L.
  1996, \aj, 111, 439

\bibitem[{Hey \& Ball(2020)}]{echelle}
Hey, D., \& Ball, W. 2020, {Echelle: Dynamic echelle diagrams for
  asteroseismology}, v1.4,  Zenodo, doi:10.5281/zenodo.3629933.
\newblock \url{https://doi.org/10.5281/zenodo.3629933}

\bibitem[{{H{\o}g} {et~al.}(2000){H{\o}g}, {Fabricius}, {Makarov}, {Urban},
  {Corbin}, {Wycoff}, {Bastian}, {Schwekendiek}, \& {Wicenec}}]{hog2000}
{H{\o}g}, E., {Fabricius}, C., {Makarov}, V.~V., {et~al.} 2000, \aap, 355, L27

\bibitem[{{Howell} {et~al.}(2011){Howell}, {Everett}, {Sherry}, {Horch}, \&
  {Ciardi}}]{howell2011}
{Howell}, S.~B., {Everett}, M.~E., {Sherry}, W., {Horch}, E., \& {Ciardi},
  D.~R. 2011, \aj, 142, 19

\bibitem[{{Huber}(2017)}]{huber2017}
{Huber}, D. 2017, {Isoclassify: V1.2}, vv1.2,  Zenodo,
  doi:10.5281/zenodo.573372

\bibitem[{{Huber} {et~al.}(2009){Huber}, {Stello}, {Bedding}, {Chaplin},
  {Arentoft}, {Quirion}, \& {Kjeldsen}}]{huber2009}
{Huber}, D., {Stello}, D., {Bedding}, T.~R., {et~al.} 2009, Communications in
  Asteroseismology, 160, 74

\bibitem[{{Huber} {et~al.}(2019){Huber}, {Chaplin}, {Chontos}, {Kjeldsen},
  {Christensen-Dalsgaard}, {Bedding}, {Ball}, {Brahm}, {Espinoza}, {Henning},
  {Jordan}, {Sarkis}, {Knudstrup}, {Albrecht}, {Grundahl}, {Fredslund
  Andersen}, {Palle}, {Crossfield}, {Fulton}, {Howard}, {Isaacson}, {Weiss},
  {Handberg}, {Lund}, {Serenelli}, {Mosumgaard}, {Stokholm}, {Bierlya},
  {Buchhave}, {Latham}, {Quinn}, {Gaidos}, {Hirano}, {Ricker}, {Vanderspek},
  {Seager}, {Jenkins}, {Winn}, {Antia}, {Appourchaux}, {Basu}, {Bell},
  {Benomar}, {Bonanno}, {Buzasi}, {Campante}, {Celik Orhan}, {Corsaro},
  {Cunha}, {Davies}, {Deheuvels}, {Grunblatt}, {Hasanzadeh}, {Di Mauro},
  {Garcia}, {Gaulme}, {Girardi}, {Guzik}, {Hon}, {Jiang}, {Kallinger},
  {Kawaler}, {Kuszlewicz}, {Lebreton}, {Li}, {Lucas}, {Lundkvist}, {Mathis},
  {Mathur}, {Mazumdar}, {Metcalfe}, {Miglio}, {Monteiro}, {Mosser}, {Noll},
  {Nsamba}, {Mann}, {Ong}, {Ortel}, {Pereira}, {Ranadive}, {Regulo},
  {Rodrigues}, {Roxburgh}, {Silva Aguirre}, {Smalley}, {Schofield}, {Sousa},
  {Stassun}, {Stello}, {Tayar}, {White}, {Verma}, {Vrard}, {Yildiz}, {Baker},
  {Bazot}, {Beichmann}, {Bergmann}, {Bugnet}, {Cale}, {Carlino}, {Cartwright},
  {Christiansen}, {Ciardi}, {Creevey}, {Dittmann}, {Dias Do Nascimento}, {van
  Eylen}, {Furesz}, {Gagne}, {Gao}, {Gazeas}, {Giddens}, {Hall}, {Hekker},
  {Ireland}, {Latouf}, {LeBrun}, {Levine}, {Matzko}, {Natinsky}, {Page},
  {Plavchan}, {Mansouri-Samani}, {McCauliff}, {Mullally}, {Orenstein}, {Soto},
  {Paegert}, {van Saders}, {Schnaible}, {Soderblom}, {Szabo}, {Tanner},
  {Tinney}, {Teske}, {Thomas}, {Trampedach}, {Wright}, \&
  {Zohrabi}}]{huber2019}
{Huber}, D., {Chaplin}, W.~J., {Chontos}, A., {et~al.} 2019, arXiv e-prints,
  arXiv:1901.01643

\bibitem[{{Jenkins} {et~al.}(2016){Jenkins}, {Twicken}, {McCauliff},
  {Campbell}, {Sanderfer}, {Lung}, {Mansouri-Samani}, {Girouard}, {Tenenbaum},
  {Klaus}, {Smith}, {Caldwell}, {Chacon}, {Henze}, {Heiges}, {Latham},
  {Morgan}, {Swade}, {Rinehart}, \& {Vanderspek}}]{jenkins2016}
{Jenkins}, J.~M., {Twicken}, J.~D., {McCauliff}, S., {et~al.} 2016, Society of
  Photo-Optical Instrumentation Engineers (SPIE) Conference Series, Vol. 9913,
  {The TESS science processing operations center}, 99133E

\bibitem[{{Kjeldsen} \& {Bedding}(1995)}]{kjeldsen1995}
{Kjeldsen}, H., \& {Bedding}, T.~R. 1995, \aap, 293, 87

\bibitem[{{Kjeldsen} {et~al.}(2008){Kjeldsen}, {Bedding}, \&
  {Christensen-Dalsgaard}}]{kjeldsen2008}
{Kjeldsen}, H., {Bedding}, T.~R., \& {Christensen-Dalsgaard}, J. 2008, \apjl,
  683, L175

\bibitem[{{Kjeldsen} {et~al.}(2003){Kjeldsen}, {Bedding}, {Baldry}, {Bruntt},
  {Butler}, {Fischer}, {Frand sen}, {Gates}, {Grundahl}, {Lang}, {Marcy},
  {Misch}, \& {Vogt}}]{kjeldsen2003}
{Kjeldsen}, H., {Bedding}, T.~R., {Baldry}, I.~K., {et~al.} 2003, \aj, 126,
  1483

\bibitem[{{Kjeldsen} {et~al.}(2005){Kjeldsen}, {Bedding}, {Butler},
  {Christensen-Dalsgaard}, {Kiss}, {McCarthy}, {Marcy}, {Tinney}, \&
  {Wright}}]{kjeldsen2005}
{Kjeldsen}, H., {Bedding}, T.~R., {Butler}, R.~P., {et~al.} 2005, \apj, 635,
  1281

\bibitem[{{Kraft}(1967)}]{kraft1967}
{Kraft}, R.~P. 1967, \apj, 150, 551

\bibitem[{{Leggett} {et~al.}(2001){Leggett}, {Allard}, {Geballe}, {Hauschildt},
  \& {Schweitzer}}]{leggett2001}
{Leggett}, S.~K., {Allard}, F., {Geballe}, T.~R., {Hauschildt}, P.~H., \&
  {Schweitzer}, A. 2001, \apj, 548, 908

\bibitem[{{Lindegren} {et~al.}(2021){Lindegren}, {Klioner}, {Hern{\'a}ndez},
  {Bombrun}, {Ramos-Lerate}, {Steidelm{\"u}ller}, {Bastian}, {Biermann}, {de
  Torres}, {Gerlach}, {Geyer}, {Hilger}, {Hobbs}, {Lammers}, {McMillan},
  {Stephenson}, {Casta{\~n}eda}, {Davidson}, {Fabricius}, {Gracia-Abril},
  {Portell}, {Rowell}, {Teyssier}, {Torra}, {Bartolom{\'e}}, {Clotet},
  {Garralda}, {Gonz{\'a}lez-Vidal}, {Torra}, {Abbas}, {Altmann}, {Anglada
  Varela}, {Balaguer-N{\'u}{\~n}ez}, {Balog}, {Barache}, {Becciani}, {Bernet},
  {Bertone}, {Bianchi}, {Bouquillon}, {Brown}, {Bucciarelli}, {Busonero},
  {Butkevich}, {Buzzi}, {Cancelliere}, {Carlucci}, {Charlot}, {Cioni},
  {Crosta}, {Crowley}, {del Peloso}, {del Pozo}, {Drimmel}, {Esquej}, {Fienga},
  {Fraile}, {Gai}, {Garcia-Reinaldos}, {Guerra}, {Hambly}, {Hauser},
  {Jan{\ss}en}, {Jordan}, {Kostrzewa-Rutkowska}, {Lattanzi}, {Liao}, {Licata},
  {Lister}, {L{\"o}ffler}, {Marchant}, {Masip}, {Mignard}, {Mints}, {Molina},
  {Mora}, {Morbidelli}, {Murphy}, {Pagani}, {Panuzzo}, {Pe{\~n}alosa Esteller},
  {Poggio}, {Re Fiorentin}, {Riva}, {Sagrist{\`a} Sell{\'e}s}, {Sanchez
  Gimenez}, {Sarasso}, {Sciacca}, {Siddiqui}, {Smart}, {Souami}, {Spagna},
  {Steele}, {Taris}, {Utrilla}, {van Reeven}, \& {Vecchiato}}]{lindegren2021}
{Lindegren}, L., {Klioner}, S.~A., {Hern{\'a}ndez}, J., {et~al.} 2021, \aap,
  649, A2

\bibitem[{{Lomb}(1976)}]{lomb1976}
{Lomb}, N.~R. 1976, \apss, 39, 447

\bibitem[{{Lovis} {et~al.}(2011){Lovis}, {Dumusque}, {Santos}, {Bouchy},
  {Mayor}, {Pepe}, {Queloz}, {S{\'e}gransan}, \& {Udry}}]{lovis2011}
{Lovis}, C., {Dumusque}, X., {Santos}, N.~C., {et~al.} 2011, arXiv e-prints,
  arXiv:1107.5325

\bibitem[{{Luck}(2018)}]{luck2018}
{Luck}, R.~E. 2018, \aj, 155, 111

\bibitem[{{Lundkvist}(2015)}]{lundkvist2015}
{Lundkvist}, M.~S. 2015, PhD thesis, Stellar Astrophysics Centre, Aarhus
  University, Denmark

\bibitem[{{Maldonado} {et~al.}(2012){Maldonado}, {Eiroa}, {Villaver},
  {Montesinos}, \& {Mora}}]{maldonado2012}
{Maldonado}, J., {Eiroa}, C., {Villaver}, E., {Montesinos}, B., \& {Mora}, A.
  2012, \aap, 541, A40

\bibitem[{{Maldonado} {et~al.}(2015){Maldonado}, {Eiroa}, {Villaver},
  {Montesinos}, \& {Mora}}]{maldonaldo2015}
---. 2015, \aap, 579, A20

\bibitem[{{Mamajek} \& {Hillenbrand}(2008)}]{mamajek2008}
{Mamajek}, E.~E., \& {Hillenbrand}, L.~A. 2008, \apj, 687, 1264

\bibitem[{{Mann} {et~al.}(2015){Mann}, {Feiden}, {Gaidos}, {Boyajian}, \& {von
  Braun}}]{mann2015}
{Mann}, A.~W., {Feiden}, G.~A., {Gaidos}, E., {Boyajian}, T., \& {von Braun},
  K. 2015, \apj, 804, 64

\bibitem[{{Mann} {et~al.}(2019){Mann}, {Dupuy}, {Kraus}, {Gaidos}, {Ansdell},
  {Ireland}, {Rizzuto}, {Hung}, {Dittmann}, {Factor}, {Feiden}, {Martinez},
  {Ru{\'\i}z-Rodr{\'\i}guez}, \& {Thao}}]{mann2019}
{Mann}, A.~W., {Dupuy}, T., {Kraus}, A.~L., {et~al.} 2019, \apj, 871, 63

\bibitem[{{Marigo} {et~al.}(2008){Marigo}, {Girardi}, {Bressan}, {Groenewegen},
  {Silva}, \& {Granato}}]{marigo2008}
{Marigo}, P., {Girardi}, L., {Bressan}, A., {et~al.} 2008, \aap, 482, 883

\bibitem[{{Marques} {et~al.}(2013){Marques}, {Goupil}, {Lebreton}, {Talon},
  {Palacios}, {Belkacem}, {Ouazzani}, {Mosser}, {Moya}, {Morel}, {Pichon},
  {Mathis}, {Zahn}, {Turck-Chi{\`e}ze}, \& {Nghiem}}]{marques2013}
{Marques}, J.~P., {Goupil}, M.~J., {Lebreton}, Y., {et~al.} 2013, \aap, 549,
  A74

\bibitem[{{Marti{\'c}} {et~al.}(1999){Marti{\'c}}, {Schmitt}, {Lebrun},
  {Barban}, {Connes}, {Bouchy}, {Michel}, {Baglin}, {Appourchaux}, \&
  {Bertaux}}]{martic1999}
{Marti{\'c}}, M., {Schmitt}, J., {Lebrun}, J.~C., {et~al.} 1999, \aap, 351, 993

\bibitem[{{Mathur} {et~al.}(2010){Mathur}, {Garc{\'\i}a}, {R{\'e}gulo},
  {Creevey}, {Ballot}, {Salabert}, {Arentoft}, {Quirion}, {Chaplin}, \&
  {Kjeldsen}}]{mathur2010}
{Mathur}, S., {Garc{\'\i}a}, R.~A., {R{\'e}gulo}, C., {et~al.} 2010, \aap, 511,
  A46

\bibitem[{{Mathur} {et~al.}(2011){Mathur}, {Hekker}, {Trampedach}, {Ballot},
  {Kallinger}, {Buzasi}, {Garc{\'\i}a}, {Huber}, {Jim{\'e}nez}, {Mosser},
  {Bedding}, {Elsworth}, {R{\'e}gulo}, {Stello}, {Chaplin}, {De Ridder},
  {Hale}, {Kinemuchi}, {Kjeldsen}, {Mullally}, \& {Thompson}}]{mathur2011}
{Mathur}, S., {Hekker}, S., {Trampedach}, R., {et~al.} 2011, \apj, 741, 119

\bibitem[{{McDonald} {et~al.}(2017){McDonald}, {Zijlstra}, \&
  {Watson}}]{mcdonald2017}
{McDonald}, I., {Zijlstra}, A.~A., \& {Watson}, R.~A. 2017, \mnras, 471, 770

\bibitem[{{McQuillan} {et~al.}(2014){McQuillan}, {Mazeh}, \&
  {Aigrain}}]{mcquillan2014}
{McQuillan}, A., {Mazeh}, T., \& {Aigrain}, S. 2014, \apjs, 211, 24

\bibitem[{{Mermilliod}(2006)}]{mermilliod2006}
{Mermilliod}, J.~C. 2006, VizieR Online Data Catalog, II/168

\bibitem[{{Metcalfe} \& {Charbonneau}(2003)}]{metcalfe2003}
{Metcalfe}, T.~S., \& {Charbonneau}, P. 2003, Journal of Computational Physics,
  185, 176

\bibitem[{{Metcalfe} {et~al.}(2009{\natexlab{a}}){Metcalfe}, {Creevey}, \&
  {Christensen-Dalsgaard}}]{metcalfe2009a}
{Metcalfe}, T.~S., {Creevey}, O.~L., \& {Christensen-Dalsgaard}, J.
  2009{\natexlab{a}}, \apj, 699, 373

\bibitem[{{Metcalfe} {et~al.}(2016){Metcalfe}, {Egeland}, \& {van
  Saders}}]{metcalfe2016}
{Metcalfe}, T.~S., {Egeland}, R., \& {van Saders}, J. 2016, \apjl, 826, L2

\bibitem[{{Metcalfe} {et~al.}(2009{\natexlab{b}}){Metcalfe}, {Judge}, {Basu},
  {Henry}, {Soderblom}, {Knoelker}, \& {Rempel}}]{metcalfe2009b}
{Metcalfe}, T.~S., {Judge}, P.~G., {Basu}, S., {et~al.} 2009{\natexlab{b}},
  arXiv e-prints, arXiv:0909.5464

\bibitem[{{Metcalfe} {et~al.}(2012{\natexlab{a}}){Metcalfe}, {Mathur},
  {Do{\u{g}}an}, \& {Woitaszek}}]{metcalfe2012b}
{Metcalfe}, T.~S., {Mathur}, S., {Do{\u{g}}an}, G., \& {Woitaszek}, M.
  2012{\natexlab{a}}, in Astronomical Society of the Pacific Conference Series,
  Vol. 462, Progress in Solar/Stellar Physics with Helio- and Asteroseismology,
  ed. H.~{Shibahashi}, M.~{Takata}, \& A.~E. {Lynas-Gray}, 213

\bibitem[{{Metcalfe} {et~al.}(2012{\natexlab{b}}){Metcalfe}, {Chaplin},
  {Appourchaux}, {Garc{\'\i}a}, {Basu}, {Brand{\~a}o}, {Creevey}, {Deheuvels},
  {Do{\v{g}}an}, {Eggenberger}, {Karoff}, {Miglio}, {Stello}, {Y{\i}ld{\i}z},
  {{\c{C}}elik}, {Antia}, {Benomar}, {Howe}, {R{\'e}gulo}, {Salabert}, {Stahn},
  {Bedding}, {Davies}, {Elsworth}, {Gizon}, {Hekker}, {Mathur}, {Mosser},
  {Bryson}, {Still}, {Christensen-Dalsgaard}, {Gilliland}, {Kawaler},
  {Kjeldsen}, {Ibrahim}, {Klaus}, \& {Li}}]{metcalfe2012a}
{Metcalfe}, T.~S., {Chaplin}, W.~J., {Appourchaux}, T., {et~al.}
  2012{\natexlab{b}}, \apjl, 748, L10

\bibitem[{{Metcalfe} {et~al.}(2020){Metcalfe}, {van Saders}, {Basu}, {Buzasi},
  {Chaplin}, {Egeland}, {Garcia}, {Gaulme}, {Huber}, {Reinhold}, {Schunker},
  {Stassun}, {Appourchaux}, {Ball}, {Bedding}, {Deheuvels},
  {Gonz{\'a}lez-Cuesta}, {Handberg}, {Jim{\'e}nez}, {Kjeldsen}, {Li}, {Lund},
  {Mathur}, {Mosser}, {Nielsen}, {Noll}, {{\c{C}}elik Orhan}, {{\"O}rtel},
  {Santos}, {Yildiz}, {Baliunas}, \& {Soon}}]{metcalfe2020}
{Metcalfe}, T.~S., {van Saders}, J.~L., {Basu}, S., {et~al.} 2020, \apj, 900,
  154

\bibitem[{{Mier}(2017)}]{mier2017}
{Mier}, P.~R. 2017, {Pablormier/Yabox: V1.0.3}, vv1.0.3,  Zenodo,
  doi:10.5281/zenodo.848679

\bibitem[{{Morel} \& {Lebreton}(2008)}]{morel2008}
{Morel}, P., \& {Lebreton}, Y. 2008, \apss, 316, 61

\bibitem[{{Mosser} {et~al.}(2008){Mosser}, {Deheuvels}, {Michel},
  {Th{\'e}venin}, {Dupret}, {Samadi}, {Barban}, \& {Goupil}}]{mosser2008}
{Mosser}, B., {Deheuvels}, S., {Michel}, E., {et~al.} 2008, \aap, 488, 635

\bibitem[{{Mosser} {et~al.}(1998){Mosser}, {Maillard}, {Mekarnia}, \&
  {Gay}}]{mosser1998}
{Mosser}, B., {Maillard}, J.~P., {Mekarnia}, D., \& {Gay}, J. 1998, \aap, 340,
  457

\bibitem[{{Nielsen} {et~al.}(2013){Nielsen}, {Gizon}, {Schunker}, \&
  {Karoff}}]{nielsen2013}
{Nielsen}, M.~B., {Gizon}, L., {Schunker}, H., \& {Karoff}, C. 2013, \aap, 557,
  L10

\bibitem[{{Nielsen} {et~al.}(2015){Nielsen}, {Schunker}, {Gizon}, \&
  {Ball}}]{nielsen2015}
{Nielsen}, M.~B., {Schunker}, H., {Gizon}, L., \& {Ball}, W.~H. 2015, \aap,
  582, A10

\bibitem[{{Nielsen} {et~al.}(2017){Nielsen}, {Schunker}, {Gizon}, {Schou}, \&
  {Ball}}]{nielsen2017}
{Nielsen}, M.~B., {Schunker}, H., {Gizon}, L., {Schou}, J., \& {Ball}, W.~H.
  2017, \aap, 603, A6

\bibitem[{{Nielsen} {et~al.}(2020){Nielsen}, {Ball}, {Standing}, {Triaud},
  {Buzasi}, {Carboneau}, {Stassun}, {Kane}, {Chaplin}, {Bellinger}, {Mosser},
  {Roxburgh}, {{\c{C}}elik Orhan}, {Y{\i}ld{\i}z}, {{\"O}rtel}, {Vrard},
  {Mazumdar}, {Ranadive}, {Deal}, {Davies}, {Campante}, {Garc{\'\i}a},
  {Mathur}, {Gonz{\'a}lez-Cuesta}, \& {Serenelli}}]{nielsen2020}
{Nielsen}, M.~B., {Ball}, W.~H., {Standing}, M.~R., {et~al.} 2020, \aap, 641,
  A25

\bibitem[{{Noyes} {et~al.}(1984){Noyes}, {Hartmann}, {Baliunas}, {Duncan}, \&
  {Vaughan}}]{noyes1984a}
{Noyes}, R.~W., {Hartmann}, L.~W., {Baliunas}, S.~L., {Duncan}, D.~K., \&
  {Vaughan}, A.~H. 1984, \apj, 279, 763

\bibitem[{{Paunzen}(2015)}]{paunzen2015}
{Paunzen}, E. 2015, \aap, 580, A23

\bibitem[{{Paxton} {et~al.}(2011){Paxton}, {Bildsten}, {Dotter}, {Herwig},
  {Lesaffre}, \& {Timmes}}]{paxton2011}
{Paxton}, B., {Bildsten}, L., {Dotter}, A., {et~al.} 2011, \apjs, 192, 3

\bibitem[{{Paxton} {et~al.}(2013){Paxton}, {Cantiello}, {Arras}, {Bildsten},
  {Brown}, {Dotter}, {Mankovich}, {Montgomery}, {Stello}, {Timmes}, \&
  {Townsend}}]{paxton2013}
{Paxton}, B., {Cantiello}, M., {Arras}, P., {et~al.} 2013, \apjs, 208, 4

\bibitem[{{Paxton} {et~al.}(2015){Paxton}, {Marchant}, {Schwab}, {Bauer},
  {Bildsten}, {Cantiello}, {Dessart}, {Farmer}, {Hu}, {Langer}, {Townsend},
  {Townsley}, \& {Timmes}}]{paxton2015}
{Paxton}, B., {Marchant}, P., {Schwab}, J., {et~al.} 2015, \apjs, 220, 15

\bibitem[{{Paxton} {et~al.}(2018){Paxton}, {Schwab}, {Bauer}, {Bildsten},
  {Blinnikov}, {Duffell}, {Farmer}, {Goldberg}, {Marchant}, {Sorokina},
  {Thoul}, {Townsend}, \& {Timmes}}]{paxton2018}
{Paxton}, B., {Schwab}, J., {Bauer}, E.~B., {et~al.} 2018, \apjs, 234, 34

\bibitem[{{Paxton} {et~al.}(2019){Paxton}, {Smolec}, {Schwab}, {Gautschy},
  {Bildsten}, {Cantiello}, {Dotter}, {Farmer}, {Goldberg}, {Jermyn}, {Kanbur},
  {Marchant}, {Thoul}, {Townsend}, {Wolf}, {Zhang}, \& {Timmes}}]{paxton2019}
{Paxton}, B., {Smolec}, R., {Schwab}, J., {et~al.} 2019, \apjs, 243, 10

\bibitem[{{Pietrinferni} {et~al.}(2004){Pietrinferni}, {Cassisi}, {Salaris}, \&
  {Castelli}}]{pietrinferni2004}
{Pietrinferni}, A., {Cassisi}, S., {Salaris}, M., \& {Castelli}, F. 2004, \apj,
  612, 168

\bibitem[{{Pope} {et~al.}(2016){Pope}, {White}, {Huber}, {Murphy}, {Bedding},
  {Caldwell}, {Sarai}, {Aigrain}, \& {Barclay}}]{pope2016}
{Pope}, B.~J.~S., {White}, T.~R., {Huber}, D., {et~al.} 2016, \mnras, 455, L36

\bibitem[{{Pope} {et~al.}(2019{\natexlab{a}}){Pope}, {Davies}, {Hawkins},
  {White}, {Stokholm}, {Bieryla}, {Latham}, {Lucey}, {Aerts}, {Aigrain},
  {Antoci}, {Bedding}, {Bowman}, {Caldwell}, {Chontos}, {Esquerdo}, {Huber},
  {Jofr{\'e}}, {Murphy}, {van Reeth}, {Silva Aguirre}, \& {Yu}}]{pope2019a}
{Pope}, B. J.~S., {Davies}, G.~R., {Hawkins}, K., {et~al.} 2019{\natexlab{a}},
  \apjs, 244, 18

\bibitem[{{Pope} {et~al.}(2019{\natexlab{b}}){Pope}, {White}, {Farr}, {Yu},
  {Greklek-McKeon}, {Huber}, {Aerts}, {Aigrain}, {Bedding}, {Boyajian},
  {Creevey}, \& {Hogg}}]{pope2019b}
{Pope}, B. J.~S., {White}, T.~R., {Farr}, W.~M., {et~al.} 2019{\natexlab{b}},
  \apjs, 245, 8

\bibitem[{{Ram{\'\i}rez} {et~al.}(2007){Ram{\'\i}rez}, {Allende Prieto}, \&
  {Lambert}}]{ramirez2007}
{Ram{\'\i}rez}, I., {Allende Prieto}, C., \& {Lambert}, D.~L. 2007, \aap, 465,
  271

\bibitem[{{Ram{\'\i}rez} {et~al.}(2012){Ram{\'\i}rez}, {Fish}, {Lambert}, \&
  {Allende Prieto}}]{ramirez2012}
{Ram{\'\i}rez}, I., {Fish}, J.~R., {Lambert}, D.~L., \& {Allende Prieto}, C.
  2012, \apj, 756, 46

\bibitem[{{Reinhold} {et~al.}(2013){Reinhold}, {Reiners}, \&
  {Basri}}]{reinhold2013}
{Reinhold}, T., {Reiners}, A., \& {Basri}, G. 2013, \aap, 560, A4

\bibitem[{{Ricker} {et~al.}(2015){Ricker}, {Winn}, {Vanderspek}, {Latham},
  {Bakos}, {Bean}, {Berta-Thompson}, {Brown}, {Buchhave}, {Butler}, {Butler},
  {Chaplin}, {Charbonneau}, {Christensen-Dalsgaard}, {Clampin}, {Deming},
  {Doty}, {De Lee}, {Dressing}, {Dunham}, {Endl}, {Fressin}, {Ge}, {Henning},
  {Holman}, {Howard}, {Ida}, {Jenkins}, {Jernigan}, {Johnson}, {Kaltenegger},
  {Kawai}, {Kjeldsen}, {Laughlin}, {Levine}, {Lin}, {Lissauer}, {MacQueen},
  {Marcy}, {McCullough}, {Morton}, {Narita}, {Paegert}, {Palle}, {Pepe},
  {Pepper}, {Quirrenbach}, {Rinehart}, {Sasselov}, {Sato}, {Seager},
  {Sozzetti}, {Stassun}, {Sullivan}, {Szentgyorgyi}, {Torres}, {Udry}, \&
  {Villasenor}}]{ricker2015}
{Ricker}, G.~R., {Winn}, J.~N., {Vanderspek}, R., {et~al.} 2015, Journal of
  Astronomical Telescopes, Instruments, and Systems, 1, 014003

\bibitem[{{Riello} {et~al.}(2021){Riello}, {De Angeli}, {Evans}, {Montegriffo},
  {Carrasco}, {Busso}, {Palaversa}, {Burgess}, {Diener}, {Davidson}, {Rowell},
  {Fabricius}, {Jordi}, {Bellazzini}, {Pancino}, {Harrison}, {Cacciari}, {van
  Leeuwen}, {Hambly}, {Hodgkin}, {Osborne}, {Altavilla}, {Barstow}, {Brown},
  {Castellani}, {Cowell}, {De Luise}, {Gilmore}, {Giuffrida}, {Hidalgo},
  {Holland}, {Marinoni}, {Pagani}, {Piersimoni}, {Pulone}, {Ragaini}, {Rainer},
  {Richards}, {Sanna}, {Walton}, {Weiler}, \& {Yoldas}}]{riello2021}
{Riello}, M., {De Angeli}, F., {Evans}, D.~W., {et~al.} 2021, \aap, 649, A3

\bibitem[{{Saar} \& {Osten}(1997)}]{saar1997}
{Saar}, S.~H., \& {Osten}, R.~A. 1997, \mnras, 284, 803

\bibitem[{{Santos} {et~al.}(2019){Santos}, {Garc{\'\i}a}, {Mathur}, {Bugnet},
  {van Saders}, {Metcalfe}, {Simonian}, \& {Pinsonneault}}]{santos2019}
{Santos}, A.~R.~G., {Garc{\'\i}a}, R.~A., {Mathur}, S., {et~al.} 2019, \apjs,
  244, 21

\bibitem[{{Santos} {et~al.}(2010){Santos}, {Gomes da Silva}, {Lovis}, \&
  {Melo}}]{santos2010}
{Santos}, N.~C., {Gomes da Silva}, J., {Lovis}, C., \& {Melo}, C. 2010, \aap,
  511, A54

\bibitem[{{Santos} {et~al.}(2001){Santos}, {Israelian}, \&
  {Mayor}}]{santos2001}
{Santos}, N.~C., {Israelian}, G., \& {Mayor}, M. 2001, \aap, 373, 1019

\bibitem[{{Santos} {et~al.}(2004){Santos}, {Israelian}, \&
  {Mayor}}]{santos2004}
---. 2004, \aap, 415, 1153

\bibitem[{{Scargle}(1982)}]{scargle1982}
{Scargle}, J.~D. 1982, \apj, 263, 835

\bibitem[{{Schofield} {et~al.}(2019){Schofield}, {Chaplin}, {Huber},
  {Campante}, {Davies}, {Miglio}, {Ball}, {Appourchaux}, {Basu}, {Bedding},
  {Christensen-Dalsgaard}, {Creevey}, {Garc{\'\i}a}, {Handberg}, {Kawaler},
  {Kjeldsen}, {Latham}, {Lund}, {Metcalfe}, {Ricker}, {Serenelli}, {Silva
  Aguirre}, {Stello}, \& {Vanderspek}}]{schofield2019}
{Schofield}, M., {Chaplin}, W.~J., {Huber}, D., {et~al.} 2019, \apjs, 241, 12

\bibitem[{Schwarz(1978)}]{schwarz1978}
Schwarz, G. 1978, The Annals of Statistics, 6, 461 .
\newblock \url{https://doi.org/10.1214/aos/1176344136}

\bibitem[{{Serenelli} {et~al.}(2017){Serenelli}, {Johnson}, {Huber},
  {Pinsonneault}, {Ball}, {Tayar}, {Silva Aguirre}, {Basu}, {Troup}, {Hekker},
  {Kallinger}, {Stello}, {Davies}, {Lund}, {Mathur}, {Mosser}, {Stassun},
  {Chaplin}, {Elsworth}, {Garc{\'\i}a}, {Handberg}, {Holtzman}, {Hearty},
  {Garc{\'\i}a-Hern{\'a}ndez}, {Gaulme}, \& {Zamora}}]{serenelli2017}
{Serenelli}, A., {Johnson}, J., {Huber}, D., {et~al.} 2017, \apjs, 233, 23

\bibitem[{{Serenelli} {et~al.}(2013){Serenelli}, {Bergemann}, {Ruchti}, \&
  {Casagrande}}]{serenelli2013}
{Serenelli}, A.~M., {Bergemann}, M., {Ruchti}, G., \& {Casagrande}, L. 2013,
  \mnras, 429, 3645

\bibitem[{{Silva Aguirre} {et~al.}(2012){Silva Aguirre}, {Casagrande}, {Basu},
  {Campante}, {Chaplin}, {Huber}, {Miglio}, {Serenelli}, {Ballot}, {Bedding},
  {Christensen-Dalsgaard}, {Creevey}, {Elsworth}, {Garc{\'\i}a}, {Gilliland},
  {Hekker}, {Kjeldsen}, {Mathur}, {Metcalfe}, {Monteiro}, {Mosser},
  {Pinsonneault}, {Stello}, {Weiss}, {Tenenbaum}, {Twicken}, \&
  {Uddin}}]{silvaaguirre2012}
{Silva Aguirre}, V., {Casagrande}, L., {Basu}, S., {et~al.} 2012, \apj, 757, 99

\bibitem[{{Silva Aguirre} {et~al.}(2013){Silva Aguirre}, {Basu}, {Brand{\~a}o},
  {Christensen-Dalsgaard}, {Deheuvels}, {Do{\u{g}}an}, {Metcalfe}, {Serenelli},
  {Ballot}, {Chaplin}, {Cunha}, {Weiss}, {Appourchaux}, {Casagrande},
  {Cassisi}, {Creevey}, {Garc{\'\i}a}, {Lebreton}, {Noels}, {Sousa}, {Stello},
  {White}, {Kawaler}, \& {Kjeldsen}}]{silvaaguirre2013}
{Silva Aguirre}, V., {Basu}, S., {Brand{\~a}o}, I.~M., {et~al.} 2013, \apj,
  769, 141

\bibitem[{{Silva Aguirre} {et~al.}(2015){Silva Aguirre}, {Davies}, {Basu},
  {Christensen-Dalsgaard}, {Creevey}, {Metcalfe}, {Bedding}, {Casagrande},
  {Handberg}, {Lund}, {Nissen}, {Chaplin}, {Huber}, {Serenelli}, {Stello}, {Van
  Eylen}, {Campante}, {Elsworth}, {Gilliland}, {Hekker}, {Karoff}, {Kawaler},
  {Kjeldsen}, \& {Lundkvist}}]{silvaaguirre2015}
{Silva Aguirre}, V., {Davies}, G.~R., {Basu}, S., {et~al.} 2015, \mnras, 452,
  2127

\bibitem[{{Silva Aguirre} {et~al.}(2017){Silva Aguirre}, {Lund}, {Antia},
  {Ball}, {Basu}, {Christensen-Dalsgaard}, {Lebreton}, {Reese}, {Verma},
  {Casagrande}, {Justesen}, {Mosumgaard}, {Chaplin}, {Bedding}, {Davies},
  {Handberg}, {Houdek}, {Huber}, {Kjeldsen}, {Latham}, {White}, {Coelho},
  {Miglio}, \& {Rendle}}]{silvaaguirre2017}
{Silva Aguirre}, V., {Lund}, M.~N., {Antia}, H.~M., {et~al.} 2017, \apj, 835,
  173

\bibitem[{{Skrutskie} {et~al.}(2006){Skrutskie}, {Cutri}, {Stiening},
  {Weinberg}, {Schneider}, {Carpenter}, {Beichman}, {Capps}, {Chester},
  {Elias}, {Huchra}, {Liebert}, {Lonsdale}, {Monet}, {Price}, {Seitzer},
  {Jarrett}, {Kirkpatrick}, {Gizis}, {Howard}, {Evans}, {Fowler}, {Fullmer},
  {Hurt}, {Light}, {Kopan}, {Marsh}, {McCallon}, {Tam}, {Van Dyk}, \&
  {Wheelock}}]{skrutskie2006}
{Skrutskie}, M.~F., {Cutri}, R.~M., {Stiening}, R., {et~al.} 2006, \aj, 131,
  1163

\bibitem[{{Skumanich}(1972)}]{skumanich1972}
{Skumanich}, A. 1972, \apj, 171, 565

\bibitem[{{Soderblom}(2010)}]{soderblom2010}
{Soderblom}, D.~R. 2010, \araa, 48, 581

\bibitem[{{Soderblom} {et~al.}(1991){Soderblom}, {Duncan}, \&
  {Johnson}}]{soderblom1991}
{Soderblom}, D.~R., {Duncan}, D.~K., \& {Johnson}, D. R.~H. 1991, \apj, 375,
  722

\bibitem[{{Spada} \& {Lanzafame}(2020)}]{spada2020}
{Spada}, F., \& {Lanzafame}, A.~C. 2020, \aap, 636, A76

\bibitem[{{Stassun} {et~al.}(2017){Stassun}, {Collins}, \&
  {Gaudi}}]{stassun2017}
{Stassun}, K.~G., {Collins}, K.~A., \& {Gaudi}, B.~S. 2017, \aj, 153, 136

\bibitem[{{Stassun} {et~al.}(2018){Stassun}, {Corsaro}, {Pepper}, \&
  {Gaudi}}]{Stassun2018a}
{Stassun}, K.~G., {Corsaro}, E., {Pepper}, J.~A., \& {Gaudi}, B.~S. 2018, \aj,
  155, 22

\bibitem[{{Stassun} \& {Torres}(2018)}]{stassun2018b}
{Stassun}, K.~G., \& {Torres}, G. 2018, \apj, 862, 61

\bibitem[{{Stassun} {et~al.}(2019){Stassun}, {Oelkers}, {Paegert}, {Torres},
  {Pepper}, {De Lee}, {Collins}, {Latham}, {Muirhead}, {Chittidi},
  {Rojas-Ayala}, {Fleming}, {Rose}, {Tenenbaum}, {Ting}, {Kane}, {Barclay},
  {Bean}, {Brassuer}, {Charbonneau}, {Ge}, {Lissauer}, {Mann}, {McLean},
  {Mullally}, {Narita}, {Plavchan}, {Ricker}, {Sasselov}, {Seager}, {Sharma},
  {Shiao}, {Sozzetti}, {Stello}, {Vanderspek}, {Wallace}, \&
  {Winn}}]{stassun2019}
{Stassun}, K.~G., {Oelkers}, R.~J., {Paegert}, M., {et~al.} 2019, \aj, 158, 138

\bibitem[{{Stevens} {et~al.}(2017){Stevens}, {Stassun}, \&
  {Gaudi}}]{stevens2017}
{Stevens}, D.~J., {Stassun}, K.~G., \& {Gaudi}, B.~S. 2017, \aj, 154, 259

\bibitem[{{Tasoulis} {et~al.}(2004){Tasoulis}, {Pavlidis}, {Plagianakos}, \&
  {Vrahatis}}]{tasoulis2004}
{Tasoulis}, D.~K., {Pavlidis}, N.~G., {Plagianakos}, V.~P., \& {Vrahatis},
  M.~N. 2004, in Proceedings of the 2004 Congress on Evolutionary Computation
  (IEEE Cat. No.04TH8753), Vol.~2, 2023--2029 Vol.2

\bibitem[{{Teixeira} {et~al.}(2009){Teixeira}, {Kjeldsen}, {Bedding}, {Bouchy},
  {Christensen-Dalsgaard}, {Cunha}, {Dall}, {Frandsen}, {Karoff}, {Monteiro},
  \& {Pijpers}}]{teixeira2009}
{Teixeira}, T.~C., {Kjeldsen}, H., {Bedding}, T.~R., {et~al.} 2009, \aap, 494,
  237

\bibitem[{{The LUVOIR Team}(2019)}]{luvoir2019}
{The LUVOIR Team}. 2019, arXiv e-prints, arXiv:1912.06219

\bibitem[{{Thoul} {et~al.}(1994){Thoul}, {Bahcall}, \& {Loeb}}]{thoul1994}
{Thoul}, A.~A., {Bahcall}, J.~N., \& {Loeb}, A. 1994, \apj, 421, 828

\bibitem[{{Tokovinin}(2014)}]{tokovinin2014}
{Tokovinin}, A. 2014, \aj, 147, 86

\bibitem[{{Townsend} \& {Teitler}(2013)}]{townsend2013}
{Townsend}, R.~H.~D., \& {Teitler}, S.~A. 2013, \mnras, 435, 3406

\bibitem[{{Valenti} \& {Fischer}(2005)}]{valenti2005}
{Valenti}, J.~A., \& {Fischer}, D.~A. 2005, \apjs, 159, 141

\bibitem[{{van Saders} {et~al.}(2016){van Saders}, {Ceillier}, {Metcalfe},
  {Silva Aguirre}, {Pinsonneault}, {Garc{\'\i}a}, {Mathur}, \&
  {Davies}}]{vansaders2016}
{van Saders}, J.~L., {Ceillier}, T., {Metcalfe}, T.~S., {et~al.} 2016, \nat,
  529, 181

\bibitem[{{van Saders} \& {Pinsonneault}(2013)}]{vansaders2013}
{van Saders}, J.~L., \& {Pinsonneault}, M.~H. 2013, \apj, 776, 67

\bibitem[{{Vauclair} {et~al.}(2008){Vauclair}, {Laymand}, {Bouchy}, {Vauclair},
  {Hui Bon Hoa}, {Charpinet}, \& {Bazot}}]{vauclair2008}
{Vauclair}, S., {Laymand}, M., {Bouchy}, F., {et~al.} 2008, \aap, 482, L5

\bibitem[{{Vernazza} {et~al.}(1981){Vernazza}, {Avrett}, \&
  {Loeser}}]{vernazza1981}
{Vernazza}, J.~E., {Avrett}, E.~H., \& {Loeser}, R. 1981, \apjs, 45, 635

\bibitem[{{Vrba} {et~al.}(2004){Vrba}, {Henden}, {Luginbuhl}, {Guetter},
  {Munn}, {Canzian}, {Burgasser}, {Kirkpatrick}, {Fan}, {Geballe},
  {Golimowski}, {Knapp}, {Leggett}, {Schneider}, \& {Brinkmann}}]{vrba2004}
{Vrba}, F.~J., {Henden}, A.~A., {Luginbuhl}, C.~B., {et~al.} 2004, \aj, 127,
  2948

\bibitem[{{Weiss} \& {Schlattl}(2008)}]{weiss2008}
{Weiss}, A., \& {Schlattl}, H. 2008, \apss, 316, 99

\bibitem[{{White} {et~al.}(2017){White}, {Pope}, {Antoci}, {P{\'a}pics},
  {Aerts}, {Gies}, {Gordon}, {Huber}, {Schaefer}, {Aigrain}, {Albrecht},
  {Barclay}, {Barentsen}, {Beck}, {Bedding}, {Fredslund Andersen}, {Grundahl},
  {Howell}, {Ireland}, {Murphy}, {Nielsen}, {Silva Aguirre}, \&
  {Tuthill}}]{white2017}
{White}, T.~R., {Pope}, B.~J.~S., {Antoci}, V., {et~al.} 2017, \mnras, 471,
  2882

\bibitem[{{Wilson}(1963)}]{wilson1963}
{Wilson}, O.~C. 1963, \apj, 138, 832

\bibitem[{{Wilson}(1966)}]{wilson1966}
---. 1966, \apj, 144, 695

\bibitem[{{Wilson}(1978)}]{wilson1978}
---. 1978, \apj, 226, 379

\bibitem[{{Wittenmyer} {et~al.}(2016){Wittenmyer}, {Butler}, {Tinney},
  {Horner}, {Carter}, {Wright}, {Jones}, {Bailey}, \&
  {O'Toole}}]{wittenmyer2016}
{Wittenmyer}, R.~A., {Butler}, R.~P., {Tinney}, C.~G., {et~al.} 2016, \apj,
  819, 28

\bibitem[{{Wright} {et~al.}(2004){Wright}, {Marcy}, {Butler}, \&
  {Vogt}}]{wright2004}
{Wright}, J.~T., {Marcy}, G.~W., {Butler}, R.~P., \& {Vogt}, S.~S. 2004, \apjs,
  152, 261

\bibitem[{{Y{\i}ld{\i}z} {et~al.}(2019){Y{\i}ld{\i}z}, {{\c{c}}elik Orhan}, \&
  {Kayhan}}]{yildiz2019}
{Y{\i}ld{\i}z}, M., {{\c{c}}elik Orhan}, Z., \& {Kayhan}, C. 2019, \mnras, 489,
  1753

\bibitem[{{Zechmeister} \& {K{\"u}rster}(2009)}]{zechmeister2009}
{Zechmeister}, M., \& {K{\"u}rster}, M. 2009, \aap, 496, 577

\bibitem[{{Zechmeister} \& {K{\"u}rster}(2018)}]{gls}
---. 2018, {GLS: Generalized Lomb-Scargle periodogram}, , , ascl:1807.019

\bibitem[{{Zechmeister} {et~al.}(2013){Zechmeister}, {K{\"u}rster}, {Endl}, {Lo
  Curto}, {Hartman}, {Nilsson}, {Henning}, {Hatzes}, \&
  {Cochran}}]{zechmeister2013}
{Zechmeister}, M., {K{\"u}rster}, M., {Endl}, M., {et~al.} 2013, \aap, 552, A78

\bibitem[{{Zhao} \& {Tinney}(2020)}]{zhao2020}
{Zhao}, J., \& {Tinney}, C.~G. 2020, \mnras, 491, 4131

\end{thebibliography}

\end{document}